\documentclass[manuscript]{acmart}
\usepackage{xcolor}
\usepackage{fontawesome}
\usepackage[most]{tcolorbox}
\usepackage{setspace}
\usepackage{xspace}
\usepackage{pgfplots}
\usepackage{pgfplotstable}
\usepackage{multirow}
\usepackage{colortbl}
\usepackage{framed}
\usepackage{changepage}
\usepackage{soul}
\AtBeginDocument{%
  }

\setcopyright{acmlicensed}
\copyrightyear{2026}
\acmYear{2026}

\begin{document}
\definecolor{PAblue}{RGB}{0,122,204}%
\definecolor{PAlightblue}{RGB}{235, 247, 255}%

\newcommand{\todo}[1]{{\color{orange} \bfseries TODO: #1}}

\newcommand{\cy}[1]{{\color{blue} \bfseries TODO (Chenyang): #1}}

\definecolor{mygray2}{RGB}{224, 224, 224}%

\newcommand{\taxonomyItem}[2]{{#1} - \scriptsize{\emph{#2}}}

\definecolor{softgreen}{RGB}{85, 149, 90}
\definecolor{softred}{RGB}{193, 86, 86}

\newcommand{\cmark}{\textcolor{softgreen}{\faCheck}}
\newcommand{\xmark}{\textcolor{softred}{\faTimes}}

\newcommand{\theme}[1]{\emph{#1}}
\newtcbox{\chip}[1][]{enhanced,
 box align=base,
 nobeforeafter,
 colback=mygray2,
 colframe=mygray2,
 fontupper=\scriptsize\ttfamily,
 left=0.2pt,
 right=0.2pt,
 top=0.2pt,
 bottom=0.2pt,
 boxsep=0.4pt,
 #1}

\definecolor{boxcolor}{RGB}{238, 223, 204} %
\DeclareRobustCommand{\mybox}[2][gray!20]{%
\begin{tcolorbox}[   %
        breakable,
        left=0pt,
        right=0pt,
        top=0pt,
        bottom=0pt,
        colback=#1,
        colframe=black,
        width=\dimexpr\columnwidth\relax, 
        enlarge left by=0mm,
        boxsep=5pt,
        outer arc=4pt,
        boxrule=.5mm
        ]
        #2
\end{tcolorbox}
}

\definecolor{green3}{RGB}{66,179,130}
\definecolor{green2}{RGB}{121,205,169}
\definecolor{green1}{RGB}{196,233,217}

\definecolor{red1}{RGB}{251,219,220}
\definecolor{red2}{RGB}{244,164,166}
\definecolor{red3}{RGB}{236,91,96}

\definecolor{blue1}{RGB}{101,173,246}
\definecolor{blue2}{RGB}{12,112,212}

\definecolor{orange1}{RGB}{250,181,97}
\definecolor{orange2}{RGB}{245,138,7}

\definecolor{purple1}{RGB}{195,176,232}
\definecolor{purple2}{RGB}{146,112,212}

\definecolor{pink1}{RGB}{236,152,223}
\definecolor{pink2}{RGB}{227,100,208}

\definecolor{gray1}{RGB}{220,220,220}

\def\savedbarchart#1#2#3{
\includegraphics[trim=0 0 0 0, clip, height=7.5pt]{#1}
\includegraphics[trim=0 0 0 0, clip, width=0.85\linewidth, height=6.5pt]{#2}
\includegraphics[trim=0 0 0 0, clip, height=7.5pt]{#3}
}

\def\savedlegend#1#2{
\includegraphics[trim=0 0 0 0, clip, width=0.02\linewidth, height=6.5pt]{#2}{#1}
}

\def\frequencybarchart#1#2#3#4#5#6#7#8{
\resizebox {1.21\linewidth} {6pt} {%
\begin{tikzpicture}[]
\begin{axis}[
      axis background/.style={fill=gray!30, draw=gray!30},
      axis line style={draw=none},
      tick style={draw=none},
      ytick=\empty,
      xtick=\empty,
      ymin=0, ymax=1,
      xmin=0, xmax=6]
\addplot [
      ybar interval=.5,
      fill=green3,
      draw=none,
]
	coordinates {(6*#1,1) (0,0.30)}; %
\addplot [
      ybar interval=.5,
      fill=green2,
      draw=none,
]
	coordinates {(6*(#1+#2),1) (6*#1,1)}; %
\addplot [
      ybar interval=.5,
      fill=gray1,
      draw=none,
]
	coordinates {(6*(#3+#2+#1),1) (6*(#2+#1),1)}; %
\addplot [
      ybar interval=.5,
      fill=red2,
      draw=none,
]
	coordinates {(6*(#4+#3+#2+#1),1) (6*(#3+#2+#1),1)}; %
\addplot [
      ybar interval=.5,
      fill=red3,
      draw=none,
]
	coordinates {(6*(#5+#4+#3+#2+#1),1) (6*(#4+#3+#2+#1),1)}; %
\addplot [
      ybar interval=.5,
      fill=red1,
      draw=none,
]
	coordinates {(6*(#6+#5+#4+#3+#2+#1),1) (6*(#5+#4+#3+#2+#1),1)}; %
\end{axis}%
\end{tikzpicture}%
}
}

\def\importancebarchart#1#2#3#4#5#6#7#8{
\resizebox {1.21\linewidth} {6pt} {%
\begin{tikzpicture}[]
\begin{axis}[
      axis background/.style={fill=gray!30, draw=gray!30},
      axis line style={draw=none},
      tick style={draw=none},
      ytick=\empty,
      xtick=\empty,
      ymin=0, ymax=0.70,
      xmin=0, xmax=6]
\addplot [
      ybar interval=.5,
      fill=blue2,
      draw=none,
]
	coordinates {(6*#1,1) (0,0.30)}; %
\addplot [
      ybar interval=.5,
      fill=blue1,
      draw=none,
]
	coordinates {(6*(#1+#2),1) (6*#1,1)}; %
\addplot [
      ybar interval=.5,
      fill=gray1,
      draw=none,
]
	coordinates {(6*(#3+#2+#1),1) (6*(#2+#1),1)}; %

\addplot [
      ybar interval=.5,
      fill=orange1,
      draw=none,
]
	coordinates {(6*(#4+#3+#2+#1),1) (6*(#3+#2+#1),1)}; %
\addplot [
      ybar interval=.5,
      fill=orange2,
      draw=none,
]
	coordinates {(6*(#5+#4+#3+#2+#1),1) (6*(#4+#3+#2+#1),1)}; %
\addplot [
      ybar interval=.5,
      fill=orange2,
      draw=none,
]
	coordinates {(6*(#6+#5+#4+#3+#2+#1),1) (6*(#5+#4+#3+#2+#1),1)}; %
\end{axis}%
\end{tikzpicture}%
}
}

\def\mylegend#1#2{
\resizebox {0.02\linewidth} {6.5pt} {%
\begin{tikzpicture}[]
\begin{axis}[
      axis background/.style={fill=white!30, draw=white!30},
      axis line style={draw=none},
      tick style={draw=none},
      ytick=\empty,
      xtick=\empty,
      ymin=0, ymax=0.70,
      xmin=0, xmax=6]
\addplot [
      ybar interval=.5,
      fill=#2,
      draw=none,
]
	coordinates {(4.5,1) (0,0.30)}; %
\end{axis}%
\end{tikzpicture}%
}%
#1
}

\definecolor{boxcolor}{RGB}{238, 223, 204} %
\DeclareRobustCommand{\mybox}[2][gray!20]{%
\begin{tcolorbox}[   %
        breakable,
        left=0pt,
        right=0pt,
        top=0pt,
        bottom=0pt,
        colback=#1,
        colframe=black,
        width=\dimexpr\columnwidth\relax, 
        enlarge left by=0mm,
        boxsep=5pt,
        outer arc=4pt,
        boxrule=.5mm
        ]
        #2
\end{tcolorbox}
}

\definecolor{boxcolor}{RGB}{238, 223, 204} %
\DeclareRobustCommand{\qbox}[2][white!20]{%
\begin{tcolorbox}[   %
        breakable,
        left=0pt,
        right=0pt,
        top=0pt,
        bottom=0pt,
        colback=#1,
        colframe=black,
        width=\dimexpr\columnwidth\relax, 
        enlarge left by=0mm,
        boxsep=5pt,
        outer arc=4pt,
        boxrule=.5mm
        ]
        #2
\end{tcolorbox}
}

\newcommand{\qResult}[2]{{\faClockO\xspace#1\% \faThumbsOUp\xspace#1\%}}

\newcommand{\numQuestions}[0]{50\xspace}

\newcommand{\numTasks}[0]{25\xspace}

\newcommand{\numLiteratureTools}[0]{45\xspace}

\newcommand{\numPracticeTools}[0]{27\xspace}

\newcommand{\numTools}[0]{72\xspace}

\newcommand{\numInterviewParticipants}[0]{16\xspace}

\newcommand{\numObservationalParticipants}[0]{8\xspace}

\newcommand{\numSurveyParticipants}[0]{50\xspace}

\newcommand{\op}[1]{OP#1}
\newcommand{\ip}[1]{IP#1}

\newcommand{\opquote}[2]{\emph{``#1''} (OP#2)}
\newcommand{\ipquote}[2]{\emph{``#1''} (IP#2)}

\newcommand{\ipBlockQuote}[2]{\begin{quote}\emph{"{#1}"} (IP#2)\end{quote}}

\newcommand{\opBlockQuote}[2]{\begin{quote}\emph{"{#1}"} (OP#2)\end{quote}}

\newcommand{\helpfulnessResult}[1]{$h$=#1\%}
\newcommand{\frequencyResult}[1]{$f$=#1\%}

\newcommand{\jointResult}[2]{\frequencyResult{#1}, \helpfulnessResult{#2}}

\newcommand{\icon}[1]{{\includegraphics[height=1.5\fontcharht\font`\B]{#1}}\xspace}
\newcommand{\meiicon}{\icon{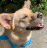}}

 \newcommand{\scenario}[1]{ 
	\def\FrameCommand{%
		\hspace{0pt}%
		{\color{PAblue}\vrule width 2pt}%
		{\color{white}\vrule width 2pt}%
		\colorbox{white}
	}%
	\MakeFramed{\advance\hsize-\width\FrameRestore}%
	\noindent\hspace{-4.55pt}%
	\begin{adjustwidth}{}{0pt}
		\emph{#1}
		\vspace{-3pt}
	\end{adjustwidth}\endMakeFramed%
}

\definecolor{gentleyellow}{HTML}{FFF2CC}
\sethlcolor{gentleyellow}
\newcommand{\highlight}[1]{\textbf{\hl{#1}}}

\title{Understanding Developer Information Needs for Software with Embedded Natural Language Prompts}

\author{Jenny T. Liang}
\email{jtliang@cs.cmu.edu}
\affiliation{%
  \institution{Carnegie Mellon University}
  \city{Pittsburgh}
  \state{Pennsylvania}
  \country{USA}
}

\author{Chenyang Yang}
\email{cyang3@cs.cmu.edu}
\affiliation{%
  \institution{Carnegie Mellon University}
  \city{Pittsburgh}
  \state{Pennsylvania}
  \country{USA}
}

\author{Agnia Sergeyuk}
\email{agnia.sergeyuk@jetbrains.com}
\affiliation{%
  \institution{JetBrains Research}
  \city{Belgrade}
  \country{Serbia}
}

\author{Travis D. Breaux}
\email{breaux@cs.cmu.edu}
\affiliation{%
  \institution{Carnegie Mellon University}
  \city{Pittsburgh}
  \state{Pennsylvania}
  \country{USA}
}

\author{Brad A. Myers}
\email{bam@cs.cmu.edu}
\affiliation{%
  \institution{Carnegie Mellon University}
  \city{Pittsburgh}
  \state{Pennsylvania}
  \country{USA}
}

\renewcommand{\shortauthors}{Liang et al.}

\begin{abstract}
Developers use foundation models (FMs) to create novel forms of intelligent software by writing natural language prompts embedded within a software's implementation---referred to as \emph{embedded prompts}---that are also passed as a string within an FM's API.
Developing software with embedded prompts requires developers to engage in iterative embedded prompt changes to obtain the desired behavior.
While the information needs of traditional software have been investigated, the information needs of modern developers for embedded prompt changes are underexplored, despite these needs shaping how developers plan and justify embedded prompt changes.
In this work, we perform a mixed-method exploratory study on the information needs of developers making changes to embedded prompts. 
We create, refine, validate, and apply a list of \numQuestions information needs that developers have while making a change to an embedded prompt by interviewing \numInterviewParticipants developers, observing \numObservationalParticipants developers make prompt changes, surveying \numSurveyParticipants developers, and evaluating \numTools research and commercial tools.
We find that embedded prompts are not well-supported by existing tools: 16 of the \numQuestions information needs, including four of the most important ones, are unaddressed.
We conclude with important design opportunities that should be addressed by future developer tools for embedded prompts.
\end{abstract}



\received{20 February 2007}
\received[revised]{12 March 2009}
\received[accepted]{5 June 2009}

\maketitle

\vspace{2em}
\section{Introduction}
\scenario{
Peter is a software developer. 
He is building a large language model (LLM) chatbot agent in Python that helps customers book travel reservations.
As the first step of building the agent, Peter must develop a natural language prompt that he will pass into OpenAI's LLM application programming interface (API).
To begin, Peter defines general purpose instructions to specify the chatbot's behavior in the prompt, which is a string: \texttt{"You are a friendly and helpful travel agent. You will help work with customers to book hotels and airplane tickets."}
To ensure the agent can personalize its responses to the customer, Peter also interpolates a list of the customer's information within the prompt that is pulled from an external database of customer data, using string concatenation.
He loads this information as a string in a variable called \texttt{customer\_info} from the code: \texttt{"Here is information about the customer: \{str(customer\_info)\}"}.
Peter uses the result from OpenAI's API to render the results on the chatbot's user interface.
}

\begin{figure}[t!]
    \begin{tcolorbox}[
    left=-14pt,right=2pt,top=2pt,bottom=2pt]
    {\sc \hspace{16pt}Definitions for Embedded Prompt Concepts}
    \begin{itemize}
        \item \emph{Prompt content}: The textual string that is the embedded prompt.
        \item \emph{Prompt parameters}: Placeholder variables that are part of the embedded prompt string. The variables are bound to specific values at runtime and can be computed from code.
        \item \emph{Prompt behavior}: A quantitative or qualitative description of the resulting output that is generated by the FM. Prompt behavior can be evaluated using different sub-criteria~\cite{kim2024evallm, joshi2025coprompter}, such as wordiness or safety.
        \item \emph{Prompt component}: The distinct structural sub-components that together compose an embedded prompt (e.g., examples, output formatting instructions)~\cite{mao2025prompts, tafreshipour2025prompting}.
        \item \emph{Run}: A single execution of the embedded prompt using one set of prompt input values.
        \item \emph{Run configuration:} Parameters to the FM's API other than the embedded prompt string.
        \item \emph{Prompt version}: A specific iteration of an embedded prompt string that captures the state of its content at a particular point in time. The same version can be run with different input values.
    \end{itemize}
    \end{tcolorbox}
    \vspace{-0.5\baselineskip}
    \caption{Definitions of important concepts for building software with embedded prompts that we developed throughout the course of the study.}
    \label{fig:definitions}
\end{figure}

Many developers like Peter are using foundation models (FMs), such as Anthropic's Claude or OpenAI's GPT-5, to create novel forms of intelligent software.
Powerful AI capabilities can now be leveraged within software just by writing natural language prompts embedded within the implementation.
In practice, this often requires developers like Peter to pass a prompt as a string through an FM's APIs.
In this work, we refer to prompts embedded \textit{within} the software implementation as \emph{embedded prompts}, which are defined as: \emph{``A prompt that accepts variable inputs and could be interpreted by a foundation model (FM) to perform specified actions and/or generate output. This prompt is executed within a software application or code by a FM.''}~\cite{liang2025prompts}; we refer to the process of creating such software as \emph{prompt programming}.
The artifacts important to developing embedded prompts are included in Figure~\ref{fig:definitions}.
We note that other works use various equivalent terms for embedded prompts such as \emph{promptware}~\cite{chen2026promptware}, \emph{AIware}~\cite{lu2026engineering}, \emph{FMware}~\cite{hassan2024rethinking}, \emph{prompt programs}~\cite{fiannaca2023programming, liang2025prompts, jiang2022promptmaker}, \emph{LLMapps}~\cite{mao2025prompts}, and \emph{AI chains}~\cite{wu2022ai, wu2022promptchainer, cheng2024prompt}.
Embedded prompts power software with a range of capabilities, from IDE-integrated coding agents~\cite{cursur2025cursor, liang2025tabletalk} to chatbots that synthesize research~\cite{chatgpt2025scholargpt}.
With increased accessibility and reduced labor of writing prompts compared to fine-tuning models---the latter of which requires expertise in machine learning, software engineering, data science, and algorithms~\cite{giray2021software, wan2019does}---software with embedded prompts is increasingly popular in practice~\cite{liang2025prompts, parnin2025building, nahar2024beyond, dolata2024development}.
A 2025 report from GitHub indicates the creation of 700,000 new GitHub projects using LLM APIs---a 178\% increase from the year prior~\cite{github2026octoverse}.

Developing software with embedded prompts shifts the development process to become a continuous experimentation loop by requiring the developer to make iterative prompt \emph{changes} to obtain the desired behavior, such as by updating the prompt directly or by modifying the code that produces the prompt~\cite{liang2022understanding}.
In traditional software development, making changes requires developers to ask questions to gather and make sense of the prerequisite information to make a change~\cite{sillito2006questions, sillito2008asking}.
While the information needs of software developers have been studied in traditional software evolution contexts~\cite{ko2010extracting, ko2008debugging, sillito2006questions, sillito2008asking, latoza2010hard, ko2007information}, the information needs of modern developers for embedded prompt changes are underexplored, despite these needs shaping how developers plan and justify embedded prompt changes.

In addition, the development of traditional software is substantially different from the development of software with embedded prompts. 
Embedded prompts are primarily expressed as long strings, rather than pure code, and the underlying FM is opaque and produces non-deterministic outputs~\cite{liang2025prompts, hassan2024rethinking, chen2026promptware}.
These characteristics make the process of developing embedded prompts distinct from traditional software engineering~\cite{liang2025prompts, parnin2025building, nahar2024beyond}.
We therefore argue that developers likewise have distinct information needs when making changes to software with embedded prompts---in fact, we observe that developers interact with different types of artifacts (see Figure~\ref{fig:definitions}) that are not present in traditional software.
Thus, developer tools should be designed to address the distinct information needs required for software with embedded prompts.
However, with the growing number of developer tools for embedded prompts across academia (e.g.,~\cite{arawjo2023chainforge, mishra2025promptaid, wu2022ai, wu2022promptchainer}) and industry (e.g.,~\cite{github2026promptfoo, github2026langsmith, github2026promptflow}), it is unclear how well these tools are designed to address developer information needs for building software with embedded prompts.
For instance, one type of popular embedded prompt development tool is a prompt playground, whose focus is to enable developers to quickly iterate on a prompt's content and view the prompt's raw textual outputs.
Playgrounds thus may not fully address developer information needs regarding prompt behavior due to the lack of systematic tracking of qualitative descriptions of prompt behaviors.
Therefore, understanding developer information needs is necessary for designing and evaluating usable developer tools~\cite{latoza2010hard}, and can assess current practice and highlight opportunities to improve tool support and developer productivity.
Further, important information needs for building software with embedded prompts are likely to persist over time, even as FM capabilities improve.
So long as humans are involved in developing software with embedded prompts, they will still need to decide on the software's requirements, evaluate to what extent the FM's outputs matches their intended requirements, and understand what actions to take when it does not.

In this work, we performed a mixed-method exploratory study on the information needs of developers making changes to embedded prompts.
Because developing embedded prompts involves many potential information needs, we studied which information needs are both important and unaddressed to help tool builders understand which ones to prioritize.
Therefore, we focused on the following research questions:
\begin{description}
\setlength{\itemsep}{0pt}
\setlength{\parskip}{2pt}
    \item[\textbf{RQ1:}] What information needs do developers have while changing embedded prompts? How do developers address them?
    \item[\textbf{RQ2:}] What information needs do developers consider as the most important for making changes to embedded prompts?
    \item[\textbf{RQ3:}] To what extent do existing tools address the information needs of developers while making changes to embedded prompts?
\end{description}
The aim of this work is not to cover all possible information needs or identify globally important needs, but rather identify sufficiently important information needs that could be addressed in tools to better support the development of software with embedded prompts.
Similarly, our investigation of existing tools aims to understand broad trends in support for developer needs, rather than perform a comprehensive survey of all tools.

In this work, we interviewed \numInterviewParticipants developers, observed \numObservationalParticipants developers change prompts, and surveyed \numSurveyParticipants developers, to create, refine, and validate a list of \numQuestions information needs that developers have while making a change to an embedded prompt (\textbf{RQ1}), and understand which of these needs are most important (\textbf{RQ2}).
Finally, we apply the list of information needs to tool features from \numTools research and commercial tools in order to evaluate these tools' abilities to address developer needs (\textbf{RQ3}).
This work builds upon our previous work~\cite{liang2025prompts}---which presents a definition for embedded prompts and examines their development process---by elucidating the specific developer information needs within that process, presenting finidngs from previously unanalyzed and unreported portions from~\cite{liang2025prompts}'s data and three additional studies.
Overall, we found that developing software with embedded prompts is not well-supported by existing tools for embedded prompts.
16 of the \numQuestions information needs, including four of the most important ones, remained unaddressed.
Further, even when tool support existed, widely used tools appeared to inadequately address developers’ information needs, as most developers addressed these needs manually rather than with tool assistance.
Based on this, we highlight the most important opportunities for future developer tools for embedded prompts.
We hope that these developer information needs will be addressed by future tools.
However, we recognize that several of these needs may never be completely addressed, as they are inherent to developing embedded prompts due to the opaque and stochastic nature of FMs~\cite{liang2025prompts} (e.g., understanding the behavior of FMs despite its outputs potentially varying between runs, even on the same inputs).

\section{Related Work}
In this section, we discuss related work about developing embedded prompts (Section~\ref{sec:prompt-programming}) as well as questions that developers ask while building software at large (Section~\ref{sec:developer-questions}).

\subsection{Developing Software Containing Embedded Prompts}
\label{sec:prompt-programming}
The software engineering community has increasingly recognized natural language prompts as important artifacts in software~\cite{villamizar2025prompts, vogelsang2024specifications, tafreshipour2025prompting, pister2024promptset}.
The development of software containing embedded prompts has been discussed in prior work under various terms, such as \emph{promptware}~\cite{chen2026promptware}, \emph{AIware}~\cite{lu2026engineering}, \emph{FMware}~\cite{hassan2024rethinking}, \emph{prompt programs}~\cite{fiannaca2023programming, liang2025prompts, jiang2022promptmaker}, \emph{LLMapps}~\cite{mao2025prompts}, and \emph{AI chains}~\cite{wu2022ai, wu2022promptchainer, cheng2024prompt}.
Compared to traditional software, which relies only on largely deterministic code-based components, software with embedded prompts differs in relying on components with natural language and non-deterministic outputs from FMs~\cite{liang2025prompts, chen2026promptware}.
The shift towards natural language prompts has placed a renewed emphasis on the specification of requirements~\cite{vogelsang2025impact, vogelsang2024specifications}.
Defining natural language specifications, however, can be difficult, since requirements are often underspecified and can conflict~\cite{tafreshipour2025prompting, yang2026prompts}.

Given the growing interest in supporting the development of embedded prompts, researchers have begun to study these artifacts themselves.
\citet{pister2024promptset} introduced PromptSet, a dataset of mined embedded prompts from GitHub.
Subsequent analyses of this data revealed that prompts tend to have a common structure, with shared components such as instructions for output formatting or providing examples for few-shot learning~\cite{tafreshipour2025prompting, mao2025prompts}.
\citet{mao2025prompts} found that embedded prompt components tend to follow a consistent ordering of first defining a role for the FM, defining instructions for the FM, providing additional context for the task, defining output formats, and then concluding with examples.

The existing literature characterizes prompt programming with embedded prompts as a trial-and-error process that requires many iterations~\cite{liang2025prompts, zamfirescu2023johnny, parnin2025building, dolata2024development, villamizar2025prompts}.
Importantly, researchers have begun identifying the systematic software engineering practices to build software with embedded prompts.
An empirical study from \citet{parnin2025building} of industry practitioners  found that software teams follow a process of exploration, implementation, evaluation, and productization.
In an interview study with 20 developers building embedded prompts, \citet{liang2025prompts} builds upon these insights by identifying that developing embedded prompts contains key software engineering activities such as requirements, design, implementation, debugging, and testing.
However, developing software with embedded prompts is challenging. 
Developers face numerous barriers such as evolving evaluation metrics, understanding best practices, and setting realistic product expectations~\cite{parnin2025building, nahar2024beyond, liang2025prompts, dolata2024development}.

A separate line of work has proposed several tools for developing embedded prompts.
Many of these tools focus on the end-to-end pipeline development of embedded prompts.
While early tools, such as PromptMaker~\cite{jiang2022promptmaker} and prototype prompt editors~\cite{fiannaca2023programming}, focused on basic editing of prompts, more recent works have focused on providing more substantial support for developing embedded prompts.
One such example is ChainForge~\cite{arawjo2024chainforge}, which
enables developers to systematically test and run their prompt across different models by enabling various interfaces such as ones for datasets, prompt templates, and metrics.
It also supports systematic comparison of prompt outputs in a table view.
Other examples include PromptAid~\cite{mishra2025promptaid} and PromptIDE~\cite{strobelt2022interactive}.
These tools enable developers to more systematically optimize prompts by automatically perturbing the prompt~\cite{mishra2025promptaid} or enabling the evaluation of multiple different prompts at the same time~\cite{strobelt2022interactive}, reducing the cognitive overhead of making prompt updates.
We note that end-user programming interfaces, such as node-based~\cite{arawjo2024chainforge, wu2022promptchainer, wu2022ai, liang2026understanding} and block-based~\cite{cheng2024prompt} notations, are a common approach for constructing embedded prompt pipelines.
In addition, some developer tools have focused on providing more specialized support of embedded prompt development.
Notable areas of support are the systematic evaluation as well as the reuse of prompts: CoPrompter~\cite{joshi2025coprompter} and EvalLM~\cite{kim2024evallm} use FMs to generate criteria to assess the prompt that are also scored using FMs, while \citet{white2023prompt, white2024chatgpt} propose reusable prompt patterns for specific tasks.
Researchers have also developed tools supporting embedded prompts for specialized domains, such as WorldSmith for fictional world-building~\cite{dang2023worldsmith}, PromptHive for content generation for subject matter expertise~\cite{reza2025prompthive}, and PromptCharm for image generation~\cite{wang2024promptcharm}.
Finally, with the rising popularity of agents powered by FMs, a new suite of tools has emerged for agent development to support tasks such as debugging~\cite{epperson2025interactive} and prompt prototyping~\cite{liang2026understanding}.

We build upon these works by enumerating the specific developer information needs while developing prompts, since existing works primarily focus on the general process of embedded prompt development (e.g.,~\cite{liang2025prompts, parnin2025building, nahar2024beyond}).
We also evaluated the aforementioned tools---including end-user programming interfaces (e.g.,~\cite{arawjo2024chainforge, wu2022promptchainer, wu2022ai}) and specialized developer tools for embedded prompts (e.g.,~\cite{white2023prompt, white2024chatgpt, joshi2025coprompter, kim2024evallm})---as well as other additional tools in Section~\ref{sec:rq3} on their coverage of the information needs enumerated in Section~\ref{sec:rq1} to better understand which needs remain under-supported.

\subsection{Developer Information Needs}
\label{sec:developer-questions}
Prior work has studied the information needs of software developers in traditional software contexts.
These studies find that developers must not only acquire information on the software system, such as writing code, triaging bugs, reasoning about design, and understanding upstream and downstream dependencies~\cite{ko2007information, breu2010information, haenni2013categorizing}, they must also maintain awareness of social aspects of software development.
Examples of social aspects include keeping up-to-date with a colleague's work~\cite{ko2007information}, understanding an end-user's intent in bug reports~\cite{breu2010information}, and becoming aware of the popularity of software libraries~\cite{haenni2013categorizing}.
With more modern sources of programming support, the information needs of developers have also changed.
An analysis of Stack Overflow search sessions by~\citet{liu2021characterizing} found that most search sessions involved searching for specific code snippets.
In the age of generative AI, \citet{al2025evolution}'s mixed-method study involving 145 developers revealed that developers used FMs to gather information on best practices, find new libraries, identify trade-offs in implementations, recall previous knowledge, understand code, and identify keywords for online search.

Another line of related research involves understanding the questions that developers ask.
One category includes an analysis of public discussions of programming information sources.
In an early analysis of public Usenet discussions, \citet{johnson1997interactive} found that developer questions were decomposed into three types: goal oriented questions to achieve task specific goals, symptom oriented questions on defective program behavior, and system oriented questions to identify system objects and functions.
More modern works have analyzed developer questions on Stack Overflow~\cite{allamanis2013and} and ChatGPT~\cite{xiao2024devgpt,chavan2024analyzing}, finding that developers use these sources to inquire about whether and why existing code does or does not work, how to implement a feature or complete a task, and how to use and learn new languages or technologies.

Other work has elicited developer questions through empirical studies of developers.
Key studies were performed by \citet{sillito2006questions, sillito2008asking}, enumerating the information needs for changing code-based software.
The authors identified 44 questions that developers ask while changing code, such as what the execution path of code was or where to find code samples.
They found that developer information needs for changing code were split into two types.
One type of information needs focused on identifying focus points of code that are relevant to the task. 
The other type centered on expanding on the focus points, often by examining relationships between entities.
Meanwhile, \citet{latoza2010hard} surveyed 179 developers and identified 94 hard-to-answer programming questions involving code changes, code elements, and their relationships.
\citet{duala2012asking} observed 20 developers working with unfamiliar APIs, eliciting 20 questions.
These included which packages were relevant to the task, how to create specific types of objects, and whether there is an API that provides certain functionality.

We extend this body of literature to understand the information needs of developers building software with generative AI components rather than traditional software, and identify which information needs are addressed by existing tooling.
By comparing these information needs with the ones identified in prior literature (see Section~\ref{sec:info-needs-software-vs-embedded-prompts}), we also outline how information needs have changed for the development of embedded prompts.

\begin{figure}[t!]
\centering
\includegraphics[trim=0 625 0 0, clip, width=0.9\linewidth, page=1, keepaspectratio]{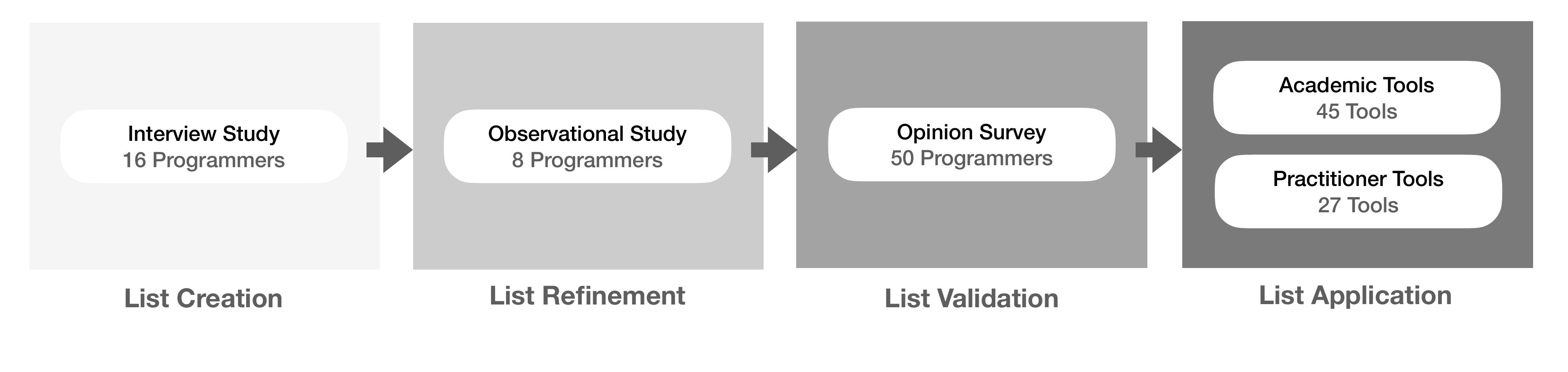} 
\caption{
An overview of the methodology used in this study.
To develop a list of information needs of embedded prompts, we conduct interviews with \numInterviewParticipants developers, observe \numObservationalParticipants developers make prompt changes, and survey \numSurveyParticipants developers.
Then, we apply the list to \numTools tools from the literature and practice.
}
\label{fig:methodology}
\end{figure}

\section{Methodology}
We employed a four-phase methodology where the list of information needs is created with \numInterviewParticipants developers (Section~\ref{sec:methodology-phase1}), refined with \numObservationalParticipants developers (Section~\ref{sec:methodology-phase2}), validated with \numSurveyParticipants developers (Section~\ref{sec:methodology-phase3}), and applied to \numTools tools (Section~\ref{sec:methodology-phase4}).
Figure~\ref{fig:methodology} summarizes the study methodology.
We collected the information needs (\textbf{RQ1}) through interviews and refined them in an observational study of developers making embedded prompt changes.
This process resulted in a reusable list of information needs for developing embedded prompts.
We then validated and quantified the importance of the information needs (\textbf{RQ2}) in a developer survey, which also helped generalize findings to a broader developer population.
Finally, we applied the list of tasks and questions to \numTools commercial and research tools to assess coverage of the information needs (\textbf{RQ3}).

Across all developer studies, a common inclusion criterion was developers who had developed software containing embedded prompts that met \citet{liang2025prompts}'s definition of an embedded prompt.
All study procedures were approved by our institution's institutional review board and were piloted on multiple developers for quality.
Following best practices for developer user studies~\cite{ko2015practical}, we piloted each protocol with three developers who were building software with embedded prompts to assess the protocol's clarity and duration.
After each pilot, we refined the protocol instructions and information needs wording based on participant feedback.
All study materials (e.g., interview scripts, surveys, codebook, tool list) are included within the supplemental materials~\cite{supplemental-materials}.
We detail the methodology below.

\begin{table*}
  \centering \small
\caption{
An overview of the participants in the first phase of the study to develop the initial list of developer information needs. 
We report the participant's development context, number of years of programming experience, number of prompt programs developed prior to the interview, prompt structure, model type, application domain, and task.
Prompt structures can be a single prompt, multiple chained prompts (multi-prompt), or a language agent (agent).
Model types can be open source (\chip{OS}) or closed source (\chip{CS}) and can be a language model (\chip{LM}) or a vision-language model (\chip{VLM}).
}

\label{tab:participants-interview}
\begin{tabular}{p{0.02\linewidth}p{0.12\linewidth}p{0.06\linewidth}p{0.09\linewidth}p{0.12\linewidth}p{0.07\linewidth}p{0.145\linewidth}p{0.175\linewidth}}
\toprule
 \textbf{ID} & \textbf{Context} & \textbf{Exp.} & \textbf{\# Prompts} & \textbf{Structure} & \textbf{Model} & \textbf{App Domain} & \textbf{Task} \\
\midrule
IP1 & Academia & 16 yrs & 30 &  Single prompt & \chip{OS} \chip{LM} & Security & Classification \\
\addlinespace[1pt]
{IP2} & \parbox[t][2em][t]{0.12\linewidth}{\setstretch{0.8} Personal project} & {10 yrs} & {40} & {Single prompt} & {\chip{CS} \chip{LM}} & {Productivity} & \parbox[t][2em][t]{2.5cm}{\setstretch{0.8} Text summarization (chatbot)} \\
\addlinespace[1pt]
IP3 & Academia & 8 yrs & 30 &  Single prompt & \chip{CS} \chip{LM} & Research analysis & Classification \\
\addlinespace[1pt]
IP4 & Big tech (R\&D) & 8 yrs & 10 &  Single prompt & \chip{OS} \chip{LM} & Software testing & Code generation \\
\addlinespace[1pt]
IP5 & Academia & 14 yrs & 35 &  Single prompt & \chip{CS} \chip{LM} & Education & Generation \\
\addlinespace[1pt]
IP6 & Academia & 10 yrs & 3 &  Multi-prompt & \chip{CS} \chip{LM} & Software testing & Code generation \\
\addlinespace[1pt]
IP7 & Big tech (Eng) & 10 yrs & 4 &  Single prompt & \chip{CS} \chip{LM} & Education & Generation \\
\addlinespace[1pt]
IP8 & Academia & 10 yrs & 20 &  Agent & \chip{CS} \chip{LM}& Software testing & Code generation \\
\addlinespace[1pt]
IP9 & Academia & 14 yrs & 20 &  Multi-prompt & \chip{OS} \chip{VLM} & Visual Q\&A & Retrieval, Q\&A \\
\addlinespace[1pt] 
IP10 & Big tech (Eng) & 7 yrs & 10 &  Multi-prompt & \chip{CS} \chip{LM} & Education & Text editing \\
\addlinespace[1pt]
IP11 & Big tech (R\&D) & 8 yrs & 5 &  Agent & \chip{CS} \chip{LM} & Productivity & Data generation \\
\addlinespace[1pt]
IP12 & Big tech (Eng) & 14 yrs & 100+ &  Multi-prompt & \chip{CS} \chip{LM} & Productivity & Code summarization \\
\addlinespace[1pt]
IP13 & Big tech (Eng) & 5 yrs & 3 &  Single prompt & \chip{CS} \chip{LM} & Productivity & Code generation \\
\addlinespace[1pt]
IP14 & Big tech (Eng) &  5 yrs & 3 &  Single prompt & \chip{CS} \chip{LM} & Career & Q\&A (chatbot) \\
\addlinespace[1pt]
{IP15} &  \parbox[t][2em][t]{1.6cm}{\setstretch{0.8} Startup (Eng), OSS} & {3.5 yrs} & {15} & {Agent} & {\chip{CS} \chip{LM}} & {Software testing} & {Test generation} \\
\addlinespace[1pt]
IP16 & Freelance & 9 yrs & 1 &  Single prompt & \chip{CS} \chip{LM} & Literature & Q\&A (chatbot) \\
\bottomrule
\end{tabular}
\end{table*}

\subsection{Phase 1: Creating the List of Information Needs for Developing Embedded Prompts}
\label{sec:methodology-phase1}
To answer \textbf{RQ1}, we developed a preliminary list of developer information needs for embedded prompts via 60-minute interviews with \numInterviewParticipants developers.
Participant interviews were conducted between February and August 2024 as part of our prior study~\cite{liang2025prompts}.
This work analyzes portions of those interviews that were not analyzed in the prior study.

\subsubsection{Sampling Strategy}
We recruited an initial sample of embedded prompt developers via snowball and convenience sampling in the authors' social networks.
Since the authors had experience working with this population of developers, their social networks included a mix of novice and experienced developers.
For this phase, the inclusion criteria required that developers had created at least one embedded prompt meeting our definition~\cite{liang2025prompts}.
After conducting six interviews, we performed maximum variation sampling~\cite{suri2011purposeful} to capture diverse experiences.
We performed sampling based on the FMs used, application domain, organization size, programming context, role, and prompt structure.

\subsubsection{Participants}
An overview of the participants is found in Table~\ref{tab:participants-interview}.
The \numInterviewParticipants interview participants (labeled IP) were experienced developers ($\mu=9.5$ years of programming experience) who were graduate students ($N=8$), software engineers ($N=7$), and researchers ($N=1$).
Participants developed embedded prompts in large technology companies ($N=7$), academia ($N=6$), personal projects ($N=1$), freelance work ($N=1$), startups ($N=1$), and open-source software ($N=1$).
They used FMs all the time ($N=8$), often ($N=4$), and sometimes ($N=4$) and wrote 1 to 5 ($N=6$), 6 to 10 ($N=2$), 11 to 20 ($N=3$), 21 to 50 ($N=4$), and 50+ ($N=1$) embedded prompts.
15 of 16 participants developed embedded prompts within a code repository.

\begin{figure}[t!]
    \begin{tcolorbox}[
    left=-14pt,right=2pt,top=2pt,bottom=2pt]
    {\sc \hspace{16pt}Interview Questions}
    \begin{itemize}
        \item What challenges do you face while trying to create a new version of a prompt?
        \item Are the existing tools you use sufficient for you while trying to create a new version a prompt? Why or why not?
        \item Suppose you had a magical system that perfectly recorded every prompt you tried and the model outputs. What would you want this system to do to help you develop a prompt?
        \item What search queries would you ask this system retrieve prompts or outputs?
    \end{itemize}
    \end{tcolorbox}
    \vspace{-0.5\baselineskip}
    \caption{A subset of the interview questions. The complete interview protocol is in the supplemental materials~\cite{supplemental-materials}.}
    \label{fig:interview}
\end{figure}

\subsubsection{Protocol}
We designed the interviews to elicit questions both within and beyond current practices to improve generalizability beyond existing tooling limitations.
Therefore, the protocol covered both the participants' present experiences in making prompt changes and potential future practices with improved tooling.
Interviews were conducted remotely on Zoom, recorded, and transcribed. 
Recordings were deleted after transcription.
Participants were compensated with a \$20 USD Amazon gift card.

To begin, participants completed a survey on their background and demographics and then discussed in depth a prompt that they recently developed (the synthesis of which is documented in~\cite{liang2025prompts}).
If a participant had access to any relevant artifacts (e.g., the prompt or its history), these items were requested.
This approach grounded the questions elicited from the interview.

The final 15 minutes of the interview were dedicated to making prompt changes to elicit specific information needs of developing embedded prompts as the basis to answer \textbf{RQ1}.
First, the participant discussed challenges in making prompt changes and whether current tools were sufficient to support updating prompts.
Then, the interviewer followed \citet{kery2021designing}'s \emph{grounded brainstorming} procedure~\cite{holtzblatt1997contextual}, asking participants to envision a magical system that perfectly tracks all prompting versions and artifacts.
Participants then brainstormed how such a system could support their embedded prompt development and what questions they would ask it to retrieve artifacts.
An overview of the interview questions is given in Figure~\ref{fig:interview}; the full interview protocol is in the supplemental materials~\cite{supplemental-materials}.

\subsubsection{Analysis}
The first two authors reviewed transcripts and identified 141 statements on questions that developers asked during prompt changes.
Qualitative analysis on this data was conducted in two rounds.
In the first round, the authors independently generated codes by performing open coding~\cite{saldana2009coding} to inductively generate codes related to developer information needs while making changes to embedded prompts.
The authors applied at least one code for each sample and gave each code a name and a brief description.

After performing the first round of coding, the two authors met to merge the independently developed codebooks.
The merged codebook was then reapplied to the dataset for a second round of coding.
We follow \citet{hammer2014confusing}'s open coding recommendations, which outlines procedures for reporting coding disagreements.
In particular, we verified the reliability of the coding by resolving disagreements through first discussion and only applied a code upon achieving unanimous agreement from the two authors.
We also interpreted coding disagreements as variance and reported the content of the disagreements.
Because the coders resolved disagreements---which resulted in 100\% agreement---inter-rater reliability metrics were not computed.
Most coding disagreements involved differences in the interpretation of participants' statements or not recalling a specific code, rather than the code's scope.

After gathering the initial codes, the two authors further grouped the codes through axial coding~\cite{saldana2009coding, corbin2015basics}.
The two authors collaboratively identified information needs related to similar development tasks and put them in a group based on unanimous agreement.
After obtaining the development tasks, the two authors performed a similar procedure to group the tasks into themes based on the entity central to the information need.
This procedure resulted in an initial set of 41 specific information needs split across 22 tasks and eight themes.

\subsection{Phase 2: Refining the List of Information Needs for Developing Embedded Prompts}
\label{sec:methodology-phase2}
We sought to further augment the findings on information needs by triangulating our data with more recent data on how developers actually update embedded prompts in practice.
We therefore refined the list of information needs by extracting information needs by observing how developers change an embedded prompt in their personal development work. 
We also obtained feedback on the initial list of information needs in a 90-minute observational study. 
The observational study sessions were conducted between April to May 2025.

\subsubsection{Sampling Strategy}
We recruited \numObservationalParticipants study participants through purposive and snowball sampling.
For this study, the inclusion criteria required developers to work on open-ended software development project and actively change embedded prompts that met \citet{liang2025prompts}'s definition.
Our initial sample was recruited from a 16-week graduate-level prompt engineering course run by our institution's software engineering graduate program.
This course covered FM design, output hazards (e.g., hallucinations, bias), and prompting techniques (e.g., chain-of-thought).
At the end of the course, the students completed an open-ended final project where they developed a software application with an embedded prompt.
We recruited participants for this study when students were actively working on the course's final project.

\begin{table*}
  \centering \small
\caption{
An overview of the participants in the second phase of the study to refine the list of developer information needs.
We report the participant's general and professional programming experience, number of prompt programs developed prior to the study, prompt structure, model type, application domain, and task.
Prompt structures can be a single prompt, multiple chained prompts (multi-prompt), or a language agent (agent).
Model types can be open source (\chip{OS}) or closed source (\chip{CS}) and can be a language model (\chip{LM}) or a vision-language model (\chip{VLM}).
}
\label{tab:participants-observational}
\begin{tabular}{p{0.02\linewidth}p{0.1\linewidth}p{0.08\linewidth}p{0.09\linewidth}p{0.12\linewidth}p{0.07\linewidth}p{0.145\linewidth}p{0.17\linewidth}}
\toprule
 \textbf{ID} & \textbf{Prog. Exp.} & \textbf{Prof. Exp.} & \textbf{\# Prompts} & \textbf{Structure} & \textbf{Model} & \textbf{App Domain} & \textbf{Task} \\
\midrule
OP1 & 10 yrs & 5 yrs & 4 & Agent & \chip{CS} \chip{VLM} & Security & Classification \\
\addlinespace[1pt]
OP2 & 11 yrs & 7 yrs & 100+ & Agent & \chip{CS} \chip{LM} & Software eng. & Code refactoring \\
\addlinespace[1pt]
OP3 & 6 yrs & 3 yrs & 1 & Agent & \chip{CS} \chip{LM} & Software eng. & Code generation \\
\addlinespace[1pt]
OP4 & 25 yrs & 0 yrs & 25 & Multi-prompt & \chip{CS} \chip{LM} & FM reasoning & Generation \\
\addlinespace[1pt]
OP5 & 5 yrs & 1 yr & 4 & Single prompt & \chip{CS} \chip{LM} & Astrology & Generation \\
\addlinespace[1pt]
OP6 & 5 yrs & 4--5 yrs & 100+ & Single prompt & \chip{CS} \chip{LM} & Education & Q\&A (chatbot) \\
\addlinespace[1pt]
OP7 & 8--9 yrs & 3 yrs & 10 & Single prompt & \chip{CS} \chip{LM} & Finance & Classification \\
\addlinespace[1pt]
OP8 & 8 yrs & 4 yrs & 5 & Agent & \chip{CS} \chip{VLM} & Gaming & Control \\
\bottomrule
\end{tabular}
\end{table*}

\subsubsection{Participants}
The \numObservationalParticipants participants were Masters ($N=6$) and Ph.D. ($N=2$) students (see Table~\ref{tab:participants-observational}). 
Study participants had general ($\mu=9.7$ years) and professional ($\mu=3.4$ years) programming experience.
Participants used FMs often ($N=2$) or always ($N=6$), and had developed 1 to 5 ($N=4$), 6 to 10 ($N=1$), 21 to 50 ($N=1$), and 50+ ($N=2$) embedded prompts.
Six of the participants developed embedded prompts within a code repository.
We note that one participant with the IDs of \ip{4} and \op{8} overlapped between the interview and observational studies.
In addition, we note that \op{4} was a Director of Product Management at a large software company and therefore had substantive general programming experience, but limited professional programming experience.

\subsubsection{Protocol}
The observational study sessions were conducted remotely on Zoom, recorded, and transcribed. 
Study participants were compensated with a \$30 USD Amazon gift card.
Each session started with observing the participant making embedded prompt changes (60 minutes) and concluded with obtaining feedback on the list of information needs for embedded prompts (30 minutes).

To begin the study session, participants described their background and demographics in a survey and verbally described their current software project with the embedded prompt.
Next, study participants shared their screen to show their natural software development environment.
They then changed an embedded prompt for an open-ended personal project and were prompted to think aloud as they made modifications.
Study participants were given a maximum of 60 minutes to make prompt changes.
The duration of the participants' tasks ranged from 38 to 59 minutes, with an average duration of 47 minutes.

Next, participants provided feedback on the initial list of information needs through a Qualtrics survey.
The survey aimed to assess the importance of information needs and elicit missing ones.
We organized the survey into sections based on the list's themes.
In each section, participants rated each individual task and information needs. 
They could also suggest missing tasks or information needs through write-in suggestions. 
The order of the sections was shuffled to minimize ordering effects.
To ensure that study participants understood the list of information needs, the interviewer began with a presentation prior to administering the survey to define important concepts in embedded prompt development utilized by the list.
In addition, for each section, the interviewer presented an overview of the theme before participants answered any survey questions.
Since the interviewer presented the list of information needs, they were also present as the participant completed the survey.

The importance of a task or information need was measured with Likert scales for two constructs defined by \citet{myers2016programmers}: the \textit{frequency} of the information need to represent importance in current practice and the \textit{helpfulness} if the information need were addressed to represent importance for tasks and questions not addressed by existing tools.
To obtain coverage of the initial list of information needs while reducing participant fatigue, study participants first rated each task by the two dimensions.
If a task was rated as \emph{``always/often''} or \emph{``very/extremely helpful''} its corresponding information needs were then rated by the study participant.

\subsubsection{Analysis}
We extended the initial list of information needs by deductively applying the codebook to observational and survey data through provisional coding~\cite{saldana2009coding}.
The first two authors who developed the initial list used the list as the starting codes to then later refine.
The first author, who conducted the interviews, coded the video footage using transcripts and noting on-screen behavior.
Due to privacy concerns of the study participants, only the first author viewed and analyzed the video recordings.
Each question code was labeled with a starting timestamp, ending timestamp, and a description of the developer's behavior.
Next, from the survey data, we extracted and independently coded three write-in suggestions of information needs from study participants. 
After identifying potential missing information needs from the initial list from the observational study and survey, the two study authors convened to discuss these needs and only added them to the list if they achieved unanimous agreement.
Consistent with the prior study in Section~\ref{sec:methodology-phase1}, the two authors independently re-applied the final codebook to the write-in suggestions, and met again to resolve disagreements through discussion and unanimous agreement.
In the end, this process resulted in nine new information needs spread across three tasks.
We achieved code saturation at \op{7}, indicating that the list had captured the breadth of concepts represented within the data.

To validate the analysis procedure on the study session recordings, we randomly extracted five samples with a starting timestamp, ending time stamp, and a description of the developer's behavior from each participant's study sessions.
This resulted in 40 total validation samples (approximately 6\% of the total samples).
The second author then coded the validation samples using the final list of information needs.
The resulting Cohen’s $\kappa$ was 0.86 at the information needs level, indicating near-perfect agreement~\cite{landis1977measurement}.

\subsection{Phase 3: Validating the List of Information Needs for Developing Embedded Prompts}
\label{sec:methodology-phase3}
The generalizability of the list of information needs for \textbf{RQ1} was validated in a survey of \numSurveyParticipants developers who built software with embedded prompts. 
This survey was a lightly modified version of the survey used to refine the list (Section~\ref{sec:methodology-phase2}).
The survey was also used to collect importance ratings for the list of tasks and information needs for \textbf{RQ2}.
The survey responses were collected in July 2025.

\subsubsection{Sampling Strategy}
Consistent with the developer studies in prior phases, the inclusion criteria for the survey study were developers who developed at least one embedded prompt that met the definition from \citet{liang2025prompts}.
In collaboration with JetBrains, we recruited participants by snowball sampling within the authors' social networks and recruiting a subset of JetBrains's software developer participant panel.
Study participants from the company participant panel were recruited through a survey link sent to a subset of developers who had previously consented to be contacted for research purposes as part of an internal participant panel maintained by JetBrains.
This subset was selected based on prior survey responses, wherein participants self-identified as users of AI-assisted development tools or were actively performing AI integrations.
These prior responses were collected as part of separate user studies conducted by JetBrains, with explicit consent for data retention and recontact.
In line with the company's internal research governance policies, no participants were contacted more than once within a 30-day period.

In the end, a total of 1,231 invitation emails were sent to eligible participants, yielding 76 link clicks and ultimately 67 initial completed survey responses prior to quality filtering, yielding a response rate of 5\%.
This response rate is consistent with those from other prior developer surveys, which range from 4\% to 8\%~\cite{liang2022understanding, liang2024large, huang2021leaving}.
Within the survey, study participants were given the opportunity to enter a sweepstakes for \$50 USD Amazon gift card or a JetBrains product pack of an equivalent value.
We set the odds of winning to be 1 in 20 (5\%).

After obtaining the initial set of survey responses, we further filtered the set to non-empty responses from developers who met the inclusion criteria, resulting in 49 participants.
Next, we removed 7 responses after verifying the attention checks, leaving 42 responses.
Since the survey was nearly identical to the one administered during the observational study, we also utilized the survey data from that study.
Following prior developer survey studies~\cite{huang2021leaving}, we validated that the survey responses from the two studies came from the same distribution and could be combined using Mann–Whitney U tests with Benjamini–Hochberg correction across all of the survey questions.
The null hypothesis of this test is that the distributions of survey responses are identical between the two studies.
After verifying the test's assumptions (e.g., sample size, independence, scale)~\cite{nachar2008mann}, we found no statistically significant differences between the survey responses from this study and the survey responses from the observational study for any of the questions.
Thus, all responses for each survey question were then combined, which yielded \numSurveyParticipants total participants for this study.

\subsubsection{Participants}
The survey participants (labeled SP) were developers located in North America ($N=19$), South America ($N=2$), Europe ($N=22$), and Asia ($N=7$).
Study participants were experienced professional programmers ($\mu=12.5$ years) who used FMs always ($N=35$), often ($N=5$), sometimes ($N=9$), or rarely ($N=1$).
They also had experience writing prompts ranging from 1 to 5 ($N=22$), 6 to 10 ($N=4$), 11 to 20 ($N=5$), 21 to 50 ($N=12$), and 50+ ($N=7$) embedded prompts.

\subsubsection{Protocol}
For the survey study, we reused the survey design from the observational study (Section~\ref{sec:methodology-phase2}) with the final list of information needs for developing embedded prompts, though we applied some modifications for the new study context.
First, to limit the effects of memory bias, participants first edited a popular summarization prompt, then recalled a recent prompt change in their own work, and were asked to use these experiences to ground their ratings.
Next, participants read definitions of important concepts introduced in the list of information needs.
For each survey section, participants read an overview of the theme and then rated all related tasks by \emph{frequency} and \emph{helpfulness} to address.
Study participants rated specific information needs only when the related task was frequent or helpful to reduce fatigue.
To protect against bot responses, and to ensure response quality from the unsupervised survey takers, study participants also completed two attention checks.
At the end of the survey, study participants reported their background and demographics, and completed a sweepstakes entry.

\subsubsection{Analysis}
We analyzed the ratings using descriptive statistics.
We analyzed the information needs introduced after the observational study separately, and the shared information needs jointly.
Following \citet{kitchenham2008personal}, we report high-frequency (\emph{``always''/``often'')} and -intensity (\emph{``very''/``extremely helpful''}) responses.
To verify that the list represents sufficient coverage of information needs for developing embedded prompts, we reviewed and coded the write-in suggestions to extend the list of information needs.
First, we identified and de-duplicated substantive suggestions from study participants, thereby obtaining a list of 27 suggested samples to code.
The two study authors who developed the list of information needs independently coded these samples and met to resolve disagreements in the codes.
Codes were only applied upon achieving unanimous agreement.
In the end, this procedure resulted in no new information needs.

\begin{table}[t]
\centering
\caption{An overview of the \numTools tools used to develop embedded prompts that we reviewed in this study, divided by the type of developer tool.}
\label{tab:tools}
\begin{tabular}{@{}p{3cm}p{12cm}@{}}
\toprule
\textbf{Category} & \textbf{Tools} \\
\midrule
\emph{Traditional office tools} & Online workspaces (e.g., Notion); Spreadsheets (e.g., Google Sheets); Text processors (e.g., Google Docs, Notepad) \\
\addlinespace
\emph{Coding tools} & Code editors; Computational notebooks (e.g., Jupyter); In-IDE programming assistants (e.g., GitHub Copilot); Version control systems (e.g., Git, GitHub) \\
\addlinespace
\emph{Prompting GUIs} & Chat interfaces (e.g., ChatGPT); Prompt playgrounds (e.g., OpenAI Chat Playground) \newline \emph{Anthropic Developer Platform}: Playground; Skills; Agents \newline \emph{OpenAI Developer Platform}: Chat; Decisions; Agents; Audio; Images \\
\addlinespace
\emph{Prompt repositories} & LangChain Hub~\cite{langsmith2026langchainhub}; Prompt Flow~\cite{microsoft2026azure} \\
\addlinespace
\emph{Popular open-source prompting tools} & Agenta~\cite{github2026agenta}; Helicone~\cite{github2026helicone}; Humanloop~\cite{github2026humanloop};  
Langfuse~\cite{github2026langfuse}; LangSmith~\cite{github2026langsmith}; Langtrace~\cite{github2026langtrace}; MLflow~\cite{github2026mlflow}; 
Prompt Flow~\cite{github2026promptflow}; PromptFoo~\cite{github2026promptfoo}; PromptLayer~\cite{github2026promptlayer} \\
\addlinespace
\emph{Research tools} &
AI Chains~\cite{wu2022ai}; AP-Adapter~\cite{fu2024ap}; AutoStruct, AutoStruct+, SelectStruct, TagStruct~\cite{fiannaca2023programming}; BotDesigner~\cite{zamfirescu2023conversation, zamfirescu2023johnny}; CoPrompt~\cite{feng2024coprompt}; CoPrompter~\cite{joshi2025coprompter}; DPO-Diff~\cite{wang2024on}; G-fair-Prompting, T-fair-Prompting~\cite{ma2023fairness}; Graphologue~\cite{jiang2023graphologue};  ChainForge~\cite{arawjo2024chainforge, arawjo2023chainforge}; ConstraintMaker~\cite{liu2024we}; EASE~\cite{wu2024prompt}; EvalLM~\cite{kim2024evallm}; MoP~\cite{wang2024one}; Partial2Global~\cite{xu2024towards}; PE2~\cite{ye-etal-2024-prompt}; ProMaC~\cite{hu2024leveraging}; PromptAid~\cite{mishra2025promptaid}; Promptbreeder\cite{fernando2024promptbreeder}; PromptCanvas~\cite{amin2025composable}; PromptChainer~\cite{wu2022promptchainer}; PromptCharm~\cite{wang2024promptcharm}; PromptIDE~\cite{strobelt2022interactive}; PromptingSheet~\cite{he2025prompting}; PromptHive~\cite{reza2025prompthive}; Promptify~\cite{brade2023promptify}; Promptist~\cite{hao2023optimizing}; PromptMaker~\cite{jiang2022promptmaker}; RePrompt~\cite{wang2023reprompt}; Reprompting~\cite{xu2024reprompting}; Responsible Prompting Recommendation~\cite{santana2025can}; RLPG~\cite{shrivastava2023repository}; Smart Prompt Advisor~\cite{phokela2023smart}; Synthetic Prompting~\cite{shao2023synthetic}; TRIPLE~\cite{shi2024efficient}; UP-DP~\cite{li2023up}; WorldSmith~\cite{dang2023worldsmith}; ZOPO~\cite{hu2024localized}; 
\citet{gero2022sensemaking, gero2024supporting};~\citet{white2023prompt, white2024chatgpt} \\
\bottomrule
\end{tabular}
\end{table}

\subsection{Phase 4: Applying the List of Information Needs for Developing Embedded Prompts}
\label{sec:methodology-phase4}
To understand the extent to which existing developer tools support developer information needs (\textbf{RQ3}), we applied the final list of information needs to \numTools tools (see Table~\ref{tab:tools}).
This set of tools comprised of \numLiteratureTools tools that we identified in the research literature, as well as \numPracticeTools commercial tools that we elicited from participant responses in the prior developer studies.
The sample of tools were collected and analyzed in May 2025 and October 2026.

\subsubsection{Tool Selection}
In order to identify tools for developing embedded prompts from the research literature, we reviewed research articles in the ACM Digital Library to obtain a sample of research tools for embedded prompt development, using its keyword search feature for a repeatable selection process.
We first retrieved 9,837 papers with the word "prompt" in the title from May 2020 to May 2025. 
We started with 2020 to capture work after the introduction of few-shot prompting~\cite{brown2020language}, which allowed FM users to select tasks at inference time even before ChatGPT's release~\cite{liu2023pre}.
We filtered this set of papers to obtain ones from major human-computer interaction (CHI, UIST, IUI, DIS) software engineering (ICSE, ASE, FSE), and machine learning (NeurIPS, ICML) venues that publish embedded prompt development tools or techniques.
This yielded 768 papers for consideration.
To identify tools or techniques that supported embedded prompt development, the first author manually reviewed titles and abstracts of these works, identifying 35 relevant papers. 
We obtained an additional 14 tools from a curated list by~\citet{liang2025prompts} by following the same manual filtering process.
After de-duplication, we identified \numLiteratureTools unique research tools that were distributed across communities as follows: 22 from human-computer interaction, three from software engineering, 18 from machine learning, and two from visualization and computer graphics.

To obtain a sample of commercial tools for developing embedded prompts, we identified publicly accessible tools that study participants used by reviewing all of the data on how developers changed embedded prompts that we collected from earlier studies.
We noted down the tools from two sources: interview transcripts from the interview study (Section~\ref{sec:methodology-phase1}), where participants self-reported developer tools, and video recordings from the observational study (Section~\ref{sec:methodology-phase2}), where we directly observed participants using specific developer tools.
We also searched for popular commercial or open-source tools in May 2025, using the authors' knowledge and experience and included tools with more than 500 GitHub stars.
We also obtained a more recent sample of commercial embedded prompt development tools in October 2026 by analyzing all of the embedded prompt tools located in the developer platforms from both OpenAI~\cite{openai2026platform} and Anthropic~\cite{anthropic2026platform} based on the tool and its documentation.
In total, this procedure yielded \numPracticeTools unique commercial tools.
We report the full list of tools in Table~\ref{tab:tools}.

\subsubsection{Analysis}
To understand which tools addressed information needs of developing embedded prompts, the first author manually noted tool features from the research tool's manuscript. 
The second author did the same for commercial tools using the tool's documentation.
This yielded a dataset of 146 features for the research tools and 227 features for commercial tools.
The authors then deductively coded the features of their respective set of tools.
If a tool directly (i.e., a specific feature aimed at addressing the information need) or indirectly (i.e., provided a general-purpose feature that could address the information need tangentially) addressed a specific information need, the authors applied the corresponding code.
To verify coding consistency on the tools, the second author independently coded a random sample of five research tools. 
We extracted the related features for each research tool, producing a validation set of 17 features (12\% of all research tool features).
The codes for each feature were masked to reduce potential biases in the coding validation procedure.
The computed Cohen’s $\kappa$ at the task level was 0.73, indicating substantial agreement~\cite{landis1977measurement}.

\begin{table*}
  \centering
\scriptsize
\caption{The list of \numQuestions information needs involved in developing embedded prompts, divided by theme and task. 
We report the percentage of developers who rated the information need as occurring frequently and helpful to be addressed with a tool, as well as whether that information need is addressed by any existing tool for developing embedded prompts (\cmark) or not (\xmark).
We \highlight{highlight} the top five most frequent or helpful information needs (including ties) across the entire list.
The wording of the tasks and information needs is adjusted for brevity.
For the original wording, refer to the supplemental materials~\cite{supplemental-materials}.}
\label{tab:information-needs}
\renewcommand{\arraystretch}{1.1}
\begin{tabular}{p{0.32\linewidth}p{0.48\linewidth}p{0.03\linewidth}p{0.03\linewidth}p{0.03\linewidth}}
\toprule
\textbf{Task (T\#)} & \textbf{Question (Q\#)} & \textbf{Freq.} & \textbf{Help.} & \textbf{Addr.}\\
\hline
\multicolumn{2}{l}{\textbf{\emph{What is this prompt?}}} \\
\hline
T1. Understanding a version's content & Q1. What is the high-level structure of the prompt content? & 63\% & 53\% & \cmark \\
& Q2. What is the specific text (e.g., characters) in the prompt content? & 56\% & 40\% & \cmark \\
& \highlight{Q3. What other parts of the prompt are related to the part I'm interested in?} & \highlight{92\%} & 80\% & \xmark \\
T2. Understanding a version's behavior & Q4. What output does this prompt generate for this example? & 87\% & 68\% & \cmark \\
& Q5. How is this prompt behaving or performing overall? & 85\% & 77\% & \cmark\\
T3. Identifying examples with interesting behavior & Q6. Which example(s) does this version not do well on? & 78\% & 85\% & \cmark\\
& Q7. Which example(s) does this version do well on? & 76\% & 72\% & \cmark \\
T4. Hypothesizing reasons for prompt behavior & Q8. Are there patterns when the prompt has this behavior? & 75\% & 83\% & \cmark \\
T5. Understanding the prompt's dependencies & \highlight{Q9. Where in the codebase does this part of the prompt rely on?} & \highlight{91\%} & 83\% & \xmark \\
& Q10. What do the prompt's related dependencies in the codebase do? & 82\% & 58\% & \xmark \\
\hline
\multicolumn{2}{l}{\textbf{\emph{What inputs should I provide to the prompt?}}} \\
\hline
T6. Understanding characteristics of a collection & Q11. What examples should I run my prompt on? & 80\% & 81\% & \cmark \\
of examples & \highlight{Q12. How representative are the examples?} & \highlight{90\%} & \highlight{90\%} & \xmark \\
& Q13. What are the data attributes of a specific example? & 53\% & 57\% & \cmark \\
\hline
\multicolumn{2}{l}{\textbf{\emph{What happened when I ran this prompt?}}} \\
\hline
T7. Understanding the configuration used in the run & Q14. What are the configuration values (e.g., temperature) used? & 42\% & 44\% & \cmark \\
& Q15. What FM is used in the run? & 71\% & 72\% & \cmark \\
T8. Understanding the inputs given during the run & Q16. Which prompt version was used for a specific run? & 68\% & 83\% & \cmark \\
 & \highlight{Q17. What example(s) or context was used for a template parameter?} & \highlight{90\%} & 76\% & \cmark \\
T9. Understanding behavior for non-functional reqs. & Q18. How many resources (e.g., tokens generated) did this run consume? & 64\% & 77\% & \cmark \\
\hline
\multicolumn{2}{l}{\textbf{\emph{Why did the prompt behave unexpectedly?}}} \\
\hline
T10. Debugging by reading the prompt's content & \highlight{Q19. What part of the content might be the cause of the observed behavior?} & \highlight{88\%} & 85\% & \cmark \\
T11. Debugging by comparing two different versions & \highlight{Q20. What part of the change might cause the behavior in version A?} & 67\% & \highlight{90\%} & \xmark \\
& \highlight{Q21. What part of the change might cause the behavior diff. in ver. A and B?} & 83\% & \highlight{86\%} & \xmark \\
T12. Debugging through FM explanations & Q22. What does the model say the reason for the observed behavior is? & 29\% & 73\% & \cmark \\
T13. Debugging by reasoning about artifacts outside of & Q23. What part of the source code might cause the behavior? & 83\% & 70\% & \xmark \\
the prompt& Q24. What part of the example, context, or input might cause the behavior? & 58\% & 75\% & \xmark \\
T14. Trying fixes given observed unwanted behavior & Q25. What FM or run configurations (e.g., temperature) should I use? & 28\% & 36\% & \cmark \\
& Q26. What changes should I make to the prompt content? & 86\% & 81\% & \cmark \\
& Q27. What changes should I make to the code? & 55\% & 53\% & \xmark \\
\hline
\multicolumn{2}{l}{\textbf{\emph{What changed between this prompt version and the one previous?}}} \\
\hline
T15. Understanding a change to the prompt's content & \highlight{Q28. What prompt content was modified?} & 78\% & \highlight{87\%} & \cmark \\
T16. Understanding prompt behavior after a change & Q29. Did the change modify the prompt's behavior? & 81\% & 83\% & \cmark \\
& \highlight{Q30. Did the change unexpectedly regress the behavior?} & 78\% & \highlight{87\%} & \cmark \\
& Q31. Did the change improve the behavior as intended? & 85\% & 83\% & \cmark \\
T17. Recalling design rationale & Q32. What was the reason for this change? & 81\% & 72\% & \xmark \\
\hline
\multicolumn{2}{l}{\textbf{\emph{How do these prompt versions compare to one another?}}} \\
\hline
T18. Understanding differences between two versions & \highlight{Q33. How does prompt version A behave differently than version B?} & 81\% & \highlight{90\%} & \cmark \\
& Q34. How does the content of prompt version A differ from version B? & 81\% & 76\% & \cmark \\
T19. Understanding differences between more than & Q35. How do more than two versions' content differ? & 75\% & 70 \% & \xmark \\
two versions & Q36. How do more than two versions' behavior differ? & 75\% & 80\% & \cmark \\
\hline
\multicolumn{2}{l}{\textbf{\emph{Where is this prompt?}}} \\
\hline
T20. Retrieving versions based on prompt content & Q37. Which versions contain a certain feature? & 36\% & 67\% & \xmark \\
& Q38. Which versions contain a keyword? & 18\% & 53\% & \cmark \\
& Q39. Which versions follow a certain structure? & 55\% & 67\% & \xmark \\
T21. Retrieving versions based on prompt behavior & Q40. Which versions contain an output with a keyword? & 56\% & 65\% & \cmark \\
& Q41. Which versions behave in a certain way? & 50\% & 53\% & \xmark \\
& Q42. Which versions make a tradeoff in its behavior? & 60\% & 59\% & \xmark \\
& Q43. Which versions do well on specific aspects of behavior? & 60\% & 71\% & \xmark \\
T22. Bookmarking specific versions & Q44. What versions are important or interesting? & 40\% & 63\% & \cmark \\
T23. Finding prompts relevant to the one being developed & \highlight{Q45. What are prompts that accomplishes a similar function as my prompt?} & \highlight{100\%} & 87\% & \cmark \\
\hline
\multicolumn{2}{l}{\textbf{\emph{What happened across all the prompt versions?}}} \\
\hline
T24. Understanding progress over multiple versions & \highlight{Q46. What are the overall best version(s) by behavior or performance?} & 86\% & \highlight{100\%} & \cmark \\
& Q47. What are the overall worst version(s) by behavior or performance? & 52\% & 52\% & \cmark \\
& \highlight{Q48. How have aspects of the behavior changed across versions?} & 68\% & \highlight{89\%} & \cmark \\
T25. Remembering what was tried previously & Q49. What are the most recent versions I have tried? & 63\% & 78\% & \cmark \\
& Q50. What have I tried so far? & 46\% & 70\% & \cmark \\
\bottomrule
\end{tabular}
\renewcommand{\arraystretch}{1.0}
\end{table*}

\section{Results}
In this section, we present our findings on what were the developer information needs while making an embedded prompt change (\textbf{RQ1}, Section~\ref{sec:rq1}), how important these information needs were (\textbf{RQ2}, Section~\ref{sec:rq2}), and to what extent existing tools addressed these information needs (\textbf{RQ3}, Section~\ref{sec:rq3}).

\subsection{Information Needs for Developing Embedded Prompts (RQ1)}
\label{sec:rq1}
Table~\ref{tab:information-needs} presents the list of information needs involved in prompt programming with embedded prompts. 
The list is made up of eight themes that are divided into development tasks (T\#).
Each task is further split into information needs that are expressed as questions (Q\#), following prior work~\cite{latoza2010hard, sillito2006questions, sillito2008asking}.
To answer \textbf{RQ1}, we report the number of participants who discussed a task or question ($\times$) and denote tasks that were performed manually or without tools by study participants (\faHandStopO).
We frame the reporting of information needs as a union of all the activities that occurred in our study. 
We also do not claim that all developers required all information needs.

\subsubsection{What is this prompt?}
Just as developers must comprehend their program and build a mental model of how it works in traditional software contexts~\cite{ko2007information, latoza2010hard, duala2012asking}, one theme of developing embedded prompts was understanding the content and behavior of a particular prompt version.
Notably, the opaque and non-deterministic nature of FMs introduced new challenges for developers to develop reliable mental models of the software~\cite{liang2025prompts}.

\paragraph{Understanding a version's content (T1, Q1-3)}
To develop embedded prompts, developers needed to obtain information on the prompt content at different granularities.
This ranged from the prompt's high-level structure (Q1, 5$\times$) to specific words and characters such as \ipquote{line breaks and the indentation}{3} (Q2, 13$\times$).
Study participants also studied the prompt by locating it in a file (4$\times$, \faHandStopO) and reading the text (7$\times$) or the code that generated the prompt (3$\times$).
For longer prompts, understanding the structure of the prompt was useful.
\ip{8} described that chat-based interfaces, which did not reveal the structure of the prompt, made it challenging to understand their 40-line prompt for their coding agent:
\ipBlockQuote{It's cool to visualize [prompts] as chat, but it can get really long at times. I would want to collapse messages and focus on just one part, which is my prompt.}{8}

In addition, participants also studied the relationships between prompt components (Q3), especially while making a change to an embedded prompt.
This is because prompt components had implicit dependencies based on having related content; if one section was updated, then changes were often required in other locations for consistency.
Participants identified related components manually by reading through the prompt (3$\times$, \faHandStopO).

\paragraph{Understanding a version's behavior (T2, Q4-5)}
Developers wanted to understand the prompt's behavior by understanding the prompt's output (Q4, 20$\times$) and performance metrics (Q5, 13$\times$).
Participants navigated files with saved outputs (2$\times$, \faHandStopO), read FM outputs and reasoning (10$\times$, \faHandStopO), observed agent behavior (2$\times$, \faHandStopO), and reviewed metrics (2$\times$).
They also developed custom systems to understand behaviors (\faHandStopO), such as tools (2$\times$), spreadsheets (2$\times$), or evaluation scripts (4$\times$).
These systems were challenging to manage and scale across prompt changes over time:
\ipBlockQuote{I was doing everything by hand...I was summarizing what change I made. I was writing that by hand in the Google Sheets. I had to copy the new notebook, make the change. Run it. Get the metrics, then paste it again on the Google Sheets. This would just be insane if there were like 100 versions of the same prompt.}{1}

Across all types of software containing embedded prompts, assessing an embedded prompt's behavior required manual verification (\faHandStopO).
For example, \op{8}, who created an agent to play a video game, assessed the agent's planned movements on a grid by using their mouse to simulate the actions on the game screen.
Because FMs, especially agents, could generate large amounts of output, participants also identified relevant sections of the output manually (2$\times$, \faHandStopO) or through the search function (1$\times$).

\paragraph{Identifying examples that result in interesting behavior (T3, Q6-7)}
Study participants needed information on instances that produced interesting negative (Q6) and positive (Q7) prompt behavior ($6\times$).
This parallels traditional software engineering contexts, where understanding the situations when failures occur is one of the most important information needs for software teams~\cite{ko2007information}.
In the case of developing embedded prompts, identifying instances with notable behavior helped participants identify \ipquote{where do I not like the results}{2}.

Depending on the type of task and/or software with embedded prompts, participants approached addressing these information needs differently.
For tasks with a more closed-ended output space (e.g., classification tasks), participants took advantage of the smaller range of potential output generations to form natural groupings.
For example, \op{7} used a heat map to identify false positives and negatives in their classification task and isolated the prompt outputs from these instances into a separate file (\faHandStopO).
However, for tasks with a more open-ended output space (e.g., text generation tasks, agents), participants were less likely to engage in this information need.
This is because their embedded prompt development process was comparatively less systematic and mostly involved trial-and-error changes to their prompt.

\paragraph{Hypothesizing why the version behaved the way it did (T4, Q8)}
Because FM behavior is opaque---which introduces unique challenges in developing embedded prompts~\cite{liang2025prompts, chen2026promptware}---study participants also hypothesized about prompt behavior by identifying behavioral patterns (Q8, 7$\times$), which was particularly helpful for debugging: 
\ipBlockQuote{I need to figure out, what is the pattern in all those problems? Finding those problems is actually challenging.}{15}

This information need occurred in different types of embedded prompt software, including single prompts and agents.
Participants identified patterns by creating files with instances of similar behavior (1$\times$, \faHandStopO), writing code for visualizations such as heat maps or tables (1$\times$, \faHandStopO), or running the prompt multiple times (1$\times$, \faHandStopO).
Hypothesizing about embedded prompt behaviors through behavioral patterns is critical for the developer to form a mental model of the embedded prompt's behavior~\cite{liang2025prompts}.
This is similar to traditional software engineering contexts, where developing hypotheses about the sources of program behavior is essential for both code comprehension and debugging~\cite{li2026grounded, maalej2014comprehension}.

\paragraph{Understanding the prompt's dependencies (T5, Q9-10)}
Similar to how developers need to be aware of a software's dependencies~\cite{ko2007information}, developers of embedded prompts needed to locate (Q9) and understand (Q10) the prompt's dependencies outside the prompt itself, like related code (5$\times$).
To our knowledge, this information need has not yet been discussed within the literature.
This information need was especially important for participants who provided the embedded prompt with context that was handled programatically, and therefore occurred across different domains and types of embedded prompt software.
For example, \op{3}, who was developing a coding agent, wanted to figure out how to shorten the context length, and spent over 55 minutes of the study session exploring and understanding the code that created the prompt and processed the context provided to the agent.
Meanwhile, to debug their horoscope chatbot, \op{5} needed to locate and understand the code calling a weather API.
Study participants manually discovered dependencies (\faHandStopO) to propagate changes to other prompts for consistency (1$\times$) and identify (2$\times$) and understand (4$\times$) code related to the generation of the prompt.

\subsubsection{What inputs should I provide to the prompt?}
Similar to how developers ask questions about testing code in traditional software contexts~\cite{latoza2010hard}, participants reflected on the inputs to the prompt when evaluating their embedded prompts.
Whereas information needs in traditional software focused on the correctness of the code base, test suite, or documentation~\cite{latoza2010hard}, we found that information needs for testing embedded prompts largely focuses on understanding the set of examples used to run and test the prompt, reflecting on the shift towards data curation observed in prompt programming~\cite{liang2025prompts}.

\paragraph{Understanding characteristics of a collection of examples (T6, Q11-13).}
Study participants considered what inputs to test (Q11, 6$\times$) by crafting ad hoc examples (1$\times$, \faHandStopO) or identifying interesting examples from a dataset (1$\times$, \faHandStopO) at runtime.
As discussed in prior work~\cite{liang2025prompts, lu2026engineering}, generating representative test cases at a large scale was difficult and time-consuming (Q12, 2$\times$):
\ipBlockQuote{Make sure you have diverse test data. Otherwise, you're going to build a prompt that's really good on the inputs you have.}{10}
To understand the dataset attributes associated with the input that was provided during the run of an embedded prompt (Q13) (e.g., the gold labels for an instance in a classification dataset), \op{4} opted to jot down and refer to the dataset instance in a code comment (\faHandStopO) rather than examine it directly from a CSV file.

\subsubsection{What happened when I ran this prompt?}
Developers building software with embedded prompts needed to understand the prompt's run configuration (i.e., the FM used and its hyperparameters).
This reflects how in traditional software engineering contexts, developers need to collect information on the runtime state of software to construct a mental model of how the software works~\cite{latoza2010hard, ko2007information, li2026grounded}. 

\paragraph{Understanding the configuration used in the run (T7, Q14-15)}
While running the prompt, developers wanted to understand which FM, FM version (Q15, 3$\times$), and hyperparameters (Q14, 3$\times$) like \ipquote{temperature}{3, \op{7}} and \opquote{max tokens}{7}.
Participants obtained this information by reading code (3$\times$, \faHandStopO) or by retrieving it in the chat UI and AI playground (2$\times$, \faHandStopO).
When the FM's configuration was embedded within code, study participants became confused on which FM was used (2$\times$).
For example, \op{8} ran their video game agent on multiple FMs to understand whether the model affected the agent's behavior.
However, when they wanted to revert the FM used to a previous run, they struggled to recall which model was used previously since they did not explicitly track this information locally.
As a result, \op{8} retrieved this information off of GitHub (\faHandStopO).

\paragraph{Understanding the inputs given to the FM during the run (T8, Q16-17)}
Developers required information on the run's FM inputs, such as the prompt (Q16) and examples or provided context (Q17).
Study participants obtained this information by printing the prompt out and reading it (2$\times$, \faHandStopO).
Only \op{8} used an external tool, Langtrace~\cite{github2026langtrace}, to view the inputs their agent was given.
For developers building embedded prompts for agents, understanding the inputs was especially challenging because combining the prompt, examples, and provided context created a long textual input, corroborating prior work that finds that agent development to be challenging due to the more complex inputs and generations~\cite{liang2026understanding, epperson2025interactive}.
For example, \op{3} struggled to view and comprehend the inputs of their coding agent due to the length of the provided code files.
In addition, complex code and managing multiple code versions created confusion about which prompt version was used during a run (2$\times$):

\ipBlockQuote{I'm worried about...using the wrong prompt... Am I using what we agreed upon?}{5}

\paragraph{Understanding prompt behavior related to non-functional requirements (T9, Q18)}
Study participants also wanted to understand the prompt's non-functional behavior (Q18).
Non-functional behaviors included \ipquote{memory}{1}, \opquote{token count}{3, \op{4}}, and time elapsed (\op{8}).
While \op{8} obtained information on the length of their agent's run in the Langtrace UI, \ip{3} collected this manually by combing through logs they had instrumented that printed the information (\faHandStopO).

\subsubsection{Why did the prompt behave unexpectedly?}
Like developers debugging code~\cite{breu2010information, ko2007information, latoza2010hard}, developers building embedded prompts needed information that could help explain why unexpected behavior occurred, which is especially challenging given the opaque and nondeterministic nature of FMs in prompt programming~\cite{liang2025prompts, chen2026promptware}.
However, debugging embedded prompts introduces new complexities beyond reading prompt content and FM output, as previous work suggests~\cite{liang2025prompts, zamfirescu2023johnny}. 
In practice, we found that developers of embedded prompts also have to reason about the data context provided to the embedded prompt as well as the code that generated prompt to identify and fix unwanted behaviors.

\paragraph{Localizing fault or debugging by reading the prompt's content (T10, Q19)}
Developers of embedded prompts used the prompt content for debugging (Q19) by reading ($2 \times$, \faHandStopO) and recalling (3$\times$, \faHandStopO) the prompt's content and relating it to observed behavior (2$\times$, \faHandStopO).
Often, developers across multiple types of software with embedded prompts repeatedly needed this information immediately after running the embedded prompt ($4 \times$), reflecting the iterative experimental loop involved in prompt programming~\cite{liang2025prompts, zamfirescu2023johnny}.
In addition, study participants wanted support on this task because they found it challenging (2$\times$): \ipquote{Identifying the areas where the fixes need to be made...was the big challenge}{14}.

\paragraph{Localizing fault or debugging by comparing two different versions (T11, Q20-21)}
Similar to traditional software engineering contexts~\cite{gilmore1991models, ko2004designing, li2026grounded}, comparing the inputs, runtime, and outputs of a buggy program was a common strategy for effective debugging embedded prompts.
We observed that developers engaged with this strategy for embedded prompts ($2\times$), though to a less frequent extent compared to just reviewing the content of a single prompt.
Study participants compared embedded prompts for debugging through mental comparisons ($2\times$, \faHandStopO) or by retrieving recent prompt changes ($1\times$, \faHandStopO).
This identifies changes causing differing behavior (Q20, Q21): \ipquote{[I need] to understand...what prompt changes caused what kind of response changes}{14}.

\paragraph{Localizing fault or debugging through FM explanations (T12, Q22)}
As noted in prior work~\cite{liang2025prompts}, developers of software with embedded prompts debugged with FM explanations (Q22) by asking an FM to provide explanations for its decisions (2$\times$, \faHandStopO) or by using FMs that generate reasoning (1$\times$).
This occurred for varying complexities of software with embedded prompts, from single prompts to agents.
For instance, \op{8} retrieved their video game agent's reasoning traces from Langtrace to debug their agent.
Meanwhile, \op{7} instructed the FM to generate explanations in their single prompt for their financial classification task.

\paragraph{Localizing fault or debugging by reasoning about artifacts outside of the prompt (T13, Q23-24)}
We also observed that defects in embedded prompts could come from outside the prompt, which is not widely acknowledged within the literature. 
For example, \op{8}'s video game agent had defects because it ingested game state with hidden levels due to the code implementation.
Developers thus reason about external artifacts, like code (Q23, 3$\times$) and the context provided to the prompt (Q24, 3$\times$) to identify the source of unwanted behavior.
Participants retrieved this information by manually reading the FM's inputs (\op{8}, \faHandStopO) and outputs (3$\times$, \faHandStopO), finding related code using IDE features (2$\times$, \faHandStopO), and using their existing knowledge of the system (\op{8}, \faHandStopO).

\paragraph{Trying fixes given observed unwanted behavior (T14, Q25-27)}
Crucial to the experimentation loop in prompt programming is updating the prompt to change the embedded prompt's unwanted behavior that was observed~\cite{liang2025prompts}, which we observed consistently across all forms of software with embedded prompts.
Study participants fixed the embedded prompt by changing the prompt's content (Q26, 3$\times$) and provided context (2$\times$), FM (Q25, 3$\times$) and hyperparameters (\ip{5}), and related code (Q27, 2$\times$).
We observed that many participants used AI programming assistants, like code completion or ChatGPT, to generate prompt (2$\times$) and code changes (2$\times$).
For \op{8}, updating the FM as a fix was particularly challenging due to unintuitive FM naming conventions (e.g., \texttt{gemini-2.5-flash-lite-preview-06-17}).
As a result, knowing what FM version to improve the embedded prompt's behavior was confusing and required manual verification on Google AI Studio (\faHandStopO).

In addition, knowing what to change was challenging (6$\times$), so multiple participants wanted suggestions on potential updates ($3\times$).
However, it was often unclear whether the change should occur in the prompt or in the code (2$\times$), reflecting a unique challenge associated with building of hybrid embedded prompt and code software systems:
\opBlockQuote{An ongoing question...is, how much of that should I be doing in my prompt engineering? How much should I do in [code]?}{4}

\subsubsection{What changed between this prompt version and the one previous?}
Due to the opaqueness of FMs in prompt programming resulting in a high volume of changes~\cite{liang2025prompts}, an important type of information need was understanding the changes between the current prompt version and the previous one in order to understand whether the prompt change resulted in meaningful modifications to embedded prompt's behavior.
Similar to traditional software contexts~\cite{latoza2010hard, ko2007information}, developers of embedded prompts also wanted to understand why prompt changes were made. 

\paragraph{Understanding a change to the prompt's 
content (T15, Q28)}
Developers of embedded prompts reported that they wanted to track changes in the prompt content (Q28, 5$\times$).
Prior studies indicate that information on what was changed is important for debugging traditional software~\cite{ko2004designing, gilmore1991models}.
However, this information was rarely explicitly captured by study participants, reflecting the existing challenges in collecting design rationale in traditional software~\cite{horvath2024meta}.
Because \ip{1} developed their embedded prompt within a research context, only \ip{1} recorded what was changed by \ipquote{summarizing what change I made}{1} in a spreadsheet (\faHandStopO) to record details of their design rationale for later use.
As a result of the lack of tracking of changes, recalling low-level changes was difficult for developers:
\ipBlockQuote{It's hard to recall what words you exactly changed, but you can usually remember at a high level what you changed.}{3}

\paragraph{Understanding prompt behavior after making a change (T16, Q29-31)}
Regardless of the type of software containing the embedded prompt, participants wanted to study the embedded prompt's behavior for differences (Q29), improvements (Q31), and regressions (Q30) after changing the prompt (6$\times$), since it was unclear \ipquote{if [the change] had any impact}{3}.
While the literature recommends that developer tooling reveal differences between program versions to achieve a better understanding of the system during debugging~\cite{liang2025prompts}, study participants detected changes manually due to a lack of such support in their development environment.
To understand the prompt behavior after a change, participants manually compared outputs ($1\times$, \faHandStopO), qualitatively assessed outputs and logs ($4\times$, \faHandStopO) and observed an agent's behavior live ($2\times$, \faHandStopO).
For example, to understand the differences between the results of their web agent, \op{1} repeatedly toggled between the generations of the previous prompt version's output with the newest version's output in the IDE.
In order to verify the change was consistent, \op{1} and \op{4} ran the change on new inputs (\faHandStopO).

\paragraph{Recalling design rationale (T17, Q32)}
Several participants reported wanting to know why a change was made (Q32, 7$\times$). 
This corroborates information needs from traditional software contexts, where developers are known to make particular design choices in software implementation~\cite{liang2023qualitative} and want to understand why the software was implemented in specific ways~\cite{ko2007information, latoza2010hard, maalej2014comprehension}.
However, only two study participants were able to retrieve this information in any form.
\op{4} recalled design rationale from memory (\faHandStopO), while \ip{1} recorded design rationales in a spreadsheet.
The design rationale helped to understand the prompt change's intention (\ip{14}), share the development process with others (\ip{13}), and inspire future changes (2$\times$):
\ipBlockQuote{This is more for...seeing previous ideas...[and] how we can adopt that with the current version of the prompt.}{10}

\subsubsection{How do these prompt versions compare to one another?}
Because developing embedded prompts requires making many experimental changes~\cite{liang2025prompts}, developers wanted to compare prompt versions across time to understand whether their prompt changes resulted in meaningful progress.
To our knowledge, this has not been previously identified in the literature as a key information need.
Understanding progress across changes was especially challenging as it often involves evaluations of qualitative criteria~\cite{kim2024evallm, liang2025prompts}.

\paragraph{Understanding the differences between two versions at a time (T18, Q33-34)}
A prominent information need was comparing the content (Q34) and behavior (Q33) between two embedded prompt versions.
While we observed this behavior within single- and multi-prompt software applications, prior work has also documented that understanding differences behavior in agent software as result of making a change is particularly challenging~\cite{epperson2025interactive}.
Study participants admitted that the process to understand differences between prompt versions was largely \opquote{a lot of manual comparison and eyeballing}{7} (\faHandStopO).
For their multi-prompt project, \op{4} retrieved this information by stacking two different prompt versions that were stored in variables on top of each other in a code file in the IDE. 
Meanwhile, \op{7} took screenshots of FM outputs of their single embedded prompt, created tables of quantitative metrics for each prompt, and compared these artifacts side by side.

\paragraph{Understanding the differences between more than two versions at a time (T19, Q35-36)}
Study participants reported wanting to compare more than two versions at a time by the prompts' content (Q35) and behavior (Q36).
We did not observe any participant perform this task in the observational study; however, interview participants noted that comparing multiple versions was challenging while developing embedded prompts:
\ipBlockQuote{I can compare [the version] with one or two different ones...but not with all the others... That is a problem that happened.}{15}

\subsubsection{Where is this prompt?}
In traditional software engineering, developers want to locate code snippets related to their task~\cite{haenni2013categorizing, ko2007information} by using sources like documentation, online communities like Stack Overflow, and more recently, AI programming assistants.
In a similar vein, developers of embedded prompts wanted to retrieve prompts and prior versions based on the prompt's content and/or behavior.
Retrieving embedded prompts can both facilitate prompt reuse~\cite{white2023prompt, white2024chatgpt} as well as support developers in recollecting what they had tried previously: \ipquote{A lot of iterations happen... [so] remember[ing] what has changed and what has not is helpful}{2}.

\paragraph{Retrieving versions based on prompt content (T20, Q37-39)}
Across multiple types of software with embedded prompts, study participants searched for previous embedded prompt versions by the prompt's content at varying granularities.
The scope of the search ranged from the embedded prompt's high-level structure (Q39, 3$\times$) and features (Q37, 5$\times$) to specific characters (Q38, 6$\times$). 
For keyword queries, \op{2} used the AI playground's search feature; otherwise, they manually combed through versions in order to search for specific versions (\op{2}, \faHandStopO). 
\ip{11} noted that for domains with many existing prompts, semantic search would be more helpful than using specific terms to capture a wider range of search results in the domain:
\ipBlockQuote{I would like to have more semantic search over the prompts rather than syntactic, so I can say, `Oh, find me all the prompts which have some stats related to code generation.'... I would like to see all the prompts I’ve written for translation or programming-related tasks.}{11}

\paragraph{Retrieving versions based on prompt behavior (T21, Q40-43)}
Study participants wanted to retrieve versions based on the embedded prompt's behavior across varying types of software with embedded prompts.
Their desired search could be carried out at different granularities, such general descriptions of prompt behavior (Q41, 2$\times$), specific aspects of prompt behavior (Q43, 6$\times$), tradeoffs in aspects of behavior (Q42, 2$\times$), or keywords in the generated output (Q40, 2$\times$).
However, no developer was able to acquire this information in the observational study because searching for embedded prompts based on their behavior not well-supported:
\ipBlockQuote{I often find myself wanting to go back to some [outputs] that may have generated...but I have no way of searching for it.}{1}

\paragraph{Bookmarking specific versions (T22, Q44)}
Developers of embedded software wanted to recall specific prompts (Q44, 24$\times$).
Bookmarking versions could focus evaluations on a small subset of prompts (\ip{4}), since some versions were more important than others: \ipquote{[I would like to] check in the specific [versions], because I don't treat every [FM output] I get equally}{9}.
Yet, in the observational study, study participants did not bookmark specific versions.
Only \op{3} recalled specific prompt versions from memory (\faHandStopO).

\paragraph{Finding different prompts that are relevant to the one being developed (T23, Q45)}
Developers of embedded prompts needed to find prompts that performed similar functionality (Q45, $4\times$), which was especially important for prompt reuse~\cite{white2023prompt, white2024chatgpt}, such as finding specific text instructions: \ipquote{I copy parts of the formatting prompt from one GPT to the other...[to share] the reusable...instructions you know work}{2}.
In the observational study, only \op{5} was able to find relevant prompts, and did so manually by opening different files on their machine (\faHandStopO).

\subsubsection{What happened across all the prompt versions?}
Developers of embedded prompts required information on their prompt change history.
Prior research indicates that developers struggle to recall this information, given the large amount of experimentation involved in prompt programming~\cite{liang2025prompts}.
Understanding the history of how software has evolved is an important information need in traditional software engineering contexts~\cite{ko2007information}, though it can be a difficult question to answer~\cite{latoza2010hard}.

\paragraph{Understanding progress over multiple versions (T24, Q46-48)}
Study participants wanted to understand how prompts progressed over multiple iterations by finding the best versions (Q46), worst versions (Q47), and various aspects of behavior (Q48).
\ip{4} tracked this information in a metrics spreadsheet (\faHandStopO). 
In addition, \ip{6} noted that AI playgrounds were \ipquote{not a great way to track progress... [Progress] is not  tracked}{6}.
Tracking prompt history could help with knowing \ipquote{what worked and [what] didn't}{3}, reverting to working versions (\ip{16}), or finding the best performing versions (2$\times$).

Many types of software with embedded prompts required this information. 
However, \op{8} noted that understanding progress for an agent that performed actions in user interfaces was particularly challenging, corroborating prior work~\cite{liang2026understanding}.
Because \op{8}'s video game agent could take many different paths to complete a game level, they noted that understanding whether the agent's behavior improved over runs was challenging:

\opBlockQuote{The issue also is because my definition of success is kind of binary for these levels. It's pretty hard to like measure progress, in some cases}{8}

\paragraph{Remembering what was tried previously (T25, Q49-50)}
Study participants retrieved their prior prompt attempts (Q49) using a variety of methods across several types of software with embedded prompts.
This included using the undo command (\op{1}, \faHandStopO), navigating through chat history (3$\times$, \faHandStopO), saving prompts on the AI playground (\ip{8}), skimming a saved file with all versions (\op{4}, \faHandStopO), uncommenting code snippets (2$\times$, \faHandStopO), using a spreadsheet (\ip{5}), and saving versions in files with custom naming conventions (2$\times$, \faHandStopO).
For developers of embedded prompts, recalling versions was challenging: there could be upwards of \ipquote{hundreds of prompts}{4} whose information was confused or lost during recall (3$\times$).
While recent versions were prioritized (Q49, 5$\times$), saving all versions could communicate the design process (\ip{13}) or inspire future prompt designs by reviewing versions (4$\times$):
\ipBlockQuote{I want to know what I tried before, regardless of how well it worked. That way I don't repeat myself.}{12}

\mybox{
\faArrowCircleRight\xspace \textbf{Key Findings (RQ1)}:
We identified \numQuestions information needs involved in developing embedded prompts, that are classified into \numTasks tasks.
A vast majority of information needs for developing embedded prompts were addressed manually by participants, rather than relying on tools.
}

\begin{table*}
  \centering
\scriptsize

\caption{
The list of tasks involved in developing embedded prompts, divided by theme.
For each task, we report how frequently developers reported performing that task and how helpful it would be for that task to be addressed by tooling.
The frequency bar charts show \% of Always/Often (left) vs. Rarely/Never (right) ratings.
The helpfulness bar charts Extremely/Very helpful (left) vs. Slightly/Not helpful (right) task ratings.
We \highlight{highlight} the top three most frequent or helpful tasks (including ties) across the entire list.
The wording of the tasks is adjusted for brevity.
For the original wording, refer to the supplemental materials~\cite{supplemental-materials}.
}
\label{tab:tasks}
\renewcommand{\arraystretch}{0.8}
\begin{tabular}{p{0.38\linewidth}|p{0.02\linewidth}p{0.14\linewidth}>{\raggedleft\arraybackslash}p{0.05\linewidth}p{0.02\linewidth}p{0.14\linewidth}>{\raggedleft\arraybackslash}p{0.05\linewidth}}
\toprule
& \multicolumn{6}{c}{\textbf{Distribution}} \\
\textbf{Task (T\#)} & \multicolumn{3}{c}{\textbf{Frequency}} & \multicolumn{3}{c}{\textbf{Helpfulness}} \\
\midrule
\multicolumn{7}{l}{\textbf{\emph{What is this prompt?}}} \\
\midrule
\addlinespace[1pt]
T1. Understanding a version's content & 49\% & \frequencybarchart{0.12}{0.36}{0.30}{0.18}{0.03}{0}{48.5\%}{21.2\%} & 21\% & 39\% & \importancebarchart{0.09}{0.30}{0.55}{0.03}{0.03}{0}{39.4\%}{6.1\%} & 6\% \\
\addlinespace[1pt]
T2. Understanding a version's behavior & 63\% & \frequencybarchart{0.12}{0.51}{0.29}{0.05}{0.02}{0}{63.4\%}{7.3\%} & 7\% & 59\% & \importancebarchart{0.22}{0.37}{0.37}{0.00}{0.05}{0}{58.5\%}{4.9\%} & 5\% \\
\addlinespace[1pt]
T3. Identifying examples with interesting behavior & 63\% & \frequencybarchart{0.20}{0.44}{0.22}{0.15}{0.00}{0}{63.4\%}{14.6\%} & 15\% & 68\% & \importancebarchart{0.37}{0.32}{0.27}{0.05}{0.00}{0}{68.3\%}{4.9\%} & 5\% \\
\addlinespace[1pt]
\highlight{T4. Hypothesizing reasons for prompt behavior} & \highlight{68\%} & \frequencybarchart{0.33}{0.35}{0.23}{0.07}{0.03}{0}{67.5\%}{10.0\%} & 10\% & 61\% & \importancebarchart{0.24}{0.37}{0.27}{0.10}{0.02}{0}{61.0\%}{12.2\%}& 12\% \\
\addlinespace[1pt]
T5. Understanding the prompt's dependencies & 45\% & \frequencybarchart{0.07}{0.38}{0.34}{0.17}{0.03}{0}{44.8\%}{20.7\%} & 20\% & 48\% & \importancebarchart{0.17}{0.31}{0.31}{0.14}{0.07}{0}{48.3\%}{20.7\%} & 21\% \\
\midrule
\multicolumn{7}{l}{\textbf{\emph{What inputs should I provide to the prompt?}}} \\
\midrule
\addlinespace[1pt]
\highlight{T6. Understanding characteristics of a collection of examples} & 52\% & \frequencybarchart{0.24}{0.29}{0.29}{0.19}{0.00}{0}{52.4\%}{19.0\%} & 19\% & \highlight{79\%} & \importancebarchart{0.38}{0.40}{0.19}{0.02}{0.00}{0}{78.6\%}{2.4\%} & 2\% \\
\midrule
\multicolumn{7}{l}{\textbf{\emph{What happened when I ran this prompt?}}} \\
\midrule
\addlinespace[1pt]
T7. Understanding the configuration used in the run & 60\% & \frequencybarchart{0.26}{0.33}{0.26}{0.12}{0.02}{0}{59.5\%}{14.3\%} & 14\% & 21\% & \importancebarchart{0.21}{0.00}{0.54}{0.17}{0.08}{0}{20.8\%}{25.0\%} & 25\% \\
\addlinespace[1pt]
T8. Understanding the inputs given during the run & 67\% & \frequencybarchart{0.35}{0.33}{0.23}{0.09}{0.00}{0}{67.4\%}{9.3\%} & 9\% & 58\% & \importancebarchart{0.58}{0.00}{0.35}{0.04}{0.04}{0}{57.7\%}{7.7\%} & 8\% \\
\addlinespace[1pt]
T9. Understanding behavior for non-functional reqs. & 49\% & \frequencybarchart{0.14}{0.35}{0.14}{0.23}{0.14}{0}{48.8\%}{37.2\%} & 37\% & 26\% & \importancebarchart{0.26}{0.00}{0.46}{0.11}{0.17}{0}{25.7\%}{28.6\%} & 29\% \\
\midrule
\multicolumn{7}{l}{\textbf{\emph{Why did the prompt behave unexpectedly?}}} \\
\midrule
\addlinespace[1pt]
T10. Debugging by reading the prompt's content & 63\% & \frequencybarchart{0.39}{0.24}{0.27}{0.10}{0.00}{0}{63.4\%}{9.8\%} & 10\% & 68\% & \importancebarchart{0.32}{0.37}{0.27}{0.02}{0.02}{0}{68.3\%}{4.9\%} & 5\% \\
\addlinespace[1pt]
T11. Debugging by comparing two different versions & 49\% & \frequencybarchart{0.12}{0.37}{0.29}{0.12}{0.10}{0}{48.8\%}{22.0\%} & 22\% & 62\% & \importancebarchart{0.21}{0.41}{0.26}{0.13}{0.00}{0}{61.5\%}{12.8\%} & 13\% \\
\addlinespace[1pt]
T12. Debugging through FM explanations & 24\% & \frequencybarchart{0.03}{0.21}{0.45}{0.17}{0.14}{0}{24.1\%}{31.0\%} & 31\% & 41\% & \importancebarchart{0.19}{0.22}{0.30}{0.22}{0.07}{0}{40.7\%}{29.6\%} & 30\% \\
\addlinespace[1pt]
T13. Debugging by reasoning about artifacts outside the prompt & 45\% & \frequencybarchart{0.14}{0.31}{0.28}{0.28}{0.00}{0}{44.8\%}{27.6\%} & 28\% & 69\% & \importancebarchart{0.21}{0.48}{0.21}{0.10}{0.00}{0}{69.0\%}{10.3\%} & 10\% \\
\addlinespace[1pt]
\highlight{T14. Trying fixes given observed unwanted behavior} & \highlight{81\%} & \frequencybarchart{0.39}{0.41}{0.10}{0.07}{0.02}{0}{80.5\%}{9.8\%} & 10\% & 75\% & \importancebarchart{0.40}{0.35}{0.20}{0.05}{0.00}{0}{75.0\%}{5.0\%} & 5\% \\
\midrule
\multicolumn{7}{l}{\textbf{\emph{What changed between this prompt version and the one previous?}}} \\
\midrule
\addlinespace[1pt]
T15. Understanding a change to the prompt's content & 61\% & \frequencybarchart{0.37}{0.24}{0.29}{0.10}{0.00}{0}{61.2\%}{10.2\%} & 10\% & 61\% & \importancebarchart{0.22}{0.39}{0.27}{0.08}{0.04}{0}{61.2\%}{12.2\%} & 12\% \\
\addlinespace[1pt]
\highlight{T16. Understanding prompt behavior after a change} & \highlight{69\%} & \frequencybarchart{0.36}{0.33}{0.24}{0.07}{0.00}{0}{69.0\%}{7.1\%} & 7\% & \highlight{76\%} & \importancebarchart{0.38}{0.38}{0.17}{0.07}{0.00}{0}{76.2\%}{7.1\%} & 7\% \\
\addlinespace[1pt]
T17. Recalling design rationale & 56\% & \frequencybarchart{0.29}{0.27}{0.20}{0.24}{0.00}{0}{56.1\%}{24.4\%} & 24\% & 67\% & \importancebarchart{0.31}{0.36}{0.21}{0.10}{0.02}{0}{66.7\%}{11.9\%} & 12\% \\
\midrule
\multicolumn{7}{l}{\textbf{\emph{How do these prompt versions compare to one another?}}} \\
\midrule
\addlinespace[1pt]
\highlight{T18. Understanding differences between two versions} & 56\% & \frequencybarchart{0.17}{0.39}{0.20}{0.20}{0.05}{0}{56.1\%}{24.4\%} & 24\% & \highlight{76\%} & \importancebarchart{0.41}{0.34}{0.12}{0.07}{0.05}{0}{75.6\%}{12.2\%} & 12\% \\
\addlinespace[1pt]
T19. Understanding differences between more than two versions & 40\% & \frequencybarchart{0.10}{0.25}{0.07}{0.33}{0.25}{0}{35.0\%}{57.5\%} & 35\% & 50\% & \importancebarchart{0.28}{0.23}{0.33}{0.07}{0.10}{0}{50.0\%}{17.5\%} & 18\% \\
\midrule
\multicolumn{7}{l}{\textbf{\emph{Where is this prompt?}}} \\
\midrule
\addlinespace[1pt]
T20. Retrieving versions based on prompt content & 35\% & \frequencybarchart{0.10}{0.25}{0.38}{0.23}{0.05}{0}{35.0\%}{27.5\%} & 28\% & 50\% & \importancebarchart{0.20}{0.30}{0.33}{0.15}{0.03}{0}{50.0\%}{17.5\%} & 18\% \\
\addlinespace[1pt]
T21. Retrieving versions based on prompt behavior & 35\% & \frequencybarchart{0.05}{0.30}{0.30}{0.23}{0.12}{0}{35.0\%}{35.0\%} & 35\% & 50\% & \importancebarchart{0.25}{0.25}{0.30}{0.10}{0.10}{0}{50.0\%}{20.0\%} & 20\% \\
\addlinespace[1pt]
T22. Bookmarking specific versions & 43\% & \frequencybarchart{0.12}{0.30}{0.17}{0.15}{0.25}{0}{42.5\%}{40.0\%} & 40\% & 53\% & \importancebarchart{0.20}{0.33}{0.20}{0.17}{0.10}{0}{52.5\%}{27.5\%} & 28\% \\
\addlinespace[1pt]
T23. Finding prompts relevant to the one being developed & 28\% & \frequencybarchart{0.07}{0.15}{0.35}{0.38}{0.05}{0}{22.5\%}{42.5\%} & 23\% & 48\% & \importancebarchart{0.28}{0.20}{0.33}{0.17}{0.03}{0}{47.5\%}{20.0\%} & 20\% \\
\midrule
\multicolumn{7}{l}{\textbf{\emph{What happened across all the prompt versions?}}} \\
\midrule
\addlinespace[1pt]
\highlight{T24. Understanding progress over multiple versions} & \highlight{69\%} & \frequencybarchart{0.36}{0.33}{0.17}{0.10}{0.05}{0}{69.0\%}{14.3\%} & 14\% & 71\% & \importancebarchart{0.38}{0.33}{0.19}{0.02}{0.07}{0}{71.4\%}{9.5\%} & 10\% \\
\addlinespace[1pt]
\highlight{T25. Remembering what was tried previously} & 67\% & \frequencybarchart{0.29}{0.38}{0.17}{0.14}{0.02}{0}{66.7\%}{16.7\%} & 17\% & \highlight{74\%} & \importancebarchart{0.40}{0.33}{0.12}{0.10}{0.05}{0}{73.8\%}{14.3\%} & 14\% \\
\midrule
\multicolumn{7}{c}{\mylegend{Always (100\% of changes)}{green3} \mylegend{Often (75\% of changes)}{green2} \mylegend{Sometimes (50\% of changes)}{gray1}\mylegend{Rarely (25\% of changes)}{red2} \mylegend{Never (0\% of changes)}{red3}} \\
\addlinespace[1pt]
\multicolumn{7}{c}{\mylegend{Extremely helpful (difficult + impactful)}{blue2} \mylegend{Very helpful}{blue1} \mylegend{Moderately helpful}{gray1}\mylegend{Slightly helpful}{orange1} \mylegend{Not helpful at all (easy or not impactful)}{orange2}} \\
\bottomrule
\end{tabular}
\renewcommand{\arraystretch}{1.0}
\end{table*}

\subsection{Importance of Information Needs (RQ2)}
\label{sec:rq2}
Table~\ref{tab:information-needs} and Table~\ref{tab:tasks} highlight the most important information needs and tasks based on their perceived frequency and helpfulness.
Given the exploratory nature of this work, we treat the important information needs and tasks as preliminary signals to guide the design and research of future tools for developing embedded prompts.
Below, we discuss prominent trends across the information needs and tasks developers perceived as most important.

\paragraph{Understanding embedded prompt changes}
Our results indicate that understanding embedded prompt changes as well as the FM behavior that they induce is a need that should be supported by developer tooling.
This corroborates the vast number of iterations that developers explore to develop an embedded prompt, which has been extensively documented by the literature~\cite{liang2025prompts, zamfirescu2023johnny, parnin2025building, dolata2024development}.
Understanding differences between two versions (T18) ranked as the second most helpful to address need (76\%).
Participants were especially interested in understanding differences in behavior (Q33), as it was one of the most helpful to address need across all information needs (90\%).

In addition, understanding prompt behavior after a change (T16) was the only task that was both one of the most frequent (69\%) and helpful to address (76\%).
For this task, participants said knowing whether their changes regressed behavior (Q30) was one of the most helpful to address needs (87\%).
Similarly, participants reported that understanding what part of the prompt was modified (Q28) was helpful (87\%).
The frequent and important nature of understanding prompt behavior after a change could be due to the stochastic nature of FMs that pose key barriers in prompt programming~\cite{liang2025prompts, zamfirescu2023johnny}, whose challenges are further exacerbated by understanding agent behaviors~\cite{liang2026understanding, epperson2025interactive}.

\paragraph{Understanding the data used to test the embedded prompt}
For developers of embedded prompts, understanding the data that the prompt was run on during testing was an important task that needed support.
This reflects how developers building embedded prompts are data practitioners and thus must learn to curate representative datasets~\cite{qian2024understanding}.
In particular, the task of understanding the characteristics of a group of examples (T6) was rated as one of the most helpful (79\%) tasks by developers of embedded prompts.
For this task, one of the most important information needs was understanding how representative the examples in the developer's dataset (Q12) were, as it was the only information need that was both frequent (90\%) and helpful for developers (90\%).
The high helpfulness ratings of these tasks and information needs indicate that building datasets is challenging, since embedded prompts often focus on narrow or specific tasks that do not have existing benchmarks or datasets~\cite{liang2025prompts, nahar2024beyond}.

Study participants also agreed that understanding the input data provided to the FM during a run (T8) was a frequent task (66\%).
For this task, participants focused on understanding which examples or context that was used for a template parameter during a run (Q17) was one of the most frequent (91\%) information needs, and was also generally regarded as helpful to address (79\%).
This could be due to the fact that the context provided to embedded prompts---especially for agents---can be difficult to interpret~\cite{liang2026understanding}.

\paragraph{
Debugging embedded prompts}
In addition to the specific information needs described above, developers needed to make sense of individual information needs in conjunction with one another to effectively debug embedded prompts; for example, in order to debug a prompt, developers needed to understand what about the prompt had changed (T15-19).
Debugging by reading the prompt (T10, 68\%) and comparing two versions (T11, 66\%) were considered as helpful to address by a majority of participants, reflecting the difficulties that developers face while working with opaque FMs~\cite{liang2025prompts}.
Understanding what part of the embedded prompt content could cause the observed behavior (Q19) occurred frequently (88\%), and understanding what part of the change caused specific behavior in a prompt (Q20, 92\%) or caused a change in behavior (Q21, 86\%) was one of the most helpful to address.

The information needs and tasks related to repairing embedded prompts with defects were consistently rated as frequent and helpful to address by developers of embedded prompts.
While generating hypotheses for a prompt's behavior (T4) was rated as one of the most frequent tasks (68\%), trying fixes given observed unwanted behavior (T14) was also rated as the most frequent task (81\%), reinforcing the iteration characteristic of prompt programming~\cite{liang2025prompts, zamfirescu2023johnny, parnin2025building, dolata2024development}.
For this task, knowing what part of the prompt to change (Q26) occurred more frequently (86\%) than changing the code (Q27, 55\%) and FM or its hyperparameters (Q25, 28\%).

\paragraph{Understanding embedded prompt version history}
Another important task that developers of embedded prompts wanted support on was understanding an embedded prompt's version history.
Study participants rated understanding progress (T24) as one of the most frequently occurring tasks (69\%) and remembering what was tried previously (T25) as one of the most helpful (74\%) tasks to address.
For understanding progress (T24), participants rated understanding which versions were the best performing across all attempts (Q46, 100\%) and how different aspects of behaviors changed over time (Q48, 89\%) was rated as some of the the most helpful to address.
This reflects challenges associated with the vast amount of experimentation involved in developing embedded prompts, which are often difficult to track and make sense of because many prompt changes are unsystematic~\cite{liang2025prompts, zamfirescu2023conversation}.

\paragraph{Understanding the external dependencies of the embedded prompt}
Finally, the last set of important information needs developers found important to address were related to understanding the dependencies of the embedded prompt, such as the code that is used to construct the prompt.
To our knowledge, this information need has not yet been identified in the literature, despite being important to developers.
Making sense of the a prompt's dependencies (T4) was considered as frequently occurring (45\%) and helpful to address (48\%) by a near majority of participants.
For this task, one of the most important information needs was understanding where in the codebase a particular part of the prompt relied on (Q9).
This information need occurred frequently (87\%), and further was a need that developers rated as being one of the most helpful to address (88\%).
Notably, developers of embedded prompts rated identifying the code dependencies (Q9, 83\%) more than understanding what the code did (Q10, 58\%) as more helpful.

\mybox{
\faArrowCircleRight\xspace \textbf{Key Findings (RQ2)}:
The most frequent tasks for developing embedded prompts was trying fixes given unwanted observed behavior (T14, 81\%), understanding embedded prompt behavior after a change (T16, 69\%), understanding progress across multiple versions (T24, 69\%), and hypothesize reasons for prompt behavior (T4, 68\%).
Among these tasks, the most important information needs were knowing whether a change regressed behavior (Q30), finding the best performing versions of a prompt (Q46), and understanding how aspects of behavior changed over time (Q48).
The most helpful tasks to address were understanding the characteristic of a collection of examples (T6, 79\%), understanding prompt behavior after making a change (T16, 74\%), understanding differences between two versions (T18, 76\%), and remembering what was tried previously (T25, 74\%).
The most important information needs for these tasks were understanding how representative examples were (Q12), knowing whether a change regressed behavior (Q30), and understanding how a prompt version's behavior compares to another (Q33).
}

\begin{figure}[t!]
\centering
\includegraphics[trim=25 0 950 0, clip, width=\linewidth, page=1, keepaspectratio]{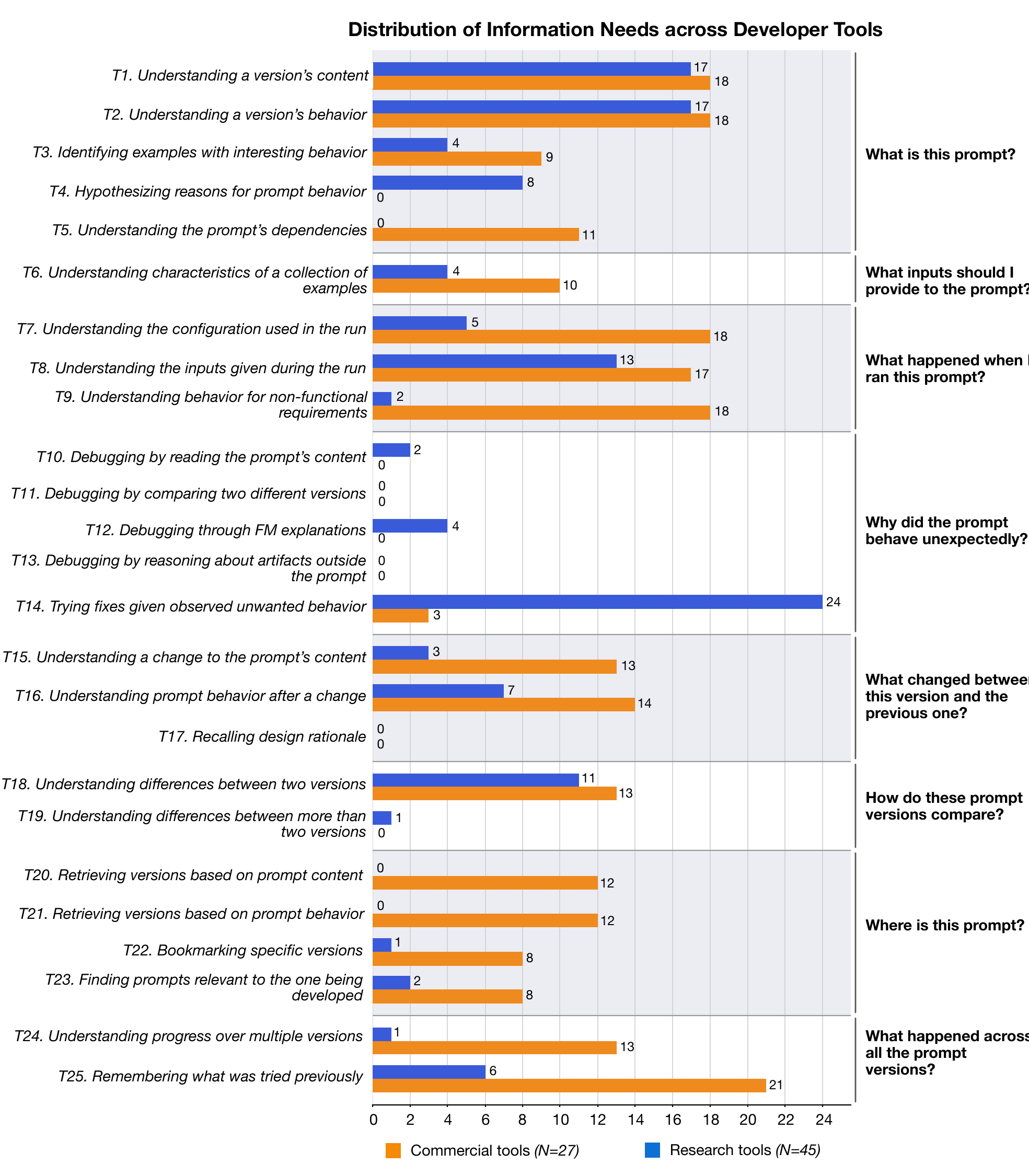} 
\caption{
The distribution of research and commercial tools that include features that address each of the tasks for developing embedded prompts.
The wording of the tasks is adjusted for brevity.
For the original wording, refer to the supplemental materials~\cite{supplemental-materials}.
}
\label{fig:tools}
\end{figure}

\subsection{Tool Coverage of Information Needs for Developing Embedded Prompts (RQ3)}
\label{sec:rq3}
Table~\ref{tab:tasks} and Figure~\ref{fig:tools} mark which information needs and tasks were not supported by existing tools.
We note that the intent of this study is to understand general trends in developer needs, rather than perform a comprehensive survey of today's tools.
Overall, we found that 16 of the \numQuestions questions are unsupported by any existing commercial or research tool within our sample of \numTools tools. 
It is also important to remember that while 34 questions are reportedly supported by some tools, our participants needed to perform some aspects of \textit{all} of these questions manually.
Below, we discuss the amount of developer tool support by theme; we use $\times$ here to represent the number of tools.

\subsubsection{What is this prompt?}
To comprehend prompt content (T1, Q1-3) and behavior (T2, Q4-6), tools often provided text editors (35$\times$) and viewers for raw prompts (Q2) and FM outputs (Q4, 29$\times$), sometimes embedded in visual programming nodes~\cite{wu2022ai, arawjo2024chainforge}.
Some supported editing of context~\cite{zamfirescu2023johnny} or output format~\cite{liu2024we}.
Given the structured nature of prompts~\cite{tafreshipour2025prompting, mao2025prompts}, several tools represented high-level structure (Q1, 6$\times$) with widgets~\cite{amin2025composable}, sub-components~\cite{jiang2022promptmaker, fiannaca2023programming}, or trees~\cite{feng2024coprompt}.
Others supported performance metrics (Q5, 18$\times$), like quantitative~\cite{arawjo2024chainforge} and FM-evaluated metrics~\cite{joshi2025coprompter, kim2024evallm} and user-defined behavioral evaluation criteria~\cite{joshi2025coprompter, kim2024evallm}.
Because FM outputs can be difficult to make sense of~\cite{jiang2023graphologue}, tools offered support to identify relevant outputs through visualizations~\cite{jiang2023graphologue, gero2022sensemaking, gero2024supporting} and text highlighting~\cite{gero2022sensemaking, gero2024supporting, kim2024evallm}.
While research tools explored rich representations for prompt understanding, commercial tools (e.g., LangSmith and MLflow) offered text editing with limited scaffolding.
Despite needing to understand what other parts were related to the current part of the prompt being changed being the most frequent information need (Q3, 94\%), no existing research or commercial tool provided support for this information need.

Next, support for identifying interesting behavior (T3, Q6-7) included manual annotations~\cite{zamfirescu2023conversation, zamfirescu2023johnny}, highlighting successful (Q7,~\cite{mishra2025promptaid}) or failing examples (Q6,~\cite{arawjo2024chainforge}), and inspecting examples via performance metrics.
These features were often supported by standard evaluation features of commercial tools.
To identify patterns (T4, Q8), research tools used confusion matrices~\cite{mishra2025promptaid, strobelt2022interactive}, highlight common substrings~\cite{zamfirescu2023conversation}, and grouping FM outputs with various visualizations~\cite{brade2023promptify, zamfirescu2023conversation}.
We note that commercial tools did not directly address this task.
Finally, to identify prompt dependencies (T5), commercial tools (11$\times$) offered observability features of how different embedded prompts depend on each other. 
This included features such as call traces (e.g., Langtrace) or graphical workflows (e.g., Prompt Flow, LangSmith).
However, no tools supported finding code dependencies (Q9-10), even though identifying code that the prompt depends on was rated as one of the most helpful information needs (Q9, 88\%).

\subsubsection{What inputs should I provide to the prompt?}
Several tools provided features to understand datasets (T6), including to define and explore datasets (Q13, $13\times$).
However, few tools suggested specific examples (Q11) beyond basic sampling~\cite{kim2024evallm}; PromptFoo was a notable exception for test case generation.
There were also no tools that assessed dataset quality (Q12), despite understanding how representative the data is (Q12) being one of the most frequently occurring (92\%) and helpful to address (89\%) information needs for developing embedded prompts.
The lack of tooling for understanding dataset quality reflects the ongoing challenges of evaluating embedded prompts at scale across many examples~\cite{kim2024evallm, liang2025prompts}, which is noted to be even more challenging in agent software~\cite{lu2026engineering}.

\subsubsection{What happened when I ran this prompt?}
We found that all information needs in this theme were supported by existing tools.
In order to nderstand prompt run configurations (T7) and non-functional behavior (T9), many tools logged FM details (Q15), hyperparameters (Q14), and runtime performance (24$\times$).
Commercial tools often tracked non-functional behavior (Q18, 18$\times$), though we found that support for this task was less present for research tools: within the set of research tools, only PromptCanvas provided information on non-functional attributes of prompts such as word count~\cite{amin2025composable}.
Several tools also supported inspecting FM inputs (T8) through text editors, with the prompt examples or context (Q17) used in the run.

\subsubsection{Why did the prompt behave unexpectedly?}
While information needs related to debugging were important, we found that support for this task was limited. 
Tasks such as reasoning about artifacts outside the prompt (T13, Q23–24), comparing prompts (T11, Q20–21), and trying fixes to the code (Q27) had no tool support.
Only fault localization via prompt content (T10, Q19, 2$\times$), FM explanations (Q22, 4$\times$), and especially prompt repair or optimization (Q26, 28$\times$) had some tool support.
Node-based tools that chained prompts could isolate defects to specific nodes in prompt pipelines~\cite{wu2022promptchainer}.
However, understanding which part of a change may have caused the behavior in a specific version (Q20), which was one of the most helpful to address information needs (92\%), was not supported by any of the research or commercial tools.
To explain FM behavior, tools used common interpretability techniques such as highlighting salient inputs~\cite{wang2024promptcharm} and leveraging FM-generated rationales for generations~\cite{jiang2023graphologue, joshi2025coprompter}. 
Prompt repair was supported by automatic methods (28$\times$), like input perturbation~\cite{mishra2025promptaid}, optimization models and techniques~\cite{ye-etal-2024-prompt, phokela2023smart, santana2025can, fiannaca2023programming}, and retrieving prompt examples~\cite{fiannaca2023programming, wang2024promptcharm}. 
Most of the machine learning papers focused on developing techniques for prompt optimization (e.g.,~\cite{fernando2024promptbreeder, hao2023optimizing, shi2024efficient}).
Commercial tools offered basic support, like one-click improvements or prompt creation experiences (e.g., OpenAI Playground, Anthropic Agents) or LLM chat suggestions (e.g., LangSmith).

\subsubsection{What changed between this prompt version and the one previous?}
Developer tools provided support for understanding embedded prompt content changes (T15, Q28, 15$\times$) by showing in-line highlights~\cite{santana2025can} or side-by-side views~\cite{feng2024coprompt}.
Behavior changes (T16, Q29-31, 20$\times$) were shown by plotting metrics~\cite{mishra2025promptaid} or comparing outputs side-by-side~\cite{zamfirescu2023johnny, kim2024evallm}.
Some tools surfaced error types while updating a prompt~\cite{zamfirescu2023johnny, zamfirescu2023conversation} to assess behavioral shifts.
However, we did not identify any research or commercial tools that supported the recording of design rationale of individual prompt changes (T17, Q32).

\subsubsection{How do these prompt versions compare to one another?}
Tasks and information needs related to comparing prompt versions were largely addressed in some form by commercial or research tools.
Many tools allowed developers to compare two prompts (T18) and their behavior (Q33) by visualizing metrics~\cite{mishra2025promptaid}, highlighting output changes~\cite{zamfirescu2023conversation}, and comparing outputs side-by-side~\cite{arawjo2024chainforge, santana2025can, brade2023promptify, reza2025prompthive, wang2024promptcharm}. 
Prompt content was compared (Q34) by side-by-side comparisons~\cite{wang2024promptcharm} and plotting the prompt content embeddings~\cite{mishra2025promptaid}.
ChainForge~\cite{arawjo2024chainforge} was the only research tool that compares three or more prompt outputs (T19, Q36) by plotting variables in a table; however, we did not observe support for comparing three or more prompts at the same time.
Other tools supported comparing more than two prompts by comparing their aggregated metrics.
We did not observe support for understanding differences in the prompt content for more than two prompts (Q35).

\subsubsection{Where is this prompt?}
Support for tasks and information needs related to retrieving prompts had notable gaps.
Many systems, especially commercial ones (8$\times$), focused on supporting these needs by providing the functionality to bookmark versions (T22, Q44).
While some commercial tools supported prompt retrieval by content (T20) and behavior (T21) via keyword searches (Q38, Q40, 12$\times$), such features were not present in research tools.
In addition, more advanced retrieval using prompt features (Q37), structure (Q39), behavioral descriptions (Q41) and tradeoffs (Q42), and aspects of behavior (Q43), were not supported by any tool in the set of research or commercial tools.
Despite being one of the most frequently needed information by developers (89\%), retrieving other relevant prompts (T23, Q45) was less supported by existing tooling and was addressed by 9 of \numTools tools total. 
For research tools, this information need was implemented through community support platforms~\cite{strobelt2022interactive} and features for categorizing prompts by task~\cite{reza2025prompthive}, while for commercial products, this came in the form of starting templates for prompts (e.g., OpenAI Agents) or prompt repositories (e.g., Langchain Hub).

\subsubsection{What happened across all the prompt versions?}
Within the sample of research and commercial tools, we found that all tasks and information needs related to understanding what occurred across multiple changes had some form of tooling support.
For example, to understand progress over versions (T24, Q46-48), developer tools often plotted performance metrics~\cite{mishra2025promptaid} (14$\times$).
Performance metrics were customized in commercial tools by defining evaluation functions.
To recall past attempts (T25, Q49-50), several tools supported history and versioning of the prompt (27$\times$), which could be represented as a global plot of changes~\cite{mishra2025promptaid} or within a graph~\cite{dang2023worldsmith}.

\mybox{
\faArrowCircleRight\xspace \textbf{Key Findings (RQ3)}:
Research and commercial tools provided some form of support for 22 of the \numTasks tasks.
The tasks that do not have any support were debugging by comparing two different versions (T11), debugging by reasoning about artifacts outside of the prompt (T13), and recalling design rationale for the embedded prompt (T17).
In addition, we find that 16 of the \numQuestions information needs, including four of the most frequent or helpful to address ones, did do not have any form of support from existing tools.
The important information needs that were not supported were understanding what other parts of the prompt are related to the current part of the prompt the developer is interested in (Q3), where in the codebase an embedded prompt depends on (Q9), how representative the evaluation dataset (Q12), and what part of a change may cause specific behavior (Q20).
}

\section{Threats to Validity}
We discuss the limitations and threats to validity of our study with respect to its internal validity, external validity, and construct validity.

\paragraph{Internal Validity}
The list of tasks and information needs for developing embedded prompts could be affected by a misunderstanding the survey or interview instructions by the study participants.
Following best practices for conducting human studies in software engineering~\cite{ko2015practical}, we addressed this threat by piloting all study protocols on three participants each to assess their clarity.

Both the interview study to elicit information needs and the opinion survey to validate information needs could have been affected by memory bias~\cite{nisbett1977telling}.
To reduce these threats, both studies required developers to consider recent embedded prompts the participant had developed.
The survey study also required participants to make a change to a popular summarization prompt.
In addition, it is possible that the participants of the survey study did not attend to questions carefully; in order to ensure the survey yielded high-quality data, participants completed attention checks that were then later used to filter out low-quality responses.
Finally, in order to respect the privacy of developers' prompts, the observational study had one coder on the video footage, which could introduce analysis bias.
We mitigated this by validating the coding quality through a second round of validation coding.
The validation coding was completed by a second coder on a random subset of the descriptions of developer behavior.
We achieved a Cohen's $\kappa$ of 0.86 through this validation procedure, indicating near-perfect agreement~\cite{landis1977measurement}.

In addition, because the developer tools for embedded prompts were selected using a combination of a keyword search in the ACM Digital Library, reports of tool usage from study participants, and known popular tools, the sampling strategy may not be representative of all developer tools.
We thus treat the results of \textbf{RQ3} as a signal towards general trends in current tooling for embedded prompts to guide the design of future development tools to solve important but unaddressed developer needs.

\paragraph{External Validity}
We note that the overall findings of the study may not generalize to all developers of embedded prompts due to limited sample sizes and sampling bias from convenience and snowball sampling, especially in the observational study to refine the list of information needs.
In addition, the interview and observational studies contained a single overlapping participant. 
While this introduces some sampling bias, we felt these samples still captured adequate variation because they represented different projects with embedded prompts that were developed one year apart.
In order to mitigate the threat of sampling bias, we included both industry and academic developers with general and professional programming experience in the interview study to create the list and opinion survey study to validate the list.
We also achieved code saturation by \op{7}, suggesting a sufficient amount of coverage of developer information needs within our sample.

For the survey study, recruiting via the JetBrains participant panel may bias the sample to users of a specific tool. 
For the tool comparison study, using one database may exclude work outside human-computer interaction, software engineering, and machine learning, limiting the list's coverage in other domains (e.g., medicine).
To reduce the threats to validity from any single method, we triangulated results across four methodologies.
For example, the survey's larger sample offset the interview and observational studies' limited size, while observational data mitigated biases in retrospective questions.

In addition, we note that the list of information needs for developing embedded prompts is not comprehensive, and therefore may not generalize to all development contexts and software domains.
For example, safety critical domains may have specific information needs (e.g., assessing the safety of the embedded prompt) that are not covered in the list of information needs.
And, some of the existing challenges of developing embedded prompts (e.g., understanding best practices, setting realistic product expectations~\cite{parnin2025building, liang2025prompts, dolata2024development}) may not be completely addressed with the list of information needs.
These additional information needs can be further explored by future work.
We reduced this threat by intentionally sampling across diverse types of embedded software, domains, organizations, and software development roles to obtain a broad range of experiences to improve generality.

Finally, the extent to which developer information needs may generalize to future FMs is unclear, given that FM performance evolves over time \cite{chen2023chatgpt}.
However, we anticipate that these information needs will persist as long as human developers are involved in building software with embedded prompts.
Regardless of FM performance, human developers will continue to specify the requirements for their software---which involves comprehending the prompt's content (T1, Q1-3), the design rationale of the software (T17, Q32)~\cite{horvath2024meta, liang2023qualitative}, and how the software has evolved over time (T24-25, Q46-50)~\cite{ko2007information, latoza2010hard}---and evaluating to what extent the prompt's behavior aligns with the requirements (T2-4, Q4-8) across diverse inputs (T6, Q11-13).
If the requirements and FM behavior do not align, developers then must also understand what to do to repair the misalignment (T10-14, Q19-27).

\paragraph{Construct Validity}
For this work, the main threat to construct validity is how the importance of information needs were measured.
Following recommendations from \citet{myers2016programmers}, we relied on the information needs' perceived frequency and helpfulness to address as a proxy for the importance to developers.
However, perceived frequency may contain memory biases.
Further, developer ratings of helpfulness to address may vary based on the participant's interpretation of the construct.
While we mitigated these threats by grounding protocols in recent development experiences and piloting the study protocols multiple times, we also emphasize that these study results should be treated as exploratory insights that should be validated with additional research.
Future work could collect larger scale \emph{in-situ} data as a more comprehensive method to understand the frequency of developer information needs.

\section{Discussion}
In this work, we cataloged a list of \numQuestions information needs to build embedded prompts (\textbf{RQ1}). 
Within this list, we identified the most important information needs that developer tools should address (\textbf{RQ2}).
Finally, we applied the list of information needs to existing developer tools for embedded prompts to understand the extent developer needs are addressed with existing tools (\textbf{RQ3}).
Overall, we found that the \numQuestions information needs were largely manually addressed by developers of embedded prompts and that current tooling did not address 16 of the information needs, including four of the ones that developers rated to be most important.
However, even the information needs that were considered addressed by the tool coverage analysis were not completely covered in practice, since study participants addressed these needs manually, rather than relying on tool support.

Below, we discuss the implications of our findings with respect to the literature (Section~\ref{sec:implications}), and conclude by enumerating important design opportunities for developer tools for embedded prompts based on the most important information needs that remain unaddressed or under-addressed (Section~\ref{sec:design-opportunities-developer-tools}).

\subsection{Implications}
\label{sec:implications}

\subsubsection{Information needs of building single or multi-prompts vs. building agents}
In our study, we found that the information needs of building software with single prompts or multi-prompts were applicable to those of building agents.
For example, we found that building agents still required a developer to understand the agent's system and tool prompts (T1, Q1-3), its behavior (T2, Q4-5), and the code that may have produced the embedded prompt (T5, Q9-10), as well as knowing what fixes to try given unwanted behavior (T14, Q25-27).
Developing agents also required developers to understand the examples and context that were used in a run (Q17)~\cite{liang2026understanding}.
We believe that the information needs are consistent between these types of software with embedded prompts, as developing agents still relies on building embedded prompts on top of stochastic and opaque FMs. 
Thus, building agents continues to produce the same challenges and information needs as developing single or multi-prompt systems.

However, developing agents introduces additional complexities to these information needs for developing embedded prompts, as agent systems are typically more complex.
Not only do agents include multiple prompts~\cite{liang2026understanding}, they generate more outputs~\cite{epperson2025interactive} compared to other forms of embedded prompt software.
This added complexity in the software can pose challenges for developers.
For example, having multiple prompts provides multiple potential sources of error, which requires the developer to understand each embedded prompt during debugging~\cite{liang2026understanding}.
In our study, while debugging their gaming agent, \op{8} had spent effort understanding whether the defect originated from their tool prompts, system prompts, or context provided to the agent.
Meanwhile, by producing long-horizon workflows, agents generate long, complex outputs that are difficult for developers to make sense of~\cite{epperson2025interactive, liang2026understanding}.
\op{8} also noted that understanding and evaluating progress for such long-horizon workflows was challenging and the definition of success was not obvious, since changes in the agent's actions to complete a game level did not necessarily change the task outcome.
Finally, the code infrastructure that powers agents, often known as harnesses, is a complicated software system, comprising of components like \emph{harness interface} code to connect the agent to reasoning, action, and live environments, as well as \emph{harness mechanism} code to enable the agent to engage with planning, memory, and tool use for long-horizon execution~\cite{ning2026code}.
The complex agent harnesses introduced more difficulties in understanding the code.
In our study, developers building agents spent much of their time trying to understand the surrounding harness---for example, OP3 spent 55 minutes making sense of the code that generated their prompt.

Given the unique development challenges that agents pose, future research could explore developer tools specific to building agents.
Existing work has examined how to debug multi-agent systems~\cite{epperson2025interactive}, prototype agent user experiences~\cite{liang2026understanding}, and designing multi-agent workflows~\cite{hao2025flowforge}.
Given that information needs related to understanding why prompts behaved unexpectedly is systematically underexplored by existing tools compared to other information needs (see Figure~\ref{fig:tools}), future tools could study techniques to help developers debug agent systems.

\subsubsection{Information needs of developers who are humans vs. developers who are agents}
As coding agents become more powerful, a line of research has recently begun investigating techniques to automatically optimize prompts~\cite{khattab2024dspy, fernando2024promptbreeder} and update code harnesses~\cite{wu2026harnessdev, lin2026agentic, yang2026better} of software with embedded prompts to improve them, especially for agents.
By and large, these self-improving approaches rely on providing an optimizer model with the raw text input prompts, the embedded prompt codebase, and/or output generations or agent traces.
Because coding agents benefit from modeling the processes of expert developers~\cite{bajpai2024let, liang2025tabletalk} and by having better context~\cite{yang2026better}, providing self-improving agents with the same information that human developers rely on to develop embedded prompts could be a promising approach to improve these systems.
For example, instead of only providing raw text prompts, codebases, and agent traces, future work could provide agents with richer or more accurate information on embedded prompt changes.
Since developers of embedded prompts struggle to recall what changes were made and why (T17, Q32)~\cite{liang2025prompts}, future developer tools could collect the provenance of all prior prompt versions, their outputs, and the design decisions behind them to enable self-improving agents to better reason about which parts of the prompt to change to prevent embedded prompt regressions~\cite{ma2024my}.
Developer tools could also consider providing code snippets that may have resulted in the prompt (Q9) to shorten the context and provide more relevant context for the agents to reason about.

However, future research is needed to understand how to present this information to developers who are humans compared to agents.
For instance, given human developers' difficulties in making sense of and recalling embedded prompts, prompt changes, and FM output generations~\cite{liang2025prompts}, human developers may prefer to rely on traditional visualizations of software such as timelines~\cite{horvath2024meta, yoon2013visualization}, graph visualizations~\cite{kery2017variolite, latoza2011visualizing}, comparison tables~\cite{arawjo2023chainforge, arawjo2024chainforge}, and window management~\cite{coblenz2006jasper}.
Yet, given that self-improving agents can fail due to being provided long or irrelevant contexts~\cite{yang2026better}, the aforementioned visualization-based approaches that represent the full change history that are useful for human developers may not be as appropriate for agents.
While human developers may prefer to view embedded prompt change data visually, self-improving agents may be better served with tools that enable them to query specific parts of the embedded prompt change history, rather than placing the entire set of data in the agent's context~\cite{yang2026better}.

Finally, given the increasing prevalence of human-coding agent collaboration in practice~\cite{liang2025tabletalk, baumann2026swechat}, which information needs are important may change in cases where humans and agents develop embedded prompts together.
For example, to review the agent's change, the developer must not only understand what natural language text was modified (T15, Q28), but also how this resulted in new behavior (T16, Q29-31).
In these scenarios, it is especially important to allow the developer to understand how the change resulted in modifications to the embedded prompt's behavior (T16, Q29-31).
Future research could investigate new representations to visualize meaningful changes to prompt behavior.
An additional open research problem is developing techniques to help maintain and share knowledge of embedded prompt and harness changes between human and agent developers, given that coding agents can impair a developer's ability to understand generated code~\cite{balepur2026paired}.

\subsubsection{Information needs of traditional software vs. embedded prompts}
\label{sec:info-needs-software-vs-embedded-prompts}
Broadly speaking, we find that many of the information needs required for embedded prompts were similar to those identified for traditional software.
For instance, information needs related to program comprehension---such as identifying and understanding related code~\cite{maalej2014comprehension, sillito2006questions, sillito2008asking}---exhibit parallels in developing embedded prompts in the form of understanding the embedded prompt's content and behavior (T1-5, Q1-10).
The need to understand test cases~\cite{latoza2010hard} and runtime behavior~\cite{ko2007information} in traditional software also corresponds to the need to understand the inputs provided to a prompt (T6, Q11-13) and what occurred during a run (T7-9, Q14-18) respectively for embedded prompts.
However, making sense of the inputs given to an embedded prompt had a special emphasis on understanding the dataset (Q12) and the plain text data given to the prompt (Q17) rather than understanding how to test specific parts of the code~\cite{latoza2010hard}.
Next, the information needs required for debugging~\cite{ko2004designing}, such as localizing the fault and comparing different versions of code and runtime data, correspond to the need to understand why an embedded prompt behaved unexpectedly (T10-14, Q19-27), changes between current and previous versions (T15-17, Q28-32), as well as changes between two or more prompt versions (T18-19, Q33-36).
For these information needs for developing embedded prompts, there was a new emphasis on understanding how changes resulted in a difference in program behavior.
This is likely due to the opaque nature of FMs that makes it difficult for developers to attribute whether prompt changes result in meaningful differences in behavior~\cite{liang2025prompts}.
In addition, both developing traditional software and embedded prompts require an awareness of how to retrieve specific code snippets~\cite{yang2016query, dit2013feature, liu2021opportunities} and prompts (T20-23, Q37-45), focusing on searching relevant code and prompts in terms of the implementation and program behavior.
Finally, both development activities require an understanding of the history of the code program~\cite{ko2007information, latoza2010hard} and embedded prompt (T24-25, Q46-50).
However, developing embedded prompts uniquely underscored the importance of understanding progress over versions (T24), rather than simply understanding what changes occurred previously.

We find that the some important information needs in both software contexts are similar.
Having the developer understand the related code and data flow of the program is essential~\cite{sillito2006questions, sillito2008asking, latoza2010hard, ko2007information}.
For example, just as knowing which parts of the codebase depended on the prompt (Q9) was an important information need within our sample (91\% rated as frequently occurring), understanding what code may have caused the program state was among the most important information needs in traditional software~\cite{ko2007information}.
Though many information needs align with traditional software, developing embedded prompts introduces new important information needs that are not as salient in developing traditional software contexts (e.g., understanding prompt behavior and how behavior is related to specific parts of the prompt).
This set of information needs corresponds to the opaque and stochastic nature of FMs, which introduces challenges in debugging embedded prompts that can make debugging in these contexts even more difficult than in traditional software~\cite{liang2025prompts}.
For example, because FM behavior is stochastic, developers struggle to understand when a change results in a meaningful shift in behavior; thus, some of the most helpful to address information needs for embedded prompts is understanding the best performing versions (Q46, 100\%), how two different prompt versions' behavior differ (Q33, 90\%), and how specific changes result in shifts in behavior (Q20, 90\%).
However, traditional debugging approaches that rely on fault localization to specific parts of the code do not apply to software with embedded prompts due to the opaque nature of FMs~\cite{liang2025prompts}.
Given this, and the fact that certain information needs are more salient in developing embedded prompts than in traditional software, future work should investigate approaches for developing tooling specific to software with embedded prompts.
We discuss future directions in Section~\ref{sec:design-opportunities-developer-tools}.

\subsection{Design Opportunities for Tools for Developing Embedded Prompts}
\label{sec:design-opportunities-developer-tools}

\subsubsection{Understanding code dependencies of embedded prompts}
Understanding the prompt's code dependencies (Q9)---such as for parsing the inputs passed to the prompt, such as those related to an agent's environment---was among one of the most helpful to address information needs for developers (88\%).
However, we find that this specific information need, or any information need related to code (Q9, Q10, Q23, Q27) was not addressed by existing research or commercial tools, revealing an important gap for developer tools for building software with embedded prompts.
Future research is needed to investigate how to connect parts of prompts to specific parts of the codebase and render these relationships when using IDEs.
One promising technique is leveraging program analysis techniques to reveal which code may have contributed to specific program behavior.
For example, the Whyline~\cite{ko2004designing} leverages both static and dynamic slicing analyses to help developers identify reasons for why or not specific program behaviors occurred, while Reacher uses static reachability analysis to reveal whether specific code is reachable from other code~\cite{latoza2011visualizing}.

\subsubsection{Debugging embedded prompts}
A prominent unaddressed information need for developing embedded prompts is debugging. 
Five of the nine information needs for debugging (Q20-21, Q23-24, Q27) were not addressed by existing tools at all.
These information needs corresponded to debugging by comparing two versions (T11, Q20-21) and reasoning about artifacts outside of the prompt (T13, Q23-24).
However, one of these information needs---understanding what part of the change might cause the behavior in specific version (Q20)---was among one of the most helpful to address needs (90\%).
Our finding corroborates prominent work, which notes the lack of debugging tools for embedded prompts~\cite{liang2025prompts, epperson2025interactive}.
A key challenge for researchers is providing developer tools specific to different types of software with embedded prompts.
Developing agents is a notably difficult setting for debugging, since such tools need to account for having multiple agents~\cite{epperson2025interactive} and agents that interact with application UIs~\cite{liang2026understanding} rather than only natural language.
Because a key challenge of debugging agents involves making sense of large amounts of information produced by generations~\cite{epperson2025interactive}, a promising approach is to leverage FMs to reason about agent outputs.
This approach has been shown to be effective in automatically improving software with embedded prompts without human intervention~\cite{yang2026better, wu2026harnessdev}.

\subsubsection{Retrieving embedded prompts}
Retrieving embedded prompts were also not well-supported.
Five of the nine information needs related to prompt retrieval were not addressed by existing tools.
Even for information needs that had existing tooling support, most of the support came in the form of rudimentary keyword search.
However, developer information needs involving more sophisticated forms of search, including retrieving prompt versions with specific features (Q37) and structures (Q39), as well as prompt versions with specific behaviors (Q41), behavior tradeoffs (Q42), and good performance on specific aspects of behavior (Q43), were not addressed.
Additional research is needed to improve prompt retrieval. 
Prior work has investigated improving prompt retrieval through fine-tuning approaches~\cite{cheng2023uprise, rubin2022learning} and leveraging the embeddings of prompts~\cite{nashid2023retrieval}.

\section{Conclusion}
In this work, we catalog a list of \numQuestions information needs related to developing embedded prompts spread across \numTasks tasks.
To obtain this list of information needs (\textbf{RQ1}) and identify which needs are the most important to developers (\textbf{RQ2}), we interviewed \numInterviewParticipants developers about their experience making changes to embedded prompts, observed \numObservationalParticipants developers make changes to embedded prompts, and surveyed \numSurveyParticipants developers on what the most important information needs are.
We then applied the list of information needs to \numTools existing commercial and research tools to understand the extent to which the information needs are addressed (\textbf{RQ3}).
Based on these findings, we conclude with a list of salient design opportunities for developer tools for developing embedded prompts.
We provide the relevant study materials (e.g., interview scripts, surveys, codebook, tool list) in the supplemental materials~\cite{supplemental-materials}.
We hope that the findings from this study will inspire future tools that better address the needs of developers who build software with embedded prompts.
Beyond these findings, we envision that the methodology from this study can offer guidance on how to elicit and understand developer needs in other software development contexts, as we found this methodology to be effective in revealing the most important issues developers are facing.

\begin{acks}
We thank our study participants for their wonderful insights.
We are also grateful to Soham Pardeshi, Millicent Li, Christopher Kang, and Manisha Mukherjee for their feedback on the work.
Next, we extend deep gratitude to JetBrains for providing funding support for the survey.
Jenny T. Liang was supported by the National Science Foundation under grants DGE1745016 and DGE2140739.
Last but not least, we give a special thanks to Mei \meiicon, an outstanding canine software engineering researcher, for providing support and motivation throughout the study.
Any opinions, findings, conclusions, or recommendations expressed in this material are those of the authors and do not necessarily reflect the views of the sponsors.
\end{acks}

\bibliographystyle{ACM-Reference-Format}
\bibliography{sample-base}


\begin{thebibliography}{142}


\ifx \showCODEN    \undefined \def \showCODEN     #1{\unskip}     \fi
\ifx \showISBNx    \undefined \def \showISBNx     #1{\unskip}     \fi
\ifx \showISBNxiii \undefined \def \showISBNxiii  #1{\unskip}     \fi
\ifx \showISSN     \undefined \def \showISSN      #1{\unskip}     \fi
\ifx \showLCCN     \undefined \def \showLCCN      #1{\unskip}     \fi
\ifx \shownote     \undefined \def \shownote      #1{#1}          \fi
\ifx \showarticletitle \undefined \def \showarticletitle #1{#1}   \fi
\ifx \showURL      \undefined \def \showURL       {\relax}        \fi
\providecommand\bibfield[2]{#2}
\providecommand\bibinfo[2]{#2}
\providecommand\natexlab[1]{#1}
\providecommand\showeprint[2][]{arXiv:#2}

\bibitem[cur(2025)]%
        {cursur2025cursor}
 \bibinfo{year}{2025}\natexlab{}.
\newblock \bibinfo{title}{Cursor - The AI Code Editor}.
\newblock
\newblock
\shownote{Retrieved September 3, 2026 from \url{https://cursor.com/en}}.


\bibitem[lan(2025)]%
        {langsmith2026langchainhub}
 \bibinfo{year}{2025}\natexlab{}.
\newblock \bibinfo{title}{Hub - LangSmith}.
\newblock
\newblock
\shownote{Retrieved September 3, 2026 from
  \url{https://smith.langchain.com/hub}}.


\bibitem[git(2025)]%
        {github2026octoverse}
 \bibinfo{year}{2025}\natexlab{}.
\newblock \bibinfo{title}{Octoverse: A new developer joins {GitHub} every
  second as AI leads TypeScript to \#1}.
\newblock
\newblock
\shownote{Retrieved July 16, 2026 from
  \url{https://github.blog/news-insights/octoverse/octoverse-a-new-developer-joins-github-every-second-as-ai-leads-typescript-to-1/}}.


\bibitem[git(2026a)]%
        {github2026agenta}
 \bibinfo{year}{2026}\natexlab{a}.
\newblock \bibinfo{title}{Agenta-AI/agenta}.
\newblock
\newblock
\shownote{Retrieved September 3, 2026 from
  \url{https://github.com/Agenta-AI/agenta}}.


\bibitem[cha(2026)]%
        {chatgpt2025scholargpt}
 \bibinfo{year}{2026}\natexlab{}.
\newblock \bibinfo{title}{ChatGPT - ScholarGPT}.
\newblock
\newblock
\shownote{Retrieved September 3, 2026 from
  \url{https://chatgpt.com/g/g-kZ0eYXlJe-scholar-gpt}}.


\bibitem[ant(2026)]%
        {anthropic2026platform}
 \bibinfo{year}{2026}\natexlab{}.
\newblock \bibinfo{title}{Claude Platform}.
\newblock
\newblock
\shownote{Retrieved October 19, 2026 from
  \url{https://platform.claude.com/dashboard}}.


\bibitem[mic(2026)]%
        {microsoft2026azure}
 \bibinfo{year}{2026}\natexlab{}.
\newblock \bibinfo{title}{Get started with prompt flow - Azure Machine
  Learning}.
\newblock
\newblock
\shownote{Retrieved September 3, 2026 from
  \url{https://learn.microsoft.com/en-us/azure/machine-learning/prompt-flow/get-started-prompt-flow}}.


\bibitem[git(2026b)]%
        {github2026helicone}
 \bibinfo{year}{2026}\natexlab{b}.
\newblock \bibinfo{title}{Helicone/helicone}.
\newblock
\newblock
\shownote{Retrieved September 3, 2026 from
  \url{https://github.com/Helicone/helicone}}.


\bibitem[git(2026c)]%
        {github2026humanloop}
 \bibinfo{year}{2026}\natexlab{c}.
\newblock \bibinfo{title}{humanloop/humanloop-python}.
\newblock
\newblock
\shownote{Retrieved September 3, 2026 from
  \url{https://github.com/humanloop/humanloop-python}}.


\bibitem[git(2026d)]%
        {github2026langsmith}
 \bibinfo{year}{2026}\natexlab{d}.
\newblock \bibinfo{title}{langchain-ai/langsmith-sdk}.
\newblock
\newblock
\shownote{Retrieved September 3, 2026 from
  \url{https://github.com/langchain-ai/langsmith-sdk}}.


\bibitem[git(2026e)]%
        {github2026langfuse}
 \bibinfo{year}{2026}\natexlab{e}.
\newblock \bibinfo{title}{langfuse/langfuse}.
\newblock
\newblock
\shownote{Retrieved September 3, 2026 from
  \url{https://github.com/langfuse/langfuse}}.


\bibitem[git(2026f)]%
        {github2026promptlayer}
 \bibinfo{year}{2026}\natexlab{f}.
\newblock \bibinfo{title}{MagnivOrg/prompt-layer-library}.
\newblock
\newblock
\shownote{Retrieved September 3, 2026 from
  \url{https://github.com/MagnivOrg/prompt-layer-library}}.


\bibitem[git(2026g)]%
        {github2026promptflow}
 \bibinfo{year}{2026}\natexlab{g}.
\newblock \bibinfo{title}{microsoft/promptflow}.
\newblock
\newblock
\shownote{Retrieved September 3, 2026 from
  \url{https://github.com/microsoft/promptflow}}.


\bibitem[git(2026h)]%
        {github2026mlflow}
 \bibinfo{year}{2026}\natexlab{h}.
\newblock \bibinfo{title}{mlflow/mlflow}.
\newblock
\newblock
\shownote{Retrieved September 3, 2026 from
  \url{https://github.com/mlflow/mlflow}}.


\bibitem[ope(2026)]%
        {openai2026platform}
 \bibinfo{year}{2026}\natexlab{}.
\newblock \bibinfo{title}{Platform Home Page - OpenAI API}.
\newblock
\newblock
\shownote{Retrieved October 19, 2026 from
  \url{https://platform.openai.com/home}}.


\bibitem[git(2026i)]%
        {github2026promptfoo}
 \bibinfo{year}{2026}\natexlab{i}.
\newblock \bibinfo{title}{promptfoo/promptfoo}.
\newblock
\newblock
\shownote{Retrieved September 3, 2026 from
  \url{https://github.com/promptfoo/promptfoo}}.


\bibitem[git(2026j)]%
        {github2026langtrace}
 \bibinfo{year}{2026}\natexlab{j}.
\newblock \bibinfo{title}{Scale3-Labs/langtrace}.
\newblock
\newblock
\shownote{Retrieved September 3, 2026 from
  \url{https://github.com/Scale3-Labs/langtrace}}.


\bibitem[Al~Haque et~al\mbox{.}(2025)]%
        {al2025evolution}
\bibfield{author}{\bibinfo{person}{Ebtesam Al~Haque}, \bibinfo{person}{Chris
  Brown}, \bibinfo{person}{Thomas~D LaToza}, {and} \bibinfo{person}{Brittany
  Johnson}.} \bibinfo{year}{2025}\natexlab{}.
\newblock \showarticletitle{The Evolution of information seeking in software
  development: Understanding the role and impact of AI assistants}. In
  \bibinfo{booktitle}{\emph{ACM International Conference on the Foundations of
  Software Engineering (FSE Companion)}}. \bibinfo{pages}{1494--1502}.
\newblock
\href{https://doi.org/10.1145/3696630.3731665}{doi:\nolinkurl{10.1145/3696630.3731665}}


\bibitem[Allamanis and Sutton(2013)]%
        {allamanis2013and}
\bibfield{author}{\bibinfo{person}{Miltiadis Allamanis} {and}
  \bibinfo{person}{Charles Sutton}.} \bibinfo{year}{2013}\natexlab{}.
\newblock \showarticletitle{Why, when, and what: analyzing stack overflow
  questions by topic, type, and code}. In
  \bibinfo{booktitle}{\emph{International Conference on Mining Software
  Repositories (MSR)}}. IEEE, \bibinfo{pages}{53--56}.
\newblock
\href{https://doi.org/10.1109/MSR.2013.6624004}{doi:\nolinkurl{10.1109/MSR.2013.6624004}}


\bibitem[Amin et~al\mbox{.}(2025)]%
        {amin2025composable}
\bibfield{author}{\bibinfo{person}{Rifat~Mehreen Amin},
  \bibinfo{person}{Oliver~Hans K{\"u}hle}, \bibinfo{person}{Daniel Buschek},
  {and} \bibinfo{person}{Andreas Butz}.} \bibinfo{year}{2025}\natexlab{}.
\newblock \showarticletitle{Composable prompting workspaces for creative
  writing: Exploration and iteration using dynamic widgets}. In
  \bibinfo{booktitle}{\emph{Extended Abstracts of the CHI Conference on Human
  Factors in Computing Systems (CHI EA)}}. \bibinfo{pages}{1--11}.
\newblock
\href{https://doi.org/10.1145/3706599.3720243}{doi:\nolinkurl{10.1145/3706599.3720243}}


\bibitem[Arawjo et~al\mbox{.}(2024)]%
        {arawjo2024chainforge}
\bibfield{author}{\bibinfo{person}{Ian Arawjo}, \bibinfo{person}{Chelse
  Swoopes}, \bibinfo{person}{Priyan Vaithilingam}, \bibinfo{person}{Martin
  Wattenberg}, {and} \bibinfo{person}{Elena~L Glassman}.}
  \bibinfo{year}{2024}\natexlab{}.
\newblock \showarticletitle{Chainforge: A visual toolkit for prompt engineering
  and llm hypothesis testing}. In \bibinfo{booktitle}{\emph{CHI Conference on
  Human Factors in Computing Systems (CHI)}}. \bibinfo{pages}{1--18}.
\newblock
\href{https://doi.org/10.1145/3613904.3642016}{doi:\nolinkurl{10.1145/3613904.3642016}}


\bibitem[Arawjo et~al\mbox{.}(2023)]%
        {arawjo2023chainforge}
\bibfield{author}{\bibinfo{person}{Ian Arawjo}, \bibinfo{person}{Priyan
  Vaithilingam}, \bibinfo{person}{Martin Wattenberg}, {and}
  \bibinfo{person}{Elena Glassman}.} \bibinfo{year}{2023}\natexlab{}.
\newblock \showarticletitle{ChainForge: An open-source visual programming
  environment for prompt engineering}. In \bibinfo{booktitle}{\emph{ACM
  Symposium on User Interface Software and Technology: Demos Track (UIST
  Demos)}}. \bibinfo{pages}{1--3}.
\newblock
\href{https://doi.org/10.1145/3586182.3616660}{doi:\nolinkurl{10.1145/3586182.3616660}}


\bibitem[Bajpai et~al\mbox{.}(2024)]%
        {bajpai2024let}
\bibfield{author}{\bibinfo{person}{Yasharth Bajpai}, \bibinfo{person}{Bhavya
  Chopra}, \bibinfo{person}{Param Biyani}, \bibinfo{person}{Cagri Aslan},
  \bibinfo{person}{Dustin Coleman}, \bibinfo{person}{Sumit Gulwani},
  \bibinfo{person}{Chris Parnin}, \bibinfo{person}{Arjun Radhakrishna}, {and}
  \bibinfo{person}{Gustavo Soares}.} \bibinfo{year}{2024}\natexlab{}.
\newblock \showarticletitle{Let’s fix this together: Conversational debugging
  with Github Copilot}. In \bibinfo{booktitle}{\emph{IEEE Symposium on Visual
  Languages and Human-Centric Computing (VL/HCC)}}. IEEE,
  \bibinfo{pages}{1--12}.
\newblock
\href{https://doi.org/10.1109/VL/HCC60511.2024.00011}{doi:\nolinkurl{10.1109/VL/HCC60511.2024.00011}}


\bibitem[Balepur et~al\mbox{.}(2026)]%
        {balepur2026paired}
\bibfield{author}{\bibinfo{person}{Nishant Balepur}, \bibinfo{person}{Connor
  Baumler}, \bibinfo{person}{Valerie Chen}, \bibinfo{person}{Eunsol Choi},
  \bibinfo{person}{Rachel Rudinger}, {and} \bibinfo{person}{Jordan~Lee
  Boyd-Graber}.} \bibinfo{year}{2026}\natexlab{}.
\newblock \showarticletitle{(Im) Paired programming: Coding agents improve
  productivity but harm understanding}.
\newblock \bibinfo{journal}{\emph{arXiv preprint arXiv:2607.26375}}
  (\bibinfo{year}{2026}).
\newblock


\bibitem[Baumann et~al\mbox{.}(2026)]%
        {baumann2026swechat}
\bibfield{author}{\bibinfo{person}{Joachim Baumann}, \bibinfo{person}{Vishakh
  Padmakumar}, \bibinfo{person}{Xiang Li}, \bibinfo{person}{John Yang},
  \bibinfo{person}{Diyi Yang}, {and} \bibinfo{person}{Sanmi Koyejo}.}
  \bibinfo{year}{2026}\natexlab{}.
\newblock \showarticletitle{{SWE}-chat: Coding agent interactions from real
  users in the wild}. In \bibinfo{booktitle}{\emph{Conference on Language
  Modeling (COLM)}}.
\newblock


\bibitem[Brade et~al\mbox{.}(2023)]%
        {brade2023promptify}
\bibfield{author}{\bibinfo{person}{Stephen Brade}, \bibinfo{person}{Bryan
  Wang}, \bibinfo{person}{Mauricio Sousa}, \bibinfo{person}{Sageev Oore}, {and}
  \bibinfo{person}{Tovi Grossman}.} \bibinfo{year}{2023}\natexlab{}.
\newblock \showarticletitle{Promptify: Text-to-image generation through
  interactive prompt exploration with large language models}. In
  \bibinfo{booktitle}{\emph{ACM Symposium on User Interface Software and
  Technology (UIST)}}. \bibinfo{pages}{1--14}.
\newblock
\href{https://doi.org/10.1145/3586183.3606725}{doi:\nolinkurl{10.1145/3586183.3606725}}


\bibitem[Breu et~al\mbox{.}(2010)]%
        {breu2010information}
\bibfield{author}{\bibinfo{person}{Silvia Breu}, \bibinfo{person}{Rahul
  Premraj}, \bibinfo{person}{Jonathan Sillito}, {and} \bibinfo{person}{Thomas
  Zimmermann}.} \bibinfo{year}{2010}\natexlab{}.
\newblock \showarticletitle{Information needs in bug reports: Improving
  cooperation between developers and users}. In \bibinfo{booktitle}{\emph{ACM
  Conference on Computer Supported Cooperative Work (CSCW)}}.
  \bibinfo{pages}{301--310}.
\newblock
\href{https://doi.org/10.1145/1718918.1718973}{doi:\nolinkurl{10.1145/1718918.1718973}}


\bibitem[Brown et~al\mbox{.}(2020)]%
        {brown2020language}
\bibfield{author}{\bibinfo{person}{Tom Brown}, \bibinfo{person}{Benjamin Mann},
  \bibinfo{person}{Nick Ryder}, \bibinfo{person}{Melanie Subbiah},
  \bibinfo{person}{Jared~D Kaplan}, \bibinfo{person}{Prafulla Dhariwal},
  \bibinfo{person}{Arvind Neelakantan}, \bibinfo{person}{Pranav Shyam},
  \bibinfo{person}{Girish Sastry}, \bibinfo{person}{Amanda Askell},
  {et~al\mbox{.}}} \bibinfo{year}{2020}\natexlab{}.
\newblock \showarticletitle{Language models are few-shot learners}.
\newblock \bibinfo{journal}{\emph{Advances in Neural Information Processing
  Systems (NeurIPS)}}  \bibinfo{volume}{33} (\bibinfo{year}{2020}),
  \bibinfo{pages}{1877--1901}.
\newblock


\bibitem[Chavan et~al\mbox{.}(2024)]%
        {chavan2024analyzing}
\bibfield{author}{\bibinfo{person}{Omkar~Sandip Chavan},
  \bibinfo{person}{Divya~Dilip Hinge}, \bibinfo{person}{Soham~Sanjay Deo},
  \bibinfo{person}{Yaxuan Wang}, {and} \bibinfo{person}{Mohamed~Wiem Mkaouer}.}
  \bibinfo{year}{2024}\natexlab{}.
\newblock \showarticletitle{Analyzing developer-ChatGPT conversations for
  software refactoring: An exploratory study}. In
  \bibinfo{booktitle}{\emph{International Conference on Mining Software
  Repositories (MSR)}}. \bibinfo{pages}{207--211}.
\newblock
\href{https://doi.org/10.1145/3643991.3645082}{doi:\nolinkurl{10.1145/3643991.3645082}}


\bibitem[Chen et~al\mbox{.}(2024)]%
        {chen2023chatgpt}
\bibfield{author}{\bibinfo{person}{Lingjiao Chen}, \bibinfo{person}{Matei
  Zaharia}, {and} \bibinfo{person}{James Zou}.}
  \bibinfo{year}{2024}\natexlab{}.
\newblock \showarticletitle{How is ChatGPT’s behavior changing over time?}
\newblock \bibinfo{journal}{\emph{Harvard Data Science Review}}
  \bibinfo{volume}{6}, \bibinfo{number}{2} (\bibinfo{year}{2024}).
\newblock
\href{https://doi.org/10.1162/99608f92.5317da47}{doi:\nolinkurl{10.1162/99608f92.5317da47}}


\bibitem[Chen et~al\mbox{.}(2026)]%
        {chen2026promptware}
\bibfield{author}{\bibinfo{person}{Zhenpeng Chen}, \bibinfo{person}{Chong
  Wang}, \bibinfo{person}{Weisong Sun}, \bibinfo{person}{Xuanzhe Liu},
  \bibinfo{person}{Jie~M Zhang}, {and} \bibinfo{person}{Yang Liu}.}
  \bibinfo{year}{2026}\natexlab{}.
\newblock \showarticletitle{Promptware engineering: Software engineering for
  prompt-enabled systems}.
\newblock \bibinfo{journal}{\emph{ACM Transactions on Software Engineering and
  Methodology (TOSEM)}} (\bibinfo{year}{2026}).
\newblock
\href{https://doi.org/10.1145/3796535}{doi:\nolinkurl{10.1145/3796535}}


\bibitem[Cheng et~al\mbox{.}(2023)]%
        {cheng2023uprise}
\bibfield{author}{\bibinfo{person}{Daixuan Cheng}, \bibinfo{person}{Shaohan
  Huang}, \bibinfo{person}{Junyu Bi}, \bibinfo{person}{Yuefeng Zhan},
  \bibinfo{person}{Jianfeng Liu}, \bibinfo{person}{Yujing Wang},
  \bibinfo{person}{Hao Sun}, \bibinfo{person}{Furu Wei},
  \bibinfo{person}{Weiwei Deng}, {and} \bibinfo{person}{Qi Zhang}.}
  \bibinfo{year}{2023}\natexlab{}.
\newblock \showarticletitle{{UPRISE}: Universal prompt retrieval for improving
  zero-shot evaluation}. In \bibinfo{booktitle}{\emph{Conference on Empirical
  Methods in Natural Language Processing (EMNLP)}}.
  \bibinfo{pages}{12318--12337}.
\newblock
\href{https://doi.org/10.18653/v1/2023.emnlp-main.758}{doi:\nolinkurl{10.18653/v1/2023.emnlp-main.758}}


\bibitem[Cheng et~al\mbox{.}(2024)]%
        {cheng2024prompt}
\bibfield{author}{\bibinfo{person}{Yu Cheng}, \bibinfo{person}{Jieshan Chen},
  \bibinfo{person}{Qing Huang}, \bibinfo{person}{Zhenchang Xing},
  \bibinfo{person}{Xiwei Xu}, {and} \bibinfo{person}{Qinghua Lu}.}
  \bibinfo{year}{2024}\natexlab{}.
\newblock \showarticletitle{Prompt {S}apper: A {LLM}-empowered production tool
  for building AI chains}.
\newblock \bibinfo{journal}{\emph{ACM Transactions on Software Engineering and
  Methodology (TOSEM)}} \bibinfo{volume}{33}, \bibinfo{number}{5}
  (\bibinfo{year}{2024}), \bibinfo{pages}{1--24}.
\newblock
\href{https://doi.org/10.1145/3638247}{doi:\nolinkurl{10.1145/3638247}}


\bibitem[Coblenz et~al\mbox{.}(2006)]%
        {coblenz2006jasper}
\bibfield{author}{\bibinfo{person}{Michael~J Coblenz}, \bibinfo{person}{Amy~J
  Ko}, {and} \bibinfo{person}{Brad~A Myers}.} \bibinfo{year}{2006}\natexlab{}.
\newblock \showarticletitle{JASPER: an Eclipse plug-in to facilitate software
  maintenance tasks}. In \bibinfo{booktitle}{\emph{Workshop on Eclipse
  Technology eXchange (eTX)}}. \bibinfo{pages}{65--69}.
\newblock
\href{https://doi.org/10.1145/1188835.1188849}{doi:\nolinkurl{10.1145/1188835.1188849}}


\bibitem[Corbin and Strauss(2015)]%
        {corbin2015basics}
\bibfield{author}{\bibinfo{person}{Juliet Corbin} {and} \bibinfo{person}{Anselm
  Strauss}.} \bibinfo{year}{2015}\natexlab{}.
\newblock \bibinfo{booktitle}{\emph{Basics of qualitative research}}.
  Vol.~\bibinfo{volume}{14}.
\newblock \bibinfo{publisher}{Sage}.
\newblock
\href{https://doi.org/10.1177/1094428108324514}{doi:\nolinkurl{10.1177/1094428108324514}}


\bibitem[Dang et~al\mbox{.}(2023)]%
        {dang2023worldsmith}
\bibfield{author}{\bibinfo{person}{Hai Dang}, \bibinfo{person}{Frederik Brudy},
  \bibinfo{person}{George Fitzmaurice}, {and} \bibinfo{person}{Fraser
  Anderson}.} \bibinfo{year}{2023}\natexlab{}.
\newblock \showarticletitle{Worldsmith: Iterative and expressive prompting for
  world building with a generative AI}. In \bibinfo{booktitle}{\emph{ACM
  Symposium on User Interface Software and Technology (UIST)}}.
  \bibinfo{pages}{1--17}.
\newblock
\href{https://doi.org/10.1145/3586183.3606772}{doi:\nolinkurl{10.1145/3586183.3606772}}


\bibitem[Dit et~al\mbox{.}(2013)]%
        {dit2013feature}
\bibfield{author}{\bibinfo{person}{Bogdan Dit}, \bibinfo{person}{Meghan
  Revelle}, \bibinfo{person}{Malcom Gethers}, {and} \bibinfo{person}{Denys
  Poshyvanyk}.} \bibinfo{year}{2013}\natexlab{}.
\newblock \showarticletitle{Feature location in source code: a taxonomy and
  survey}.
\newblock \bibinfo{journal}{\emph{Journal of software: Evolution and Process}}
  \bibinfo{volume}{25}, \bibinfo{number}{1} (\bibinfo{year}{2013}),
  \bibinfo{pages}{53--95}.
\newblock
\href{https://doi.org/10.1002/smr.567}{doi:\nolinkurl{10.1002/smr.567}}


\bibitem[Dolata et~al\mbox{.}(2024)]%
        {dolata2024development}
\bibfield{author}{\bibinfo{person}{Mateusz Dolata}, \bibinfo{person}{Norbert
  Lange}, {and} \bibinfo{person}{Gerhard Schwabe}.}
  \bibinfo{year}{2024}\natexlab{}.
\newblock \showarticletitle{Development in times of hype: How freelancers
  explore Generative AI?}. In \bibinfo{booktitle}{\emph{IEEE/ACM International
  Conference on Software Engineering (ICSE)}}. \bibinfo{pages}{1--13}.
\newblock
\href{https://doi.org/10.1145/3597503.3639111}{doi:\nolinkurl{10.1145/3597503.3639111}}


\bibitem[Duala-Ekoko and Robillard(2012)]%
        {duala2012asking}
\bibfield{author}{\bibinfo{person}{Ekwa Duala-Ekoko} {and}
  \bibinfo{person}{Martin~P Robillard}.} \bibinfo{year}{2012}\natexlab{}.
\newblock \showarticletitle{Asking and answering questions about unfamiliar
  APIs: An exploratory study}. In \bibinfo{booktitle}{\emph{International
  Conference on Software Engineering (ICSE)}}. IEEE, \bibinfo{pages}{266--276}.
\newblock
\href{https://doi.org/10.1109/ICSE.2012.6227187}{doi:\nolinkurl{10.1109/ICSE.2012.6227187}}


\bibitem[Epperson et~al\mbox{.}(2025)]%
        {epperson2025interactive}
\bibfield{author}{\bibinfo{person}{Will Epperson}, \bibinfo{person}{Gagan
  Bansal}, \bibinfo{person}{Victor~C Dibia}, \bibinfo{person}{Adam Fourney},
  \bibinfo{person}{Jack Gerrits}, \bibinfo{person}{Erkang Zhu}, {and}
  \bibinfo{person}{Saleema Amershi}.} \bibinfo{year}{2025}\natexlab{}.
\newblock \showarticletitle{Interactive debugging and steering of multi-agent
  ai systems}. In \bibinfo{booktitle}{\emph{CHI Conference on Human Factors in
  Computing Systems}}. \bibinfo{pages}{1--15}.
\newblock
\href{https://doi.org/10.1145/3706598.3713581}{doi:\nolinkurl{10.1145/3706598.3713581}}


\bibitem[Feng et~al\mbox{.}(2024)]%
        {feng2024coprompt}
\bibfield{author}{\bibinfo{person}{Li Feng}, \bibinfo{person}{Ryan Yen},
  \bibinfo{person}{Yuzhe You}, \bibinfo{person}{Mingming Fan},
  \bibinfo{person}{Jian Zhao}, {and} \bibinfo{person}{Zhicong Lu}.}
  \bibinfo{year}{2024}\natexlab{}.
\newblock \showarticletitle{Coprompt: Supporting prompt sharing and referring
  in collaborative natural language programming}. In
  \bibinfo{booktitle}{\emph{CHI Conference on Human Factors in Computing
  Systems}}. \bibinfo{pages}{1--21}.
\newblock
\href{https://doi.org/10.1145/3613904.3642212}{doi:\nolinkurl{10.1145/3613904.3642212}}


\bibitem[Fernando et~al\mbox{.}(2024)]%
        {fernando2024promptbreeder}
\bibfield{author}{\bibinfo{person}{Chrisantha Fernando},
  \bibinfo{person}{Dylan~Sunil Banarse}, \bibinfo{person}{Henryk Michalewski},
  \bibinfo{person}{Simon Osindero}, {and} \bibinfo{person}{Tim
  Rockt{\"a}schel}.} \bibinfo{year}{2024}\natexlab{}.
\newblock \showarticletitle{Promptbreeder: Self-referential self-improvement
  via prompt evolution}. In \bibinfo{booktitle}{\emph{International Conference
  on Machine Learning (ICML)}}.
\newblock


\bibitem[Fiannaca et~al\mbox{.}(2023)]%
        {fiannaca2023programming}
\bibfield{author}{\bibinfo{person}{Alexander~J Fiannaca},
  \bibinfo{person}{Chinmay Kulkarni}, \bibinfo{person}{Carrie~J Cai}, {and}
  \bibinfo{person}{Michael Terry}.} \bibinfo{year}{2023}\natexlab{}.
\newblock \showarticletitle{Programming without a programming language:
  Challenges and opportunities for designing developer tools for prompt
  programming}. In \bibinfo{booktitle}{\emph{CHI Conference on Human Factors in
  Computing Systems Extended Abstracts (CHI EA)}}. \bibinfo{pages}{1--7}.
\newblock
\href{https://doi.org/10.1145/3544549.3585737}{doi:\nolinkurl{10.1145/3544549.3585737}}


\bibitem[Fu et~al\mbox{.}(2024)]%
        {fu2024ap}
\bibfield{author}{\bibinfo{person}{Yuchen Fu}, \bibinfo{person}{Zhiwei Jiang},
  \bibinfo{person}{Yuliang Liu}, \bibinfo{person}{Cong Wang},
  \bibinfo{person}{Zexuan Deng}, \bibinfo{person}{Zhaoling Chen}, {and}
  \bibinfo{person}{Qing Gu}.} \bibinfo{year}{2024}\natexlab{}.
\newblock \showarticletitle{AP-Adapter: Improving generalization of automatic
  prompts on unseen text-to-image diffusion models}.
\newblock \bibinfo{journal}{\emph{Advances in Neural Information Processing
  Systems (NeurIPS)}}  \bibinfo{volume}{37} (\bibinfo{year}{2024}),
  \bibinfo{pages}{98320--98346}.
\newblock
\href{https://doi.org/10.52202/079017-3120}{doi:\nolinkurl{10.52202/079017-3120}}


\bibitem[Gero et~al\mbox{.}(2022)]%
        {gero2022sensemaking}
\bibfield{author}{\bibinfo{person}{Katy~Ilonka Gero},
  \bibinfo{person}{Jonathan~K Kummerfeld}, {and} \bibinfo{person}{Elena~L
  Glassman}.} \bibinfo{year}{2022}\natexlab{}.
\newblock \showarticletitle{Sensemaking interfaces for human evaluation of
  language model outputs}. In \bibinfo{booktitle}{\emph{NeurIPS Workshop on
  Human Evaluation of Generative Models}}.
\newblock


\bibitem[Gero et~al\mbox{.}(2024)]%
        {gero2024supporting}
\bibfield{author}{\bibinfo{person}{Katy~Ilonka Gero}, \bibinfo{person}{Chelse
  Swoopes}, \bibinfo{person}{Ziwei Gu}, \bibinfo{person}{Jonathan~K
  Kummerfeld}, {and} \bibinfo{person}{Elena~L Glassman}.}
  \bibinfo{year}{2024}\natexlab{}.
\newblock \showarticletitle{Supporting sensemaking of large language model
  outputs at scale}. In \bibinfo{booktitle}{\emph{CHI Conference on Human
  Factors in Computing Systems (CHI)}}. \bibinfo{pages}{1--21}.
\newblock
\href{https://doi.org/10.1145/3613904.3642139}{doi:\nolinkurl{10.1145/3613904.3642139}}


\bibitem[Gilmore(1991)]%
        {gilmore1991models}
\bibfield{author}{\bibinfo{person}{David~J Gilmore}.}
  \bibinfo{year}{1991}\natexlab{}.
\newblock \showarticletitle{Models of debugging}.
\newblock \bibinfo{journal}{\emph{Acta psychologica}} \bibinfo{volume}{78},
  \bibinfo{number}{1-3} (\bibinfo{year}{1991}), \bibinfo{pages}{151--172}.
\newblock
\href{https://doi.org/10.1016/0001-6918(91)90009-O}{doi:\nolinkurl{10.1016/0001-6918(91)90009-O}}


\bibitem[Giray(2021)]%
        {giray2021software}
\bibfield{author}{\bibinfo{person}{G{\"o}rkem Giray}.}
  \bibinfo{year}{2021}\natexlab{}.
\newblock \showarticletitle{A software engineering perspective on engineering
  machine learning systems: State of the art and challenges}.
\newblock \bibinfo{journal}{\emph{Journal of Systems and Software (JSS)}}
  \bibinfo{volume}{180} (\bibinfo{year}{2021}), \bibinfo{pages}{111031}.
\newblock
\href{https://doi.org/10.1016/j.jss.2021.111031}{doi:\nolinkurl{10.1016/j.jss.2021.111031}}


\bibitem[Haenni et~al\mbox{.}(2013)]%
        {haenni2013categorizing}
\bibfield{author}{\bibinfo{person}{Nicole Haenni}, \bibinfo{person}{Mircea
  Lungu}, \bibinfo{person}{Niko Schwarz}, {and} \bibinfo{person}{Oscar
  Nierstrasz}.} \bibinfo{year}{2013}\natexlab{}.
\newblock \showarticletitle{Categorizing developer information needs in
  software ecosystems}. In \bibinfo{booktitle}{\emph{International Workshop on
  Ecosystem Architectures (WEA)}}. \bibinfo{pages}{1--5}.
\newblock
\href{https://doi.org/10.1145/2501585.2501586}{doi:\nolinkurl{10.1145/2501585.2501586}}


\bibitem[Hammer and Berland(2014)]%
        {hammer2014confusing}
\bibfield{author}{\bibinfo{person}{David Hammer} {and} \bibinfo{person}{Leema~K
  Berland}.} \bibinfo{year}{2014}\natexlab{}.
\newblock \showarticletitle{Confusing claims for data: A critique of common
  practices for presenting qualitative research on learning}.
\newblock \bibinfo{journal}{\emph{Journal of the Learning Sciences}}
  \bibinfo{volume}{23}, \bibinfo{number}{1} (\bibinfo{year}{2014}),
  \bibinfo{pages}{37--46}.
\newblock
\href{https://doi.org/10.1080/10508406.2013.802652}{doi:\nolinkurl{10.1080/10508406.2013.802652}}


\bibitem[Hao et~al\mbox{.}(2025)]%
        {hao2025flowforge}
\bibfield{author}{\bibinfo{person}{Pan Hao}, \bibinfo{person}{Dongyeop Kang},
  \bibinfo{person}{Nicholas Hinds}, {and} \bibinfo{person}{Qianwen Wang}.}
  \bibinfo{year}{2025}\natexlab{}.
\newblock \showarticletitle{FlowForge: Guiding the creation of multi-agent
  workflows with design space visualization as a thinking scaffold}.
\newblock \bibinfo{journal}{\emph{IEEE Transactions on Visualization and
  Computer Graphics (TVCG)}} (\bibinfo{year}{2025}).
\newblock
\href{https://doi.org/10.1109/TVCG.2025.3634627}{doi:\nolinkurl{10.1109/TVCG.2025.3634627}}


\bibitem[Hao et~al\mbox{.}(2023)]%
        {hao2023optimizing}
\bibfield{author}{\bibinfo{person}{Yaru Hao}, \bibinfo{person}{Zewen Chi},
  \bibinfo{person}{Li Dong}, {and} \bibinfo{person}{Furu Wei}.}
  \bibinfo{year}{2023}\natexlab{}.
\newblock \showarticletitle{Optimizing prompts for text-to-image generation}.
\newblock \bibinfo{journal}{\emph{Advances in Neural Information Processing
  Systems (NeurIPS)}}  \bibinfo{volume}{36} (\bibinfo{year}{2023}),
  \bibinfo{pages}{66923--66939}.
\newblock
\href{https://doi.org/10.52202/075280-2923}{doi:\nolinkurl{10.52202/075280-2923}}


\bibitem[Hassan et~al\mbox{.}(2024)]%
        {hassan2024rethinking}
\bibfield{author}{\bibinfo{person}{Ahmed~E Hassan}, \bibinfo{person}{Dayi Lin},
  \bibinfo{person}{Gopi~Krishnan Rajbahadur}, \bibinfo{person}{Keheliya
  Gallaba}, \bibinfo{person}{Filipe~Roseiro Cogo}, \bibinfo{person}{Boyuan
  Chen}, \bibinfo{person}{Haoxiang Zhang}, \bibinfo{person}{Kishanthan
  Thangarajah}, \bibinfo{person}{Gustavo Oliva}, \bibinfo{person}{Jiahuei Lin},
  {et~al\mbox{.}}} \bibinfo{year}{2024}\natexlab{}.
\newblock \showarticletitle{Rethinking software engineering in the era of
  foundation models: A curated catalogue of challenges in the development of
  trustworthy {FM}ware}. In \bibinfo{booktitle}{\emph{ACM International
  Conference on the Foundations of Software Engineering (FSE)---Industry
  Papers}}. \bibinfo{pages}{294--305}.
\newblock
\href{https://doi.org/10.1145/3663529.3663849}{doi:\nolinkurl{10.1145/3663529.3663849}}


\bibitem[He et~al\mbox{.}(2025)]%
        {he2025prompting}
\bibfield{author}{\bibinfo{person}{Zeyu He}, \bibinfo{person}{Saniya Naphade},
  {and} \bibinfo{person}{Ting-Hao~Kenneth Huang}.}
  \bibinfo{year}{2025}\natexlab{}.
\newblock \showarticletitle{Prompting in the dark: Assessing human performance
  in prompt engineering for data labeling when gold labels are absent}. In
  \bibinfo{booktitle}{\emph{ACM CHI Conference on Human Factors in Computing
  Systems (CHI)}}. \bibinfo{pages}{1--33}.
\newblock
\href{https://doi.org/10.1145/3706598.3714319}{doi:\nolinkurl{10.1145/3706598.3714319}}


\bibitem[Holtzblatt and Beyer(1997)]%
        {holtzblatt1997contextual}
\bibfield{author}{\bibinfo{person}{Karen Holtzblatt} {and}
  \bibinfo{person}{Hugh Beyer}.} \bibinfo{year}{1997}\natexlab{}.
\newblock \bibinfo{booktitle}{\emph{Contextual design: defining
  customer-centered systems}}.
\newblock \bibinfo{publisher}{Elsevier}.
\newblock


\bibitem[Horvath et~al\mbox{.}(2024)]%
        {horvath2024meta}
\bibfield{author}{\bibinfo{person}{Amber Horvath}, \bibinfo{person}{Andrew
  Macvean}, {and} \bibinfo{person}{Brad~A Myers}.}
  \bibinfo{year}{2024}\natexlab{}.
\newblock \showarticletitle{Meta-manager: A tool for collecting and exploring
  meta information about code}. In \bibinfo{booktitle}{\emph{ACM CHI Conference
  on Human Factors in Computing Systems (CHI)}}. \bibinfo{pages}{1--17}.
\newblock
\href{https://doi.org/10.1145/3613904.3642676}{doi:\nolinkurl{10.1145/3613904.3642676}}


\bibitem[Hu et~al\mbox{.}(2024a)]%
        {hu2024leveraging}
\bibfield{author}{\bibinfo{person}{Jian Hu}, \bibinfo{person}{Jiayi Lin},
  \bibinfo{person}{Junchi Yan}, {and} \bibinfo{person}{Shaogang Gong}.}
  \bibinfo{year}{2024}\natexlab{a}.
\newblock \showarticletitle{Leveraging hallucinations to reduce manual prompt
  dependency in promptable segmentation}.
\newblock \bibinfo{journal}{\emph{Advances in Neural Information Processing
  Systems (NeurIPS)}}  \bibinfo{volume}{37} (\bibinfo{year}{2024}),
  \bibinfo{pages}{107171--107197}.
\newblock
\href{https://doi.org/10.52202/079017-3403}{doi:\nolinkurl{10.52202/079017-3403}}


\bibitem[Hu et~al\mbox{.}(2024b)]%
        {hu2024localized}
\bibfield{author}{\bibinfo{person}{Wenyang Hu}, \bibinfo{person}{Yao Shu},
  \bibinfo{person}{Zongmin Yu}, \bibinfo{person}{Zhaoxuan Wu},
  \bibinfo{person}{Xiaoqiang Lin}, \bibinfo{person}{Zhongxiang Dai},
  \bibinfo{person}{See-Kiong Ng}, {and} \bibinfo{person}{Bryan Kian~Hsiang
  Low}.} \bibinfo{year}{2024}\natexlab{b}.
\newblock \showarticletitle{Localized zeroth-order prompt optimization}.
\newblock \bibinfo{journal}{\emph{Advances in Neural Information Processing
  Systems (NeurIPS)}}  \bibinfo{volume}{37} (\bibinfo{year}{2024}),
  \bibinfo{pages}{86309--86345}.
\newblock
\href{https://doi.org/10.52202/079017-2740}{doi:\nolinkurl{10.52202/079017-2740}}


\bibitem[Huang et~al\mbox{.}(2021)]%
        {huang2021leaving}
\bibfield{author}{\bibinfo{person}{Yu Huang}, \bibinfo{person}{Denae Ford},
  {and} \bibinfo{person}{Thomas Zimmermann}.} \bibinfo{year}{2021}\natexlab{}.
\newblock \showarticletitle{Leaving my fingerprints: Motivations and challenges
  of contributing to OSS for social good}. In
  \bibinfo{booktitle}{\emph{IEEE/ACM International Conference on Software
  Engineering (ICSE)}}. IEEE, \bibinfo{pages}{1020--1032}.
\newblock
\href{https://doi.org/10.1109/ICSE43902.2021.0009}{doi:\nolinkurl{10.1109/ICSE43902.2021.0009}}


\bibitem[Jiang et~al\mbox{.}(2022)]%
        {jiang2022promptmaker}
\bibfield{author}{\bibinfo{person}{Ellen Jiang}, \bibinfo{person}{Kristen
  Olson}, \bibinfo{person}{Edwin Toh}, \bibinfo{person}{Alejandra Molina},
  \bibinfo{person}{Aaron Donsbach}, \bibinfo{person}{Michael Terry}, {and}
  \bibinfo{person}{Carrie~J Cai}.} \bibinfo{year}{2022}\natexlab{}.
\newblock \showarticletitle{Promptmaker: Prompt-based prototyping with large
  language models}. In \bibinfo{booktitle}{\emph{CHI Conference on Human
  Factors in Computing Systems Extended Abstracts (CHI EA)}}.
  \bibinfo{pages}{1--8}.
\newblock
\href{https://doi.org/10.1145/3491101.3503564}{doi:\nolinkurl{10.1145/3491101.3503564}}


\bibitem[Jiang et~al\mbox{.}(2023)]%
        {jiang2023graphologue}
\bibfield{author}{\bibinfo{person}{Peiling Jiang}, \bibinfo{person}{Jude
  Rayan}, \bibinfo{person}{Steven~P Dow}, {and} \bibinfo{person}{Haijun Xia}.}
  \bibinfo{year}{2023}\natexlab{}.
\newblock \showarticletitle{Graphologue: Exploring large language model
  responses with interactive diagrams}. In \bibinfo{booktitle}{\emph{ACM
  Symposium on User Interface Software and Technology}}.
  \bibinfo{pages}{1--20}.
\newblock
\href{https://doi.org/10.1145/3586183.3606737}{doi:\nolinkurl{10.1145/3586183.3606737}}


\bibitem[Johnson and Erdem(1997)]%
        {johnson1997interactive}
\bibfield{author}{\bibinfo{person}{W~Lewis Johnson} {and} \bibinfo{person}{Ali
  Erdem}.} \bibinfo{year}{1997}\natexlab{}.
\newblock \showarticletitle{Interactive explanation of software systems}.
\newblock \bibinfo{journal}{\emph{Automated Software Engineering}}
  \bibinfo{volume}{4}, \bibinfo{number}{1} (\bibinfo{year}{1997}),
  \bibinfo{pages}{53--75}.
\newblock
\href{https://doi.org/10.1023/A:1008655629091}{doi:\nolinkurl{10.1023/A:1008655629091}}


\bibitem[Joshi et~al\mbox{.}(2025)]%
        {joshi2025coprompter}
\bibfield{author}{\bibinfo{person}{Ishika Joshi}, \bibinfo{person}{Simra
  Shahid}, \bibinfo{person}{Shreeya~Manasvi Venneti},
  \bibinfo{person}{Manushree Vasu}, \bibinfo{person}{Yantao Zheng},
  \bibinfo{person}{Yunyao Li}, \bibinfo{person}{Balaji Krishnamurthy}, {and}
  \bibinfo{person}{Gromit Yeuk-Yin Chan}.} \bibinfo{year}{2025}\natexlab{}.
\newblock \showarticletitle{Coprompter: User-centric evaluation of LLM
  instruction alignment for improved prompt engineering}. In
  \bibinfo{booktitle}{\emph{International Conference on Intelligent User
  Interfaces (IUI)}}. \bibinfo{pages}{341--365}.
\newblock
\href{https://doi.org/10.1145/3708359.3712102}{doi:\nolinkurl{10.1145/3708359.3712102}}


\bibitem[Kery(2021)]%
        {kery2021designing}
\bibfield{author}{\bibinfo{person}{Mary~Beth Kery}.}
  \bibinfo{year}{2021}\natexlab{}.
\newblock \emph{\bibinfo{title}{Designing effective history support for
  exploratory programming data work}}.
\newblock \bibinfo{thesistype}{Ph.\,D. Dissertation}. \bibinfo{school}{Ph. D.
  Dissertation. Carnegie Mellon University, Pittsburgh, Pennsylvania}.
\newblock


\bibitem[Kery et~al\mbox{.}(2017)]%
        {kery2017variolite}
\bibfield{author}{\bibinfo{person}{Mary~Beth Kery}, \bibinfo{person}{Amber
  Horvath}, {and} \bibinfo{person}{Brad~A Myers}.}
  \bibinfo{year}{2017}\natexlab{}.
\newblock \showarticletitle{Variolite: Supporting exploratory programming by
  data scientists}. In \bibinfo{booktitle}{\emph{ACM CHI Conference on Human
  Factors in Computing Systems (CHI)}}, Vol.~\bibinfo{volume}{10}.
  \bibinfo{pages}{3025453--3025626}.
\newblock
\href{https://doi.org/10.1145/3025453.3025626}{doi:\nolinkurl{10.1145/3025453.3025626}}


\bibitem[Khattab et~al\mbox{.}(2024)]%
        {khattab2024dspy}
\bibfield{author}{\bibinfo{person}{Omar Khattab}, \bibinfo{person}{Arnav
  Singhvi}, \bibinfo{person}{Paridhi Maheshwari}, \bibinfo{person}{Zhiyuan
  Zhang}, \bibinfo{person}{Keshav Santhanam}, \bibinfo{person}{Saiful Haq},
  \bibinfo{person}{Ashutosh Sharma}, \bibinfo{person}{Thomas Joshi},
  \bibinfo{person}{Hanna Moazam}, \bibinfo{person}{Heather Miller},
  {et~al\mbox{.}}} \bibinfo{year}{2024}\natexlab{}.
\newblock \showarticletitle{Dspy: Compiling declarative language model calls
  into state-of-the-art pipelines}. In \bibinfo{booktitle}{\emph{International
  Conference on Learning Representations (ICLR)}}, Vol.~\bibinfo{volume}{2024}.
  \bibinfo{pages}{54928--54958}.
\newblock


\bibitem[Kim et~al\mbox{.}(2024)]%
        {kim2024evallm}
\bibfield{author}{\bibinfo{person}{Tae~Soo Kim}, \bibinfo{person}{Yoonjoo Lee},
  \bibinfo{person}{Jamin Shin}, \bibinfo{person}{Young-Ho Kim}, {and}
  \bibinfo{person}{Juho Kim}.} \bibinfo{year}{2024}\natexlab{}.
\newblock \showarticletitle{Evallm: Interactive evaluation of large language
  model prompts on user-defined criteria}. In \bibinfo{booktitle}{\emph{CHI
  Conference on Human Factors in Computing Systems (CHI)}}.
  \bibinfo{pages}{1--21}.
\newblock
\href{https://doi.org/10.1145/3613904.3642216}{doi:\nolinkurl{10.1145/3613904.3642216}}


\bibitem[Kitchenham and Pfleeger(2008)]%
        {kitchenham2008personal}
\bibfield{author}{\bibinfo{person}{Barbara~A Kitchenham} {and}
  \bibinfo{person}{Shari~L Pfleeger}.} \bibinfo{year}{2008}\natexlab{}.
\newblock \showarticletitle{Personal opinion surveys}.
\newblock In \bibinfo{booktitle}{\emph{Guide to Advanced Empirical Software
  Engineering}}. \bibinfo{publisher}{Springer}, \bibinfo{pages}{63--92}.
\newblock
\href{https://doi.org/10.1007/978-1-84800-044-5_3}{doi:\nolinkurl{10.1007/978-1-84800-044-5_3}}


\bibitem[Ko et~al\mbox{.}(2007)]%
        {ko2007information}
\bibfield{author}{\bibinfo{person}{Amy~J Ko}, \bibinfo{person}{Robert DeLine},
  {and} \bibinfo{person}{Gina Venolia}.} \bibinfo{year}{2007}\natexlab{}.
\newblock \showarticletitle{Information needs in collocated software
  development teams}. In \bibinfo{booktitle}{\emph{International Conference on
  Software Engineering (ICSE)}}. IEEE, \bibinfo{pages}{344--353}.
\newblock
\href{https://doi.org/10.1109/ICSE.2007.45}{doi:\nolinkurl{10.1109/ICSE.2007.45}}


\bibitem[Ko et~al\mbox{.}(2015)]%
        {ko2015practical}
\bibfield{author}{\bibinfo{person}{Amy~J Ko}, \bibinfo{person}{Thomas~D
  LaToza}, {and} \bibinfo{person}{Margaret~M Burnett}.}
  \bibinfo{year}{2015}\natexlab{}.
\newblock \showarticletitle{A practical guide to controlled experiments of
  software engineering tools with human participants}.
\newblock \bibinfo{journal}{\emph{Empirical Software Engineering (ESE)}}
  \bibinfo{volume}{20} (\bibinfo{year}{2015}), \bibinfo{pages}{110--141}.
\newblock
\href{https://doi.org/10.1007/s10664-013-9279-3}{doi:\nolinkurl{10.1007/s10664-013-9279-3}}


\bibitem[Ko and Myers(2004)]%
        {ko2004designing}
\bibfield{author}{\bibinfo{person}{Amy~J Ko} {and} \bibinfo{person}{Brad~A
  Myers}.} \bibinfo{year}{2004}\natexlab{}.
\newblock \showarticletitle{Designing the whyline: A debugging interface for
  asking questions about program behavior}. In \bibinfo{booktitle}{\emph{ACM
  CHI Conference on Human Factors in Computing Systems (CHI)}}.
  \bibinfo{pages}{151--158}.
\newblock
\href{https://doi.org/10.1145/985692.985712}{doi:\nolinkurl{10.1145/985692.985712}}


\bibitem[Ko and Myers(2008)]%
        {ko2008debugging}
\bibfield{author}{\bibinfo{person}{Amy~J Ko} {and} \bibinfo{person}{Brad~A
  Myers}.} \bibinfo{year}{2008}\natexlab{}.
\newblock \showarticletitle{Debugging reinvented: Asking and answering why and
  why not questions about program behavior}. In
  \bibinfo{booktitle}{\emph{International Conference on Software Engineering
  (ICSE)}}. \bibinfo{pages}{301--310}.
\newblock
\href{https://doi.org/10.1145/1368088.1368130}{doi:\nolinkurl{10.1145/1368088.1368130}}


\bibitem[Ko and Myers(2010)]%
        {ko2010extracting}
\bibfield{author}{\bibinfo{person}{Amy~J Ko} {and} \bibinfo{person}{Brad~A
  Myers}.} \bibinfo{year}{2010}\natexlab{}.
\newblock \showarticletitle{Extracting and answering why and why not questions
  about Java program output}.
\newblock \bibinfo{journal}{\emph{ACM Transactions on Software Engineering and
  Methodology (TOSEM)}} \bibinfo{volume}{20}, \bibinfo{number}{2}
  (\bibinfo{year}{2010}), \bibinfo{pages}{1--36}.
\newblock
\href{https://doi.org/10.1145/1824760.1824761}{doi:\nolinkurl{10.1145/1824760.1824761}}


\bibitem[Landis and Koch(1977)]%
        {landis1977measurement}
\bibfield{author}{\bibinfo{person}{J~Richard Landis} {and}
  \bibinfo{person}{Gary~G Koch}.} \bibinfo{year}{1977}\natexlab{}.
\newblock \showarticletitle{The measurement of observer agreement for
  categorical data}.
\newblock \bibinfo{journal}{\emph{biometrics}} (\bibinfo{year}{1977}),
  \bibinfo{pages}{159--174}.
\newblock
\href{https://doi.org/10.2307/2529310}{doi:\nolinkurl{10.2307/2529310}}


\bibitem[LaToza and Myers(2010)]%
        {latoza2010hard}
\bibfield{author}{\bibinfo{person}{Thomas~D LaToza} {and}
  \bibinfo{person}{Brad~A Myers}.} \bibinfo{year}{2010}\natexlab{}.
\newblock \showarticletitle{Hard-to-answer questions about code}.
\newblock In \bibinfo{booktitle}{\emph{Evaluation and Usability of Programming
  Languages and Tools (PLATEAU)}}. \bibinfo{pages}{1--6}.
\newblock
\href{https://doi.org/10.1145/1937117.1937125}{doi:\nolinkurl{10.1145/1937117.1937125}}


\bibitem[LaToza and Myers(2011)]%
        {latoza2011visualizing}
\bibfield{author}{\bibinfo{person}{Thomas~D LaToza} {and}
  \bibinfo{person}{Brad~A Myers}.} \bibinfo{year}{2011}\natexlab{}.
\newblock \showarticletitle{Visualizing call graphs}. In
  \bibinfo{booktitle}{\emph{IEEE Symposium on Visual Languages and
  Human-Centric Computing (VL/HCC)}}. IEEE, \bibinfo{pages}{117--124}.
\newblock
\href{https://doi.org/10.1109/VLHCC.2011.6070388}{doi:\nolinkurl{10.1109/VLHCC.2011.6070388}}


\bibitem[Li and Coblenz(2026)]%
        {li2026grounded}
\bibfield{author}{\bibinfo{person}{Haolin Li} {and} \bibinfo{person}{Michael
  Coblenz}.} \bibinfo{year}{2026}\natexlab{}.
\newblock \showarticletitle{A Grounded theory of debugging in professional
  software engineering practice}.
\newblock \bibinfo{journal}{\emph{Proceedings of the ACM on Software
  Engineering}} \bibinfo{volume}{3}, \bibinfo{number}{FSE}
  (\bibinfo{year}{2026}), \bibinfo{pages}{1082--1103}.
\newblock
\href{https://doi.org/10.1145/3797077}{doi:\nolinkurl{10.1145/3797077}}


\bibitem[Li et~al\mbox{.}(2023)]%
        {li2023up}
\bibfield{author}{\bibinfo{person}{Xin Li}, \bibinfo{person}{Sima Behpour},
  \bibinfo{person}{Thang~Long Doan}, \bibinfo{person}{Wenbin He},
  \bibinfo{person}{Liang Gou}, {and} \bibinfo{person}{Liu Ren}.}
  \bibinfo{year}{2023}\natexlab{}.
\newblock \showarticletitle{UP-DP: Unsupervised prompt learning for data
  pre-selection with vision-language models}.
\newblock \bibinfo{journal}{\emph{Advances in Neural Information Processing
  Systems (NeurIPS)}}  \bibinfo{volume}{36} (\bibinfo{year}{2023}),
  \bibinfo{pages}{2188--2206}.
\newblock
\href{https://doi.org/10.52202/075280-0104}{doi:\nolinkurl{10.52202/075280-0104}}


\bibitem[Liang et~al\mbox{.}(2023)]%
        {liang2023qualitative}
\bibfield{author}{\bibinfo{person}{Jenny~T Liang}, \bibinfo{person}{Maryam
  Arab}, \bibinfo{person}{Minhyuk Ko}, \bibinfo{person}{Amy~J Ko}, {and}
  \bibinfo{person}{Thomas~D LaToza}.} \bibinfo{year}{2023}\natexlab{}.
\newblock \showarticletitle{A qualitative study on the implementation design
  decisions of developers}. In \bibinfo{booktitle}{\emph{IEEE/ACM 45th
  International Conference on Software Engineering (ICSE)}}.
  \bibinfo{pages}{435--447}.
\newblock
\href{https://doi.org/10.1109/ICSE48619.2023.00047}{doi:\nolinkurl{10.1109/ICSE48619.2023.00047}}


\bibitem[Liang et~al\mbox{.}(2026a)]%
        {liang2026understanding}
\bibfield{author}{\bibinfo{person}{Jenny~T Liang}, \bibinfo{person}{Titus
  Barik}, \bibinfo{person}{Jeffrey Nichols}, \bibinfo{person}{Eldon Schoop},
  {and} \bibinfo{person}{Ruijia Cheng}.} \bibinfo{year}{2026}\natexlab{a}.
\newblock \showarticletitle{Understanding user experiences of computer use
  agents: Design space and opportunities for building agent UX prototypes}.
\newblock \bibinfo{journal}{\emph{arXiv preprint arXiv:2510.04452}}
  (\bibinfo{year}{2026}).
\newblock


\bibitem[Liang et~al\mbox{.}(2025a)]%
        {liang2025tabletalk}
\bibfield{author}{\bibinfo{person}{Jenny~T Liang}, \bibinfo{person}{Aayush
  Kumar}, \bibinfo{person}{Yasharth Bajpai}, \bibinfo{person}{Sumit Gulwani},
  \bibinfo{person}{Vu Le}, \bibinfo{person}{Chris Parnin},
  \bibinfo{person}{Arjun Radhakrishna}, \bibinfo{person}{Ashish Tiwari},
  \bibinfo{person}{Emerson Murphy-Hill}, {and} \bibinfo{person}{Gustavo
  Soares}.} \bibinfo{year}{2025}\natexlab{a}.
\newblock \showarticletitle{Tabletalk: Scaffolding spreadsheet development with
  a language agent}.
\newblock \bibinfo{journal}{\emph{ACM Transactions on Computer-Human
  Interaction (TOCHI)}} \bibinfo{volume}{32}, \bibinfo{number}{6}
  (\bibinfo{year}{2025}), \bibinfo{pages}{1--49}.
\newblock
\href{https://doi.org/10.1145/3765286}{doi:\nolinkurl{10.1145/3765286}}


\bibitem[Liang et~al\mbox{.}(2025b)]%
        {liang2025prompts}
\bibfield{author}{\bibinfo{person}{Jenny~T Liang}, \bibinfo{person}{Melissa
  Lin}, \bibinfo{person}{Nikitha Rao}, {and} \bibinfo{person}{Brad~A Myers}.}
  \bibinfo{year}{2025}\natexlab{b}.
\newblock \showarticletitle{Prompts are programs too! Understanding how
  developers build software containing prompts}.
\newblock \bibinfo{journal}{\emph{Proceedings of the ACM on Software
  Engineering}} \bibinfo{volume}{2}, \bibinfo{number}{FSE}
  (\bibinfo{year}{2025}), \bibinfo{pages}{1591--1614}.
\newblock
\href{https://doi.org/10.1145/3729342}{doi:\nolinkurl{10.1145/3729342}}


\bibitem[Liang et~al\mbox{.}(2024)]%
        {liang2024large}
\bibfield{author}{\bibinfo{person}{Jenny~T Liang}, \bibinfo{person}{Chenyang
  Yang}, {and} \bibinfo{person}{Brad~A Myers}.}
  \bibinfo{year}{2024}\natexlab{}.
\newblock \showarticletitle{A large-scale survey on the usability of AI
  programming assistants: Successes and challenges}. In
  \bibinfo{booktitle}{\emph{IEEE/ACM International Conference on Software
  Engineering (ICSE)}}. \bibinfo{pages}{1--13}.
\newblock
\href{https://doi.org/10.1145/3597503.3608128}{doi:\nolinkurl{10.1145/3597503.3608128}}


\bibitem[Liang et~al\mbox{.}(2026b)]%
        {supplemental-materials}
\bibfield{author}{\bibinfo{person}{Jenny~T. Liang}, \bibinfo{person}{Chenyang
  Yang}, \bibinfo{person}{Agnia Sergeyuk}, \bibinfo{person}{Travis~D. Breaux},
  {and} \bibinfo{person}{Brad~A. Myers}.} \bibinfo{year}{2026}\natexlab{b}.
\newblock \bibinfo{title}{Supplemental Materials to "Understanding Developer
  Information Needs for Software with Embedded Natural Language Prompts"}.
\newblock
\newblock
\shownote{The supplemental materials are available on Figshare at
  \url{https://figshare.com/s/9418c4b1acfbc4e181a7}. We will make the
  supplemental materials publicly available upon acceptance}.


\bibitem[Liang et~al\mbox{.}(2022)]%
        {liang2022understanding}
\bibfield{author}{\bibinfo{person}{Jenny~T Liang}, \bibinfo{person}{Thomas
  Zimmermann}, {and} \bibinfo{person}{Denae Ford}.}
  \bibinfo{year}{2022}\natexlab{}.
\newblock \showarticletitle{Understanding skills for OSS communities on
  github}. In \bibinfo{booktitle}{\emph{ACM Joint European Software Engineering
  Conference and Symposium on the Foundations of Software Engineering
  (ESEC/FSE)}}. \bibinfo{pages}{170--182}.
\newblock
\href{https://doi.org/10.1145/3540250.3549082}{doi:\nolinkurl{10.1145/3540250.3549082}}


\bibitem[Lin et~al\mbox{.}(2026)]%
        {lin2026agentic}
\bibfield{author}{\bibinfo{person}{Jiahang Lin}, \bibinfo{person}{Shichun Liu},
  \bibinfo{person}{Chengjun Pan}, \bibinfo{person}{Lizhi Lin},
  \bibinfo{person}{Shihan Dou}, \bibinfo{person}{Zhiheng Xi},
  \bibinfo{person}{Xuanjing Huang}, \bibinfo{person}{Hang Yan},
  \bibinfo{person}{Zhenhua Han}, \bibinfo{person}{Tao Gui}, {et~al\mbox{.}}}
  \bibinfo{year}{2026}\natexlab{}.
\newblock \showarticletitle{Agentic harness engineering: Observability-driven
  automatic evolution of coding-agent harnesses}.
\newblock \bibinfo{journal}{\emph{arXiv preprint arXiv:2604.25850}}
  (\bibinfo{year}{2026}).
\newblock


\bibitem[Liu et~al\mbox{.}(2021b)]%
        {liu2021opportunities}
\bibfield{author}{\bibinfo{person}{Chao Liu}, \bibinfo{person}{Xin Xia},
  \bibinfo{person}{David Lo}, \bibinfo{person}{Cuiyun Gao},
  \bibinfo{person}{Xiaohu Yang}, {and} \bibinfo{person}{John Grundy}.}
  \bibinfo{year}{2021}\natexlab{b}.
\newblock \showarticletitle{Opportunities and challenges in code search tools}.
\newblock \bibinfo{journal}{\emph{ACM Computing Surveys (CSUR)}}
  \bibinfo{volume}{54}, \bibinfo{number}{9} (\bibinfo{year}{2021}),
  \bibinfo{pages}{1--40}.
\newblock
\href{https://doi.org/10.1145/3480027}{doi:\nolinkurl{10.1145/3480027}}


\bibitem[Liu et~al\mbox{.}(2021a)]%
        {liu2021characterizing}
\bibfield{author}{\bibinfo{person}{Jiakun Liu}, \bibinfo{person}{Sebastian
  Baltes}, \bibinfo{person}{Christoph Treude}, \bibinfo{person}{David Lo},
  \bibinfo{person}{Yun Zhang}, {and} \bibinfo{person}{Xin Xia}.}
  \bibinfo{year}{2021}\natexlab{a}.
\newblock \showarticletitle{Characterizing search activities on stack
  overflow}. In \bibinfo{booktitle}{\emph{ACM Joint Meeting on European
  Software Engineering Conference and Symposium on the Foundations of Software
  Engineering (FSE)}}. \bibinfo{pages}{919--931}.
\newblock
\href{https://doi.org/10.1145/3468264.3468582}{doi:\nolinkurl{10.1145/3468264.3468582}}


\bibitem[Liu et~al\mbox{.}(2024)]%
        {liu2024we}
\bibfield{author}{\bibinfo{person}{Michael~Xieyang Liu},
  \bibinfo{person}{Frederick Liu}, \bibinfo{person}{Alexander~J Fiannaca},
  \bibinfo{person}{Terry Koo}, \bibinfo{person}{Lucas Dixon},
  \bibinfo{person}{Michael Terry}, {and} \bibinfo{person}{Carrie~J Cai}.}
  \bibinfo{year}{2024}\natexlab{}.
\newblock \showarticletitle{"We need structured output": Towards user-centered
  constraints on large language model output}. In
  \bibinfo{booktitle}{\emph{Extended Abstracts of the CHI Conference on Human
  Factors in Computing Systems (CHI EA)}}. \bibinfo{pages}{1--9}.
\newblock
\href{https://doi.org/10.1145/3613905.3650756}{doi:\nolinkurl{10.1145/3613905.3650756}}


\bibitem[Liu et~al\mbox{.}(2023)]%
        {liu2023pre}
\bibfield{author}{\bibinfo{person}{Pengfei Liu}, \bibinfo{person}{Weizhe Yuan},
  \bibinfo{person}{Jinlan Fu}, \bibinfo{person}{Zhengbao Jiang},
  \bibinfo{person}{Hiroaki Hayashi}, {and} \bibinfo{person}{Graham Neubig}.}
  \bibinfo{year}{2023}\natexlab{}.
\newblock \showarticletitle{Pre-train, prompt, and predict: A systematic survey
  of prompting methods in natural language processing}.
\newblock \bibinfo{journal}{\emph{ACM Computing Surveys (CSUR)}}
  \bibinfo{volume}{55}, \bibinfo{number}{9} (\bibinfo{year}{2023}),
  \bibinfo{pages}{1--35}.
\newblock
\href{https://doi.org/10.1145/3560815}{doi:\nolinkurl{10.1145/3560815}}


\bibitem[Lu et~al\mbox{.}(2026)]%
        {lu2026engineering}
\bibfield{author}{\bibinfo{person}{Qinghua Lu}, \bibinfo{person}{Ahmed~E
  Hassan}, \bibinfo{person}{David Lo}, {and} \bibinfo{person}{Thomas
  Zimmermann}.} \bibinfo{year}{2026}\natexlab{}.
\newblock \showarticletitle{Engineering agentic systems: Toward
  verification-first {AI}ware engineering}.
\newblock \bibinfo{journal}{\emph{IEEE Software}} \bibinfo{volume}{43},
  \bibinfo{number}{5} (\bibinfo{year}{2026}), \bibinfo{pages}{22--29}.
\newblock
\href{https://doi.org/10.1109/MS.2026.3697911}{doi:\nolinkurl{10.1109/MS.2026.3697911}}


\bibitem[Ma et~al\mbox{.}(2023)]%
        {ma2023fairness}
\bibfield{author}{\bibinfo{person}{Huan Ma}, \bibinfo{person}{Changqing Zhang},
  \bibinfo{person}{Yatao Bian}, \bibinfo{person}{Lemao Liu},
  \bibinfo{person}{Zhirui Zhang}, \bibinfo{person}{Peilin Zhao},
  \bibinfo{person}{Shu Zhang}, \bibinfo{person}{Huazhu Fu},
  \bibinfo{person}{Qinghua Hu}, {and} \bibinfo{person}{Bingzhe Wu}.}
  \bibinfo{year}{2023}\natexlab{}.
\newblock \showarticletitle{Fairness-guided few-shot prompting for large
  language models}.
\newblock \bibinfo{journal}{\emph{Advances in Neural Information Processing
  Systems (NeurIPS)}}  \bibinfo{volume}{36} (\bibinfo{year}{2023}),
  \bibinfo{pages}{43136--43155}.
\newblock
\href{https://doi.org/10.52202/075280-1870}{doi:\nolinkurl{10.52202/075280-1870}}


\bibitem[Ma et~al\mbox{.}(2024)]%
        {ma2024my}
\bibfield{author}{\bibinfo{person}{Wanqin Ma}, \bibinfo{person}{Chenyang Yang},
  {and} \bibinfo{person}{Christian K{\"a}stner}.}
  \bibinfo{year}{2024}\natexlab{}.
\newblock \showarticletitle{(why) is my prompt getting worse? Rethinking
  regression testing for evolving llm apis}. In
  \bibinfo{booktitle}{\emph{Proceedings of the IEEE/ACM 3rd international
  conference on AI engineering-software engineering for AI}}.
  \bibinfo{pages}{166--171}.
\newblock
\href{https://doi.org/10.1145/3644815.3644950}{doi:\nolinkurl{10.1145/3644815.3644950}}


\bibitem[Maalej et~al\mbox{.}(2014)]%
        {maalej2014comprehension}
\bibfield{author}{\bibinfo{person}{Walid Maalej}, \bibinfo{person}{Rebecca
  Tiarks}, \bibinfo{person}{Tobias Roehm}, {and} \bibinfo{person}{Rainer
  Koschke}.} \bibinfo{year}{2014}\natexlab{}.
\newblock \showarticletitle{On the comprehension of program comprehension}.
\newblock \bibinfo{journal}{\emph{ACM Transactions on Software Engineering and
  Methodology (TOSEM)}} \bibinfo{volume}{23}, \bibinfo{number}{4}
  (\bibinfo{year}{2014}), \bibinfo{pages}{1--37}.
\newblock
\href{https://doi.org/10.1145/2622669}{doi:\nolinkurl{10.1145/2622669}}


\bibitem[Mao et~al\mbox{.}(2025)]%
        {mao2025prompts}
\bibfield{author}{\bibinfo{person}{Yuetian Mao}, \bibinfo{person}{Junjie He},
  {and} \bibinfo{person}{Chunyang Chen}.} \bibinfo{year}{2025}\natexlab{}.
\newblock \showarticletitle{From prompts to templates: A systematic prompt
  template analysis for real-world LLMapps}.
\newblock \bibinfo{journal}{\emph{arXiv preprint arXiv:2504.02052}}
  (\bibinfo{year}{2025}).
\newblock


\bibitem[Mishra et~al\mbox{.}(2025)]%
        {mishra2025promptaid}
\bibfield{author}{\bibinfo{person}{Aditi Mishra}, \bibinfo{person}{Bretho
  Danzy}, \bibinfo{person}{Utkarsh Soni}, \bibinfo{person}{Anjana Arunkumar},
  \bibinfo{person}{Jinbin Huang}, \bibinfo{person}{Bum~Chul Kwon}, {and}
  \bibinfo{person}{Chris Bryan}.} \bibinfo{year}{2025}\natexlab{}.
\newblock \showarticletitle{PromptAid: Visual prompt exploration, perturbation,
  testing and iteration for large language models}.
\newblock \bibinfo{journal}{\emph{IEEE Transactions on Visualization and
  Computer Graphics (TVCG)}} (\bibinfo{year}{2025}).
\newblock
\href{https://doi.org/10.1109/TVCG.2025.3535332}{doi:\nolinkurl{10.1109/TVCG.2025.3535332}}


\bibitem[Myers et~al\mbox{.}(2016)]%
        {myers2016programmers}
\bibfield{author}{\bibinfo{person}{Brad~A Myers}, \bibinfo{person}{Amy~J Ko},
  \bibinfo{person}{Thomas~D LaToza}, {and} \bibinfo{person}{YoungSeok Yoon}.}
  \bibinfo{year}{2016}\natexlab{}.
\newblock \showarticletitle{Programmers are users too: Human-centered methods
  for improving programming tools}.
\newblock \bibinfo{journal}{\emph{Computer}} \bibinfo{volume}{49},
  \bibinfo{number}{7} (\bibinfo{year}{2016}), \bibinfo{pages}{44--52}.
\newblock
\href{https://doi.org/10.1109/MC.2016.200}{doi:\nolinkurl{10.1109/MC.2016.200}}


\bibitem[Nachar et~al\mbox{.}(2008)]%
        {nachar2008mann}
\bibfield{author}{\bibinfo{person}{Nadim Nachar} {et~al\mbox{.}}}
  \bibinfo{year}{2008}\natexlab{}.
\newblock \showarticletitle{The Mann-Whitney U: A test for assessing whether
  two independent samples come from the same distribution}.
\newblock \bibinfo{journal}{\emph{Tutorials in Quantitative Methods for
  Psychology}} \bibinfo{volume}{4}, \bibinfo{number}{1} (\bibinfo{year}{2008}),
  \bibinfo{pages}{13--20}.
\newblock
\href{https://doi.org/10.20982/tqmp.04.1.p013}{doi:\nolinkurl{10.20982/tqmp.04.1.p013}}


\bibitem[Nahar et~al\mbox{.}(2024)]%
        {nahar2024beyond}
\bibfield{author}{\bibinfo{person}{Nadia Nahar}, \bibinfo{person}{Christian
  K{\"a}stner}, \bibinfo{person}{Jenna Butler}, \bibinfo{person}{Chris Parnin},
  \bibinfo{person}{Thomas Zimmermann}, {and} \bibinfo{person}{Christian Bird}.}
  \bibinfo{year}{2024}\natexlab{}.
\newblock \showarticletitle{Beyond the comfort zone: Emerging solutions to
  overcome challenges in integrating LLMs into software products}.
\newblock \bibinfo{journal}{\emph{arXiv preprint arXiv:2410.12071}}
  (\bibinfo{year}{2024}).
\newblock


\bibitem[Nashid et~al\mbox{.}(2023)]%
        {nashid2023retrieval}
\bibfield{author}{\bibinfo{person}{Noor Nashid}, \bibinfo{person}{Mifta
  Sintaha}, {and} \bibinfo{person}{Ali Mesbah}.}
  \bibinfo{year}{2023}\natexlab{}.
\newblock \showarticletitle{Retrieval-based prompt selection for code-related
  few-shot learning}. In \bibinfo{booktitle}{\emph{IEEE/ACM International
  Conference on Software Engineering (ICSE)}}. IEEE,
  \bibinfo{pages}{2450--2462}.
\newblock
\href{https://doi.org/10.1109/ICSE48619.2023.00205}{doi:\nolinkurl{10.1109/ICSE48619.2023.00205}}


\bibitem[Ning et~al\mbox{.}(2026)]%
        {ning2026code}
\bibfield{author}{\bibinfo{person}{Xuying Ning}, \bibinfo{person}{Katherine
  Tieu}, \bibinfo{person}{Dongqi Fu}, \bibinfo{person}{Tianxin Wei},
  \bibinfo{person}{Zihao Li}, \bibinfo{person}{Yuanchen Bei},
  \bibinfo{person}{Jiaru Zou}, \bibinfo{person}{Mengting Ai},
  \bibinfo{person}{Zhining Liu}, \bibinfo{person}{Ting-Wei Li},
  {et~al\mbox{.}}} \bibinfo{year}{2026}\natexlab{}.
\newblock \showarticletitle{Code as agent harness}.
\newblock \bibinfo{journal}{\emph{arXiv preprint arXiv:2605.18747}}
  (\bibinfo{year}{2026}).
\newblock


\bibitem[Nisbett and Wilson(1977)]%
        {nisbett1977telling}
\bibfield{author}{\bibinfo{person}{Richard~E Nisbett} {and}
  \bibinfo{person}{Timothy~D Wilson}.} \bibinfo{year}{1977}\natexlab{}.
\newblock \showarticletitle{Telling more than we can know: Verbal reports on
  mental processes.}
\newblock \bibinfo{journal}{\emph{Psychological review}} \bibinfo{volume}{84},
  \bibinfo{number}{3} (\bibinfo{year}{1977}), \bibinfo{pages}{231}.
\newblock
\href{https://doi.org/10.1037/0033-295X.84.3.231}{doi:\nolinkurl{10.1037/0033-295X.84.3.231}}


\bibitem[Parnin et~al\mbox{.}(2025)]%
        {parnin2025building}
\bibfield{author}{\bibinfo{person}{Chris Parnin}, \bibinfo{person}{Gustavo
  Soares}, \bibinfo{person}{Rahul Pandita}, \bibinfo{person}{Sumit Gulwani},
  \bibinfo{person}{Jessica Rich}, {and} \bibinfo{person}{Austin~Z Henley}.}
  \bibinfo{year}{2025}\natexlab{}.
\newblock \showarticletitle{Building your own product copilot: Challenges,
  opportunities, and needs}. In \bibinfo{booktitle}{\emph{IEEE International
  Conference on Software Analysis, Evolution and Reengineering (SANER)}}. IEEE,
  \bibinfo{pages}{338--348}.
\newblock
\href{https://doi.org/10.1109/SANER64311.2025.00039}{doi:\nolinkurl{10.1109/SANER64311.2025.00039}}


\bibitem[Phokela et~al\mbox{.}(2023)]%
        {phokela2023smart}
\bibfield{author}{\bibinfo{person}{Kanchanjot~Kaur Phokela},
  \bibinfo{person}{Samarth Sikand}, \bibinfo{person}{Kapil Singi},
  \bibinfo{person}{Kuntal Dey}, \bibinfo{person}{Vibhu~Saujanya Sharma}, {and}
  \bibinfo{person}{Vikrant Kaulgud}.} \bibinfo{year}{2023}\natexlab{}.
\newblock \showarticletitle{Smart prompt advisor: Multi-objective prompt
  framework for consistency and best practices}. In
  \bibinfo{booktitle}{\emph{IEEE/ACM International Conference on Automated
  Software Engineering (ASE)}}. \bibinfo{pages}{1846--1848}.
\newblock
\href{https://doi.org/10.1109/ASE56229.2023.00019}{doi:\nolinkurl{10.1109/ASE56229.2023.00019}}


\bibitem[Pister et~al\mbox{.}(2024)]%
        {pister2024promptset}
\bibfield{author}{\bibinfo{person}{Kaiser Pister},
  \bibinfo{person}{Dhruba~Jyoti Paul}, \bibinfo{person}{Ishan Joshi}, {and}
  \bibinfo{person}{Patrick Brophy}.} \bibinfo{year}{2024}\natexlab{}.
\newblock \showarticletitle{Promptset: A programmer's prompting dataset}. In
  \bibinfo{booktitle}{\emph{International Workshop on Large Language Models for
  Code (LLM4Code)}}. \bibinfo{pages}{62--69}.
\newblock
\href{https://doi.org/10.1145/3643795.3648395}{doi:\nolinkurl{10.1145/3643795.3648395}}


\bibitem[Qian et~al\mbox{.}(2024)]%
        {qian2024understanding}
\bibfield{author}{\bibinfo{person}{Crystal Qian}, \bibinfo{person}{Emily Reif},
  {and} \bibinfo{person}{Minsuk Kahng}.} \bibinfo{year}{2024}\natexlab{}.
\newblock \showarticletitle{Understanding the dataset practitioners behind
  large language models}. In \bibinfo{booktitle}{\emph{ACM CHI Conference on
  Human Factors in Computing Systems Extended Abstracts (CHI EA)}}.
  \bibinfo{pages}{1--7}.
\newblock
\href{https://doi.org/10.1145/3613905.3651007}{doi:\nolinkurl{10.1145/3613905.3651007}}


\bibitem[Reza et~al\mbox{.}(2025)]%
        {reza2025prompthive}
\bibfield{author}{\bibinfo{person}{Mohi Reza}, \bibinfo{person}{Ioannis
  Anastasopoulos}, \bibinfo{person}{Shreya Bhandari}, {and}
  \bibinfo{person}{Zachary~A Pardos}.} \bibinfo{year}{2025}\natexlab{}.
\newblock \showarticletitle{PromptHive: Bringing subject matter experts back to
  the forefront with collaborative prompt engineering for educational content
  creation}. In \bibinfo{booktitle}{\emph{CHI Conference on Human Factors in
  Computing Systems (CHI)}}. \bibinfo{pages}{1--22}.
\newblock
\href{https://doi.org/10.1145/3706598.3714051}{doi:\nolinkurl{10.1145/3706598.3714051}}


\bibitem[Rubin et~al\mbox{.}(2022)]%
        {rubin2022learning}
\bibfield{author}{\bibinfo{person}{Ohad Rubin}, \bibinfo{person}{Jonathan
  Herzig}, {and} \bibinfo{person}{Jonathan Berant}.}
  \bibinfo{year}{2022}\natexlab{}.
\newblock \showarticletitle{Learning to retrieve prompts for in-context
  learning}. In \bibinfo{booktitle}{\emph{Conference of the North American
  Chapter of the Association for Computational Linguistics: Human Language
  Technologies (NAACL)}}. \bibinfo{pages}{2655--2671}.
\newblock
\href{https://doi.org/10.18653/v1/2022.naacl-main.191}{doi:\nolinkurl{10.18653/v1/2022.naacl-main.191}}


\bibitem[Salda{\~n}a(2009)]%
        {saldana2009coding}
\bibfield{author}{\bibinfo{person}{Johnny Salda{\~n}a}.}
  \bibinfo{year}{2009}\natexlab{}.
\newblock \bibinfo{booktitle}{\emph{The Coding Manual for Qualitative
  Researchers}}.
\newblock \bibinfo{publisher}{SAGE Publications}.
\newblock
\showISBNx{9781529731750}


\bibitem[Santana et~al\mbox{.}(2025)]%
        {santana2025can}
\bibfield{author}{\bibinfo{person}{Vagner Figueredo~de Santana},
  \bibinfo{person}{Sara Berger}, \bibinfo{person}{Tiago Machado},
  \bibinfo{person}{Maysa Malfiza~Garcia de Macedo},
  \bibinfo{person}{Cassia~Sampaio Sanctos}, \bibinfo{person}{Lemara Williams},
  {and} \bibinfo{person}{Zhaoqing Wu}.} \bibinfo{year}{2025}\natexlab{}.
\newblock \showarticletitle{Can LLMs recommend more responsible prompts?}. In
  \bibinfo{booktitle}{\emph{International Conference on Intelligent User
  Interfaces (IUI)}}. \bibinfo{pages}{298--313}.
\newblock
\href{https://doi.org/10.1145/3708359.3712137}{doi:\nolinkurl{10.1145/3708359.3712137}}


\bibitem[Shao et~al\mbox{.}(2023)]%
        {shao2023synthetic}
\bibfield{author}{\bibinfo{person}{Zhihong Shao}, \bibinfo{person}{Yeyun Gong},
  \bibinfo{person}{Yelong Shen}, \bibinfo{person}{Minlie Huang},
  \bibinfo{person}{Nan Duan}, {and} \bibinfo{person}{Weizhu Chen}.}
  \bibinfo{year}{2023}\natexlab{}.
\newblock \showarticletitle{Synthetic prompting: Generating chain-of-thought
  demonstrations for large language models}. In
  \bibinfo{booktitle}{\emph{International Conference on Machine Learning
  (ICML)}}. PMLR, \bibinfo{pages}{30706--30775}.
\newblock


\bibitem[Shi et~al\mbox{.}(2024)]%
        {shi2024efficient}
\bibfield{author}{\bibinfo{person}{Chengshuai Shi}, \bibinfo{person}{Kun Yang},
  \bibinfo{person}{Zihan Chen}, \bibinfo{person}{Jundong Li},
  \bibinfo{person}{Jing Yang}, {and} \bibinfo{person}{Cong Shen}.}
  \bibinfo{year}{2024}\natexlab{}.
\newblock \showarticletitle{Efficient prompt optimization through the lens of
  best arm identification}.
\newblock \bibinfo{journal}{\emph{Advances in Neural Information Processing
  Systems (NeurIPS)}}  \bibinfo{volume}{37} (\bibinfo{year}{2024}),
  \bibinfo{pages}{99646--99685}.
\newblock
\href{https://doi.org/10.52202/079017-3161}{doi:\nolinkurl{10.52202/079017-3161}}


\bibitem[Shrivastava et~al\mbox{.}(2023)]%
        {shrivastava2023repository}
\bibfield{author}{\bibinfo{person}{Disha Shrivastava}, \bibinfo{person}{Hugo
  Larochelle}, {and} \bibinfo{person}{Daniel Tarlow}.}
  \bibinfo{year}{2023}\natexlab{}.
\newblock \showarticletitle{Repository-level prompt generation for large
  language models of code}. In \bibinfo{booktitle}{\emph{International
  Conference on Machine Learning (ICML)}}. PMLR, \bibinfo{pages}{31693--31715}.
\newblock


\bibitem[Sillito et~al\mbox{.}(2006)]%
        {sillito2006questions}
\bibfield{author}{\bibinfo{person}{Jonathan Sillito}, \bibinfo{person}{Gail~C
  Murphy}, {and} \bibinfo{person}{Kris De~Volder}.}
  \bibinfo{year}{2006}\natexlab{}.
\newblock \showarticletitle{Questions programmers ask during software evolution
  tasks}. In \bibinfo{booktitle}{\emph{ACM International Symposium on
  Foundations of Software Engineering (FSE)}}. \bibinfo{pages}{23--34}.
\newblock
\href{https://doi.org/10.1145/1181775.1181779}{doi:\nolinkurl{10.1145/1181775.1181779}}


\bibitem[Sillito et~al\mbox{.}(2008)]%
        {sillito2008asking}
\bibfield{author}{\bibinfo{person}{Jonathan Sillito}, \bibinfo{person}{Gail~C
  Murphy}, {and} \bibinfo{person}{Kris De~Volder}.}
  \bibinfo{year}{2008}\natexlab{}.
\newblock \showarticletitle{Asking and answering questions during a programming
  change task}.
\newblock \bibinfo{journal}{\emph{IEEE Transactions on Software Engineering
  (TSE)}} \bibinfo{volume}{34}, \bibinfo{number}{4} (\bibinfo{year}{2008}),
  \bibinfo{pages}{434--451}.
\newblock
\href{https://doi.org/10.1109/TSE.2008.26}{doi:\nolinkurl{10.1109/TSE.2008.26}}


\bibitem[Strobelt et~al\mbox{.}(2022)]%
        {strobelt2022interactive}
\bibfield{author}{\bibinfo{person}{Hendrik Strobelt}, \bibinfo{person}{Albert
  Webson}, \bibinfo{person}{Victor Sanh}, \bibinfo{person}{Benjamin Hoover},
  \bibinfo{person}{Johanna Beyer}, \bibinfo{person}{Hanspeter Pfister}, {and}
  \bibinfo{person}{Alexander~M Rush}.} \bibinfo{year}{2022}\natexlab{}.
\newblock \showarticletitle{Interactive and visual prompt engineering for
  ad-hoc task adaptation with large language models}.
\newblock \bibinfo{journal}{\emph{IEEE Transactions on Visualization and
  Computer Graphics (TVCG)}} \bibinfo{volume}{29}, \bibinfo{number}{1}
  (\bibinfo{year}{2022}), \bibinfo{pages}{1146--1156}.
\newblock
\href{https://doi.org/10.1109/TVCG.2022.3209479}{doi:\nolinkurl{10.1109/TVCG.2022.3209479}}


\bibitem[Suri(2011)]%
        {suri2011purposeful}
\bibfield{author}{\bibinfo{person}{Harsh Suri}.}
  \bibinfo{year}{2011}\natexlab{}.
\newblock \showarticletitle{Purposeful sampling in qualitative research
  synthesis}.
\newblock \bibinfo{journal}{\emph{Qualitative research journal}}
  \bibinfo{volume}{11}, \bibinfo{number}{2} (\bibinfo{year}{2011}),
  \bibinfo{pages}{63--75}.
\newblock
\href{https://doi.org/10.3316/QRJ1102063}{doi:\nolinkurl{10.3316/QRJ1102063}}


\bibitem[Tafreshipour et~al\mbox{.}(2025)]%
        {tafreshipour2025prompting}
\bibfield{author}{\bibinfo{person}{Mahan Tafreshipour}, \bibinfo{person}{Aaron
  Imani}, \bibinfo{person}{Eric Huang}, \bibinfo{person}{Eduardo~Santana de
  Almeida}, \bibinfo{person}{Thomas Zimmermann}, {and}
  \bibinfo{person}{Iftekhar Ahmed}.} \bibinfo{year}{2025}\natexlab{}.
\newblock \showarticletitle{Prompting in the wild: An empirical study of prompt
  evolution in software repositories}. In \bibinfo{booktitle}{\emph{IEEE/ACM
  22nd International Conference on Mining Software Repositories (MSR)}}. IEEE,
  \bibinfo{pages}{686--698}.
\newblock
\href{https://doi.org/10.1109/MSR66628.2025.00106}{doi:\nolinkurl{10.1109/MSR66628.2025.00106}}


\bibitem[Villamizar et~al\mbox{.}(2025)]%
        {villamizar2025prompts}
\bibfield{author}{\bibinfo{person}{Hugo Villamizar}, \bibinfo{person}{Jannik
  Fischbach}, \bibinfo{person}{Alexander Korn}, \bibinfo{person}{Andreas
  Vogelsang}, {and} \bibinfo{person}{Daniel Mendez}.}
  \bibinfo{year}{2025}\natexlab{}.
\newblock \showarticletitle{Prompts as software engineering artifacts: A
  research agenda and preliminary findings}. In
  \bibinfo{booktitle}{\emph{International Conference on Product-Focused
  Software Process Improvement}}. Springer, \bibinfo{pages}{470--478}.
\newblock
\href{https://doi.org/10.1007/978-3-032-12089-2_32}{doi:\nolinkurl{10.1007/978-3-032-12089-2_32}}


\bibitem[Vogelsang(2024)]%
        {vogelsang2024specifications}
\bibfield{author}{\bibinfo{person}{Andreas Vogelsang}.}
  \bibinfo{year}{2024}\natexlab{}.
\newblock \showarticletitle{From specifications to prompts: On the future of
  generative large language models in requirements engineering}.
\newblock \bibinfo{journal}{\emph{IEEE Software}} \bibinfo{volume}{41},
  \bibinfo{number}{5} (\bibinfo{year}{2024}), \bibinfo{pages}{9--13}.
\newblock
\href{https://doi.org/10.1109/MS.2024.3410712}{doi:\nolinkurl{10.1109/MS.2024.3410712}}


\bibitem[Vogelsang et~al\mbox{.}(2025)]%
        {vogelsang2025impact}
\bibfield{author}{\bibinfo{person}{Andreas Vogelsang},
  \bibinfo{person}{Alexander Korn}, \bibinfo{person}{Giovanna Broccia},
  \bibinfo{person}{Alessio Ferrari}, \bibinfo{person}{Jannik Fischbach}, {and}
  \bibinfo{person}{Chetan Arora}.} \bibinfo{year}{2025}\natexlab{}.
\newblock \showarticletitle{On the impact of requirements smells in prompts:
  The case of automated traceability}. In \bibinfo{booktitle}{\emph{IEEE/ACM
  International Conference on Software Engineering: New Ideas and Emerging
  Results (ICSE-NIER)}}. IEEE, \bibinfo{pages}{51--55}.
\newblock
\href{https://doi.org/10.1109/ICSE-NIER66352.2025.00016}{doi:\nolinkurl{10.1109/ICSE-NIER66352.2025.00016}}


\bibitem[Wan et~al\mbox{.}(2019)]%
        {wan2019does}
\bibfield{author}{\bibinfo{person}{Zhiyuan Wan}, \bibinfo{person}{Xin Xia},
  \bibinfo{person}{David Lo}, {and} \bibinfo{person}{Gail~C Murphy}.}
  \bibinfo{year}{2019}\natexlab{}.
\newblock \showarticletitle{How does machine learning change software
  development practices?}
\newblock \bibinfo{journal}{\emph{IEEE Transactions on Software Engineering
  (TSE)}} \bibinfo{volume}{47}, \bibinfo{number}{9} (\bibinfo{year}{2019}),
  \bibinfo{pages}{1857--1871}.
\newblock
\href{https://doi.org/10.1109/TSE.2019.2937083}{doi:\nolinkurl{10.1109/TSE.2019.2937083}}


\bibitem[Wang et~al\mbox{.}(2024a)]%
        {wang2024one}
\bibfield{author}{\bibinfo{person}{Ruochen Wang}, \bibinfo{person}{Sohyun An},
  \bibinfo{person}{Minhao Cheng}, \bibinfo{person}{Tianyi Zhou},
  \bibinfo{person}{Sung~Ju Hwang}, {and} \bibinfo{person}{Cho-Jui Hsieh}.}
  \bibinfo{year}{2024}\natexlab{a}.
\newblock \showarticletitle{One prompt is not enough: Automated construction of
  a mixture-of-expert prompts}. In \bibinfo{booktitle}{\emph{International
  Conference on Machine Learning (ICML)}}.
\newblock


\bibitem[Wang et~al\mbox{.}(2024c)]%
        {wang2024on}
\bibfield{author}{\bibinfo{person}{Ruochen Wang}, \bibinfo{person}{Ting Liu},
  \bibinfo{person}{Cho-Jui Hsieh}, {and} \bibinfo{person}{Boqing Gong}.}
  \bibinfo{year}{2024}\natexlab{c}.
\newblock \showarticletitle{On discrete prompt optimization for diffusion
  models}. In \bibinfo{booktitle}{\emph{International Conference on Machine
  Learning (ICML)}}.
\newblock


\bibitem[Wang et~al\mbox{.}(2023)]%
        {wang2023reprompt}
\bibfield{author}{\bibinfo{person}{Yunlong Wang}, \bibinfo{person}{Shuyuan
  Shen}, {and} \bibinfo{person}{Brian~Y Lim}.} \bibinfo{year}{2023}\natexlab{}.
\newblock \showarticletitle{Reprompt: Automatic prompt editing to refine
  ai-generative art towards precise expressions}. In
  \bibinfo{booktitle}{\emph{ACM CHI Conference on Human Factors in Computing
  Systems (CHI)}}. \bibinfo{pages}{1--29}.
\newblock
\href{https://doi.org/10.1145/3544548.3581402}{doi:\nolinkurl{10.1145/3544548.3581402}}


\bibitem[Wang et~al\mbox{.}(2024b)]%
        {wang2024promptcharm}
\bibfield{author}{\bibinfo{person}{Zhijie Wang}, \bibinfo{person}{Yuheng
  Huang}, \bibinfo{person}{Da Song}, \bibinfo{person}{Lei Ma}, {and}
  \bibinfo{person}{Tianyi Zhang}.} \bibinfo{year}{2024}\natexlab{b}.
\newblock \showarticletitle{Promptcharm: Text-to-image generation through
  multi-modal prompting and refinement}. In \bibinfo{booktitle}{\emph{CHI
  Conference on Human Factors in Computing Systems (CHI)}}.
  \bibinfo{pages}{1--21}.
\newblock
\href{https://doi.org/10.1145/3613904.3642803}{doi:\nolinkurl{10.1145/3613904.3642803}}


\bibitem[White et~al\mbox{.}(2023)]%
        {white2023prompt}
\bibfield{author}{\bibinfo{person}{Jules White}, \bibinfo{person}{Quchen Fu},
  \bibinfo{person}{Sam Hays}, \bibinfo{person}{Michael Sandborn},
  \bibinfo{person}{Carlos Olea}, \bibinfo{person}{Henry Gilbert},
  \bibinfo{person}{Ashraf Elnashar}, \bibinfo{person}{Jesse Spencer-Smith},
  {and} \bibinfo{person}{Douglas~C Schmidt}.} \bibinfo{year}{2023}\natexlab{}.
\newblock \showarticletitle{A prompt pattern catalog to enhance prompt
  engineering with chatgpt}.
\newblock \bibinfo{journal}{\emph{arXiv preprint arXiv:2302.11382}}
  (\bibinfo{year}{2023}).
\newblock


\bibitem[White et~al\mbox{.}(2024)]%
        {white2024chatgpt}
\bibfield{author}{\bibinfo{person}{Jules White}, \bibinfo{person}{Sam Hays},
  \bibinfo{person}{Quchen Fu}, \bibinfo{person}{Jesse Spencer-Smith}, {and}
  \bibinfo{person}{Douglas~C Schmidt}.} \bibinfo{year}{2024}\natexlab{}.
\newblock \showarticletitle{Chatgpt prompt patterns for improving code quality,
  refactoring, requirements elicitation, and software design}.
\newblock In \bibinfo{booktitle}{\emph{Generative ai for effective software
  development}}. \bibinfo{publisher}{Springer}, \bibinfo{pages}{71--108}.
\newblock
\href{https://doi.org/10.1007/978-3-031-55642-5_4}{doi:\nolinkurl{10.1007/978-3-031-55642-5_4}}


\bibitem[Wu et~al\mbox{.}(2022a)]%
        {wu2022promptchainer}
\bibfield{author}{\bibinfo{person}{Tongshuang Wu}, \bibinfo{person}{Ellen
  Jiang}, \bibinfo{person}{Aaron Donsbach}, \bibinfo{person}{Jeff Gray},
  \bibinfo{person}{Alejandra Molina}, \bibinfo{person}{Michael Terry}, {and}
  \bibinfo{person}{Carrie~J Cai}.} \bibinfo{year}{2022}\natexlab{a}.
\newblock \showarticletitle{Promptchainer: Chaining large language model
  prompts through visual programming}. In \bibinfo{booktitle}{\emph{Extended
  Abstracts of CHI Conference on Human Factors in Computing Systems (CHI EA)}}.
  \bibinfo{pages}{1--10}.
\newblock
\href{https://doi.org/10.1145/3491101.3519729}{doi:\nolinkurl{10.1145/3491101.3519729}}


\bibitem[Wu et~al\mbox{.}(2022b)]%
        {wu2022ai}
\bibfield{author}{\bibinfo{person}{Tongshuang Wu}, \bibinfo{person}{Michael
  Terry}, {and} \bibinfo{person}{Carrie~Jun Cai}.}
  \bibinfo{year}{2022}\natexlab{b}.
\newblock \showarticletitle{Ai chains: Transparent and controllable human-ai
  interaction by chaining large language model prompts}. In
  \bibinfo{booktitle}{\emph{CHI Conference on Human Factors in Computing
  Systems (CHI)}}. \bibinfo{pages}{1--22}.
\newblock
\href{https://doi.org/10.1145/3491102.3517582}{doi:\nolinkurl{10.1145/3491102.3517582}}


\bibitem[Wu et~al\mbox{.}(2026)]%
        {wu2026harnessdev}
\bibfield{author}{\bibinfo{person}{Yuhao Wu}, \bibinfo{person}{Jingyuan Zhang},
  \bibinfo{person}{Jiajun Shi}, \bibinfo{person}{Xinping Lei},
  \bibinfo{person}{Qingshui Gu}, \bibinfo{person}{Yuxuan Zhang},
  \bibinfo{person}{Zexuan Wang}, \bibinfo{person}{Chen He},
  \bibinfo{person}{Chen Huang}, \bibinfo{person}{Maojia Song}, {et~al\mbox{.}}}
  \bibinfo{year}{2026}\natexlab{}.
\newblock \showarticletitle{HarnessDev: Can {LLMs} create and evolve their own
  agent harness?}
\newblock \bibinfo{journal}{\emph{arXiv preprint arXiv:2609.01437}}
  (\bibinfo{year}{2026}).
\newblock


\bibitem[Wu et~al\mbox{.}(2024)]%
        {wu2024prompt}
\bibfield{author}{\bibinfo{person}{Zhaoxuan Wu}, \bibinfo{person}{Xiaoqiang
  Lin}, \bibinfo{person}{Zhongxiang Dai}, \bibinfo{person}{Wenyang Hu},
  \bibinfo{person}{Yao Shu}, \bibinfo{person}{See-Kiong Ng},
  \bibinfo{person}{Patrick Jaillet}, {and} \bibinfo{person}{Bryan Kian~Hsiang
  Low}.} \bibinfo{year}{2024}\natexlab{}.
\newblock \showarticletitle{Prompt optimization with ease? efficient
  ordering-aware automated selection of exemplars}.
\newblock \bibinfo{journal}{\emph{Advances in Neural Information Processing
  Systems (NeurIPS)}}  \bibinfo{volume}{37} (\bibinfo{year}{2024}),
  \bibinfo{pages}{122706--122740}.
\newblock
\href{https://doi.org/10.52202/079017-3899}{doi:\nolinkurl{10.52202/079017-3899}}


\bibitem[Xiao et~al\mbox{.}(2024)]%
        {xiao2024devgpt}
\bibfield{author}{\bibinfo{person}{Tao Xiao}, \bibinfo{person}{Christoph
  Treude}, \bibinfo{person}{Hideaki Hata}, {and} \bibinfo{person}{Kenichi
  Matsumoto}.} \bibinfo{year}{2024}\natexlab{}.
\newblock \showarticletitle{Devgpt: Studying developer-chatgpt conversations}.
  In \bibinfo{booktitle}{\emph{International Conference on Mining Software
  Repositories (MSR)}}. \bibinfo{pages}{227--230}.
\newblock
\href{https://doi.org/10.1145/3643991.3648400}{doi:\nolinkurl{10.1145/3643991.3648400}}


\bibitem[Xu et~al\mbox{.}(2024b)]%
        {xu2024towards}
\bibfield{author}{\bibinfo{person}{Chengming Xu}, \bibinfo{person}{Chen Liu},
  \bibinfo{person}{Yikai Wang}, \bibinfo{person}{Yuan Yao}, {and}
  \bibinfo{person}{Yanwei Fu}.} \bibinfo{year}{2024}\natexlab{b}.
\newblock \showarticletitle{Towards global optimal visual in-context learning
  prompt selection}.
\newblock \bibinfo{journal}{\emph{Advances in Neural Information Processing
  Systems (NeurIPS)}}  \bibinfo{volume}{37} (\bibinfo{year}{2024}),
  \bibinfo{pages}{74945--74965}.
\newblock
\href{https://doi.org/10.52202/079017-2384}{doi:\nolinkurl{10.52202/079017-2384}}


\bibitem[Xu et~al\mbox{.}(2024a)]%
        {xu2024reprompting}
\bibfield{author}{\bibinfo{person}{Weijia Xu}, \bibinfo{person}{Andrzej
  Banburski}, {and} \bibinfo{person}{Nebojsa Jojic}.}
  \bibinfo{year}{2024}\natexlab{a}.
\newblock \showarticletitle{Reprompting: Automated chain-of-thought prompt
  inference through Gibbs sampling}. In \bibinfo{booktitle}{\emph{International
  Conference on Machine Learning (ICML)}}.
\newblock


\bibitem[Yang et~al\mbox{.}(2026a)]%
        {yang2026prompts}
\bibfield{author}{\bibinfo{person}{Chenyang Yang}, \bibinfo{person}{Yike Shi},
  \bibinfo{person}{Qianou Ma}, \bibinfo{person}{Michael~Xieyang Liu},
  \bibinfo{person}{Christian K{\"a}stner}, {and} \bibinfo{person}{Tongshuang
  Wu}.} \bibinfo{year}{2026}\natexlab{a}.
\newblock \showarticletitle{What prompts don’t say: Understanding and
  managing underspecification in llm prompts}. In
  \bibinfo{booktitle}{\emph{Findings of the Association for Computational
  Linguistics: ACL}}. \bibinfo{pages}{9072--9101}.
\newblock
\href{https://doi.org/10.18653/v1/2026.findings-acl.441}{doi:\nolinkurl{10.18653/v1/2026.findings-acl.441}}


\bibitem[Yang et~al\mbox{.}(2026b)]%
        {yang2026better}
\bibfield{author}{\bibinfo{person}{Chenyang Yang}, \bibinfo{person}{Xinran
  Zhao}, \bibinfo{person}{Tongshuang Wu}, {and} \bibinfo{person}{Christian
  K{\"a}stner}.} \bibinfo{year}{2026}\natexlab{b}.
\newblock \showarticletitle{Better harnesses, smaller models: Building 90\%
  cheaper agents via automated harness adaptation}.
\newblock \bibinfo{journal}{\emph{arXiv preprint arXiv:2607.08938}}
  (\bibinfo{year}{2026}).
\newblock


\bibitem[Yang et~al\mbox{.}(2016)]%
        {yang2016query}
\bibfield{author}{\bibinfo{person}{Di Yang}, \bibinfo{person}{Aftab Hussain},
  {and} \bibinfo{person}{Cristina~Videira Lopes}.}
  \bibinfo{year}{2016}\natexlab{}.
\newblock \showarticletitle{From query to usable code: an analysis of stack
  overflow code snippets}. In \bibinfo{booktitle}{\emph{Proceedings of the 13th
  International Conference on Mining Software Repositories}}.
  \bibinfo{pages}{391--402}.
\newblock
\href{https://doi.org/10.1145/2901739.2901767}{doi:\nolinkurl{10.1145/2901739.2901767}}


\bibitem[Ye et~al\mbox{.}(2024)]%
        {ye-etal-2024-prompt}
\bibfield{author}{\bibinfo{person}{Qinyuan Ye}, \bibinfo{person}{Mohamed
  Ahmed}, \bibinfo{person}{Reid Pryzant}, {and} \bibinfo{person}{Fereshte
  Khani}.} \bibinfo{year}{2024}\natexlab{}.
\newblock \showarticletitle{Prompt engineering a prompt engineer}. In
  \bibinfo{booktitle}{\emph{Findings of the Association for Computational
  Linguistics}}, \bibfield{editor}{\bibinfo{person}{Lun-Wei Ku},
  \bibinfo{person}{Andre Martins}, {and} \bibinfo{person}{Vivek Srikumar}}
  (Eds.). \bibinfo{pages}{355--385}.
\newblock
\href{https://doi.org/10.18653/v1/2024.findings-acl.21}{doi:\nolinkurl{10.18653/v1/2024.findings-acl.21}}


\bibitem[Yoon et~al\mbox{.}(2013)]%
        {yoon2013visualization}
\bibfield{author}{\bibinfo{person}{YoungSeok Yoon}, \bibinfo{person}{Brad~A
  Myers}, {and} \bibinfo{person}{Sebon Koo}.} \bibinfo{year}{2013}\natexlab{}.
\newblock \showarticletitle{Visualization of fine-grained code change history}.
  In \bibinfo{booktitle}{\emph{IEEE Symposium on Visual Languages and
  Human-Centric Computing (VL/HCC)}}. IEEE, \bibinfo{pages}{119--126}.
\newblock
\href{https://doi.org/10.1109/VLHCC.2013.6645254}{doi:\nolinkurl{10.1109/VLHCC.2013.6645254}}


\bibitem[Zamfirescu-Pereira et~al\mbox{.}(2023a)]%
        {zamfirescu2023conversation}
\bibfield{author}{\bibinfo{person}{JD Zamfirescu-Pereira},
  \bibinfo{person}{Bjoern Hartmann}, {and} \bibinfo{person}{Qian Yang}.}
  \bibinfo{year}{2023}\natexlab{a}.
\newblock \showarticletitle{Conversation regression testing: A design technique
  for prototyping generalizable prompt strategies for pre-trained language
  models}.
\newblock \bibinfo{journal}{\emph{arXiv preprint arXiv:2302.03154}}
  (\bibinfo{year}{2023}).
\newblock


\bibitem[Zamfirescu-Pereira et~al\mbox{.}(2023b)]%
        {zamfirescu2023johnny}
\bibfield{author}{\bibinfo{person}{J~Diego Zamfirescu-Pereira},
  \bibinfo{person}{Richmond~Y Wong}, \bibinfo{person}{Bjoern Hartmann}, {and}
  \bibinfo{person}{Qian Yang}.} \bibinfo{year}{2023}\natexlab{b}.
\newblock \showarticletitle{Why Johnny can’t prompt: How non-AI experts try
  (and fail) to design LLM prompts}. In \bibinfo{booktitle}{\emph{CHI
  Conference on Human Factors in Computing Systems (CHI)}}.
  \bibinfo{pages}{1--21}.
\newblock
\href{https://doi.org/10.1145/3544548.3581388}{doi:\nolinkurl{10.1145/3544548.3581388}}


\end{thebibliography}

\end{document}